\documentclass[12pt,journal,draftcls,letterpaper,onecolumn]{IEEEtran}
\usepackage{ucs}
\usepackage[utf8x]{inputenc} 
\usepackage[cmex10]{amsmath}
\usepackage{cases}
\usepackage{cite, amsfonts, amssymb, amsthm, bm, bbm, graphicx, relsize, multirow, booktabs, color, colortbl,url,soul}
\usepackage[american]{babel}
\usepackage[T1]{fontenc}
\usepackage{algorithmic, algorithm}
\usepackage{tracklang}
\usepackage{glossaries}
\usepackage[tight,TABTOPCAP]{subfigure}
\usepackage[withpage, printonlyused]{acronym}
\usepackage{pgfplots}
\usepackage{etoolbox}
\usepackage{dtsc-creafig}
\usepackage{notationV12}
\usepackage{epsfig,epstopdf}
\usepackage{amsbsy,amsfonts}
\usepackage{cancel}
\usepackage{expdlist}
\usepackage{rotating}
\usepackage{multirow} 
\usepackage{tabularx}
\usepackage{pdflscape}
\usepackage{bmpsize}
\usepackage{mathrsfs}
\acrodef{MAE}{mean absolute error}
\acrodef{AD}{arbitrary deployment}
\acrodef{aDEP}{approximate DEP}
\acrodef{aDEP-GS}{approximate DEP based on GS}
\acrodef{aMMSE-GS}{approximate MMSE based on GS}
\acrodef{aDEP-NSE}{approximate DEP based on NSE}
\acrodef{aDEP-CG}{approximate DEP based on CG}
\acrodef{AN}{atomic norm}
\acrodef{APP}{a posteriori probabilities}
\acrodef{AoA}{angle of arrival}
\acrodef{AoD}{angle of departure}
\acrodef{AWGN}{additive white Gaussian noise}
\acrodef{BEP}{block expectation propagation}
\acrodef{BER}{bit error rate}
\acrodef{BF}{beamforming}
\acrodef{BP}{belief propagation}
\acrodef{BP-EP}{belief propagation expectation propagation}
\acrodef{CG}{conjugate gradient}
\acrodef{CPI}{coherent processing interval}
\acrodef{CRLB}{Cram\'er-Rao lower bound}
\acrodef{CSI}{channel state information}
\acrodef{CHEMP}{channel hardening-exploiting message passing}
\acrodef{DEP}{\textit{double}-EP}
\acrodef{DEP-GS}{approximate DEP based on GS}
\acrodef{DEP-NSE}{approximate DEP based on NSE}
\acrodef{DEP-CG}{approximate DEP based on CG}
\acrodef{DFE}{decision feedback equalization}
\acrodef{EC}{expectation consistency}
\acrodef{ECC}{error correction code}
\acrodef{EMB}{enhanced mobile broadband}
\acrodef{EP}{expectation propagation}
\acrodef{EP-NSE}{approximate EP based on NSE}
\acrodef{ESPRIT}{estimation of signal parameters via rotational invariance techniques}	
\acrodef{EXIT}{extrinsic information transfer}
\acrodef{FEC}{forward error correction}
\acrodef{FIR}{finite impulse response}
\acrodef{GMP}{Gaussian message passing} 
\acrodef{GS}{Gauss-Seidel}
\acrodef{GTA}{Gaussian tree approximation}
\acrodef{IDD}{iterative detection and decoding}
\acrodef{i.i.d.}{independent and identically distributed} 
\acrodef{IIC-AMF}{adaptive matched filter with iterative interference cancellation}
\acrodef{ICI}{intercarrier interference}
\acrodef{ISI}{intersymbol interference}
\acrodef{KL}{Kullback-Leibler} 
\acrodef{KSEP}{Kalman smoothing expectation propagation}
\acrodef{LDPC}{low-density parity-check}
\acrodef{LTI}{linear time invariant}
\acrodef{LMMSE}{linear minimum mean square error}
\acrodef{LS}{least squares}
\acrodef{LMMSE-GS}{approximate LMMSE based on GS}
\acrodef{LMMSE-NSE}{approximate LMMSE based on NSE}
\acrodef{LLRs}{log-likelihood ratios}
\acrodef{LLR}{log-likelihood ratio}
\acrodef{LTI}{linear time invariant}
\acrodef{MAP}{maximum a posteriori}
\acrodef{MCMC}{Markov chain Monte Carlo}
\acrodef{ML}{multi-level}
\acrodef{MLT}{multi--level Toeplitz}
\acrodef{MIMO}{multiple-input multiple-output}
\acrodef{MMV}{multiple measurement vector}
\acrodef{MMSE}{minimum mean square error}
\acrodef{mmWave}{millimiter--wave}
\acrodef{MMT}{massive machine type}
\acrodef{MAE}{mean-absolute-error}
\acrodef{MSE}{mean-square-error}
\acrodef{MU-MIMO}{multi-user MIMO}
\acrodef{MUSIC}{multiple signal classification}
\acrodef{MD-MUSIC}{multidimensional \ac{MUSIC}}
\acrodef{NLS}{nonlinear least squares}
\acrodef{NSE}{Neumann series expansion}
\acrodef{nuBEP}{non-uniform block expectation propagation}
\acrodef{pmf}{probability mass functions}
\acrodef{pdf}{probability density function}
\acrodef{OMP}{orthogonal matching pursuit}
 \acrodef{PRT}{pulse repetition time}
\acrodef{PSD}{positive semidefinite}
\acrodef{RCS}{radar cross section}
\acrodef{RL}{Rayleigh length}
\acrodef{RF}{radio frequency}
\acrodef{SD}{sphere decoding}
\acrodef{SDP}{semidefinite programming}
\acrodef{SEP}{smoothing expectation propagation}
\acrodef{SER}{symbol error rate}
\acrodef{SISO}{single-input single-output}
\acrodef{SNR}{signal-to-noise ratio}
\acrodef{SPA}{sum-product algorithm} 
\acrodef{S/P}{serial to paralell} 
\acrodef{TBEP}{turbo block expectation propagation}
\acrodef{TLMMSE}{turbo linear minimum mean square error}
\acrodef{URLL}{ultra-reliable low-latency}
\acrodef{ZF}{zero forcing}
\usepackage{upgreek}
\DeclareGraphicsExtensions{.jpg,.pdf,.png,.gif,.eps}

\DeclareFontFamily{OT1}{pzc}{}
\DeclareFontShape{OT1}{pzc}{m}{it}{<-> s * [0.900] pzcmi7t}{}
\DeclareMathAlphabet{\mathpzc}{OT1}{pzc}{m}{it}

\graphicspath{{figures/}}

\usepackage{pgfplots}

\makeatletter
\def\addlegendimage{\pgfplots@addlegendimage}
\makeatother

\newtheorem{theorem}{Theorem}
\newtheorem{lemma}{Lemma}
\newtheorem{proposition}{Proposition}
\newtheorem{corollary}{Corollary}
\newtheorem{remark}{Remark}
\newtheorem{definition}{Definition}
\newtheorem{problem}{Problem}

\newtheorem{assumption}{Assumption}

\def\b{\mathbf{b}}
\def\e{{\mathsf e}}

\def\x{\mathbf{x}}

\def\dd{\mathsf d}
\def\ddl{{\mathsf d}_{{\mathsf L}}}
\def\kappaL{\kappa_{\mathsf L}}

\def\t{\mathsf{tx}}
\def\r{\mathsf{rx}}
\def\Atx{\miA^\t}
\def\Arx{\miA^\r}
\def\AL{\miA^\mathsf{L}}

\def\H{\dagger}
\def\opt{\mathsf o}

\def\miu{{\mathbf u}}
\def\miy{{\mathbf y}}
\def\miH{{\mathbf H}}

\def\miHa{\mathbf{H}}
\def\mihu{\mathbf{h}_u}
\def\miha{\mathbf{h}}
\def\miJ{{\mathbf J}}

\def\mib{{\mathbf b}}
\def\mii{\boldsymbol{\varphi}}
\def\mib{{\mathbf b}}
\def\mic{{\mathbf c}}

\def\mig{{\mathbf g}}

\def\mih{{\mathbf h}}

\def\mix{{\mathbf x}}
\def\miv{{\mathbf v}}

\def\j{{\mathtt j}}

\def\a{{\mathsf a}}
\def\b{{\mathsf b}}

\def\x{{\mathsf x}}

\def\nn{{\boldsymbol{\mathsf n}}}
\def\n{\mathsf n}

\def\Nsf{{{\mathsf N}}}
\def\Msf{{{\mathsf M}}}
\def\Ssf{{{\mathsf S}}}
\def\Rsf{{{\mathsf R}}}
\def\Tsf{{\mathsf T}}
\def\Rsfm#1{{{\mathsf R}_{#1}^{\mathsf{max}}}}

\def\MM{{\boldsymbol{\mathsf M}}}

\def\NN{{\boldsymbol{\mathsf N}}}
\def\SS{{\boldsymbol{\mathsf S}}}
\def\RR{{\boldsymbol{\mathsf R}}}
\def\RRm{{{\boldsymbol{\mathsf R}}^{\mathsf{max}}}}

\def\TT{{\boldsymbol{\mathsf L}}}
\def\Tsf{{\mathsf L}}
\def\LL{{\boldsymbol{\mathsf T}}}

\def\mil{{\mathbf l}}

\def\mif{{\boldsymbol{\mathsf f}}}
\def\f{\mathsf f}
\def\ff{{\boldsymbol{\ell}}}
\def\ll{{\boldsymbol{\nu}}}
\def\ii{{\boldsymbol{\iota}}}
\def\mig{{\boldsymbol{\mathsf g}}}

\def\miv{{\mathbf v}}
\def\miw{{\mathbf w}}

\def\miPi{{\bf \Pi}}

\def\miA{{\mathbf A}}
\def\miG{{\mathbf G}}
\def\miB{{\mathbf B}}

\def\miD{{\mathbf D}}

\def\miK{{\mathbf K}}
\def\miM{{\mathbf M}}

\def\miP{{\mathbf P}}
\def\miO{{\mathbf O}}
\def\miR{{\mathbf R}}

\def\miX{{\mathbf X}}
\def\miQ{{\mathbf Q}}
\def\miC{{\mathbf C}}
\def\miI{{\mathbf I}}
\def\miU{{\mathbf U}}
\def\miT{{\mathbf T}}
\def\miV{{\mathbf V}}
\def\miW{{\mathbf W}}
\def\miZ{{\mathbf Z}}
\def\miY{{\mathbf Y}}
\def\C{{\mathbb C}}
\def\Pcal{{\mathcal P}}
\def\Z{{\mathbb Z}}
\def\R{{\mathbb R}}
\def\T{{\mathbb T}}

\def\F{{\mathcal F}}
\def\U{{\mathcal U}}

\def\SNR{\mathsf{SNR}}

\def\E{{\mathbb E}}

\def\K{K}
\def\P{P}
\def\d{\mathsf{d}}

\def\Nu{\Rsf_u}

\def\Na{{N}}
\def\Ta{{L}}
\def\Nu{{N_u}}
\def\Ma{{M}}
\def\Mu{{M_u}}
\def\Tu{{L_u}}

\def\MM{{\boldsymbol{\mathsf M}}}

\def\NN{{\boldsymbol{\mathsf N}}}

\def\error{\frac{1}{\ddl}\E\big\{\|\ff_k-\hat{\ff}_k\|^2_2\big\}}

\DeclareMathOperator{\tr}{Tr}
\DeclareMathOperator{\diag}{diag}

\DeclareMathOperator{\rank}{rank}

\DeclareMathOperator{\spn}{span}
\DeclareMathOperator{\vc}{vec}
\DeclareMathOperator*{\Motimes}{\raisebox{-0.15ex}{\scalebox{0.9}{$\bigotimes$}}}

\def\bin{$\Pcal_\text{BPSK}$}
\def\binc{$\Pcal_\text{QPSK}$}
\def\rand{$\Pcal_{\text{Gauss}}$}
\def\randc{$\Pcal_{\text{$\C$-Gauss}}$}

\def\ANnoisy{\tiny \eqref{eq:Tracenoisy}}
\def\music{\tiny MUSIC\\ \tiny\cite{Liao15}}
\def\omp{\tiny OMP \\ \tiny \cite{Swapna22}}
\def\iic{\tiny IIC-AMF \\ \tiny \cite{Grossi20}}
\def\mmse{\tiny LMMSE \\ \tiny \cite{Assalini09}}
\def\ls{\tiny LS \\ \tiny \cite{Nan15}}

\def\Pint{{\bf \Pi}_{\text{n-triv}}}

\newcommand{\arxiv}{\color[rgb]{0,0,0}}

\def\sparsem{{\em sparse model}}
\def\sparsems{{\em sparse models}}
\def\atom{{\em atom}}
\def\atomset{{\em atom set}}
\def\atoms{{\em atoms}}
\def\ntrivial{{\em non-trivial}}
\def\rankQ{{\rho}}
\def\rd{{\em reconstruction degree}}
\def\sd{{\em sparsity degree}}
\def\rds{{\em reconstruction degrees}}
\def\channelvec{{\em Sparse channel model}}
\def\freqvec{{\em Frequency parameters}}
\def\refAlg1{{\arxiv Algorithm~\ref{Alg:PrettyLemma}}}

\usepackage{soul}

\long\def\symbolfootnote[#1]#2{\begingroup%
\def\thefootnote{\fnsymbol{footnote}}\footnote[#1]{#2}\endgroup}

\makeatletter
\newenvironment{opteq}{%
	\let\c@equation\c@defcounter
	
	\begin{eqnarray}}
{\end{eqnarray}}

\makeatother

\IEEEoverridecommandlockouts

\begin{document}
%
\title{Nuclear Atomic Norm for parametric estimation of sparse channels}

\author{Álvaro Callejas-Ramos, Matilde~S\'{a}nchez-Fern\'{a}ndez,~Antonia~Tulino, and~Jaime~Llorca 
\thanks{A. Callejas-Ramos and M. Sánchez Fernández are with the Signal Theory \& Communications Dept., Universidad Carlos III de Madrid, Spain.} 
\thanks{Jaime Llorca is with the Electrical and Computer Engineering Department, New York University, Brooklyn, NY 11201 USA.}
\thanks{A. Tulino is with the Universit\`a degli Studi di Napoli Federico II, Italy.}
\thanks{This work has been partly funded by the Spanish Government through projects IRENE-EARTH and SOFIA-AIR (PID2020-115323RB-C33/ AEI / 10.13039/501100011033) (PID2023-147305OB-C31/AEI)  (Ministerio de Ciencia e Innovaci\'on/AEI/FEDER, UE).}
}

\maketitle

\begin{abstract}
Parametric channel estimation in \acs{mmWave} not only enables the anticipated large spectral efficiency gains of \acs{MIMO} systems but also reveals important propagation parameters, allowing for a low complexity representation of the channel matrix. In this work, we propose to use atomic norm as a gridless multidimensional spectral estimation approach to address parametric channel estimation where both \acs{AoD}  and \acs{AoA} are identified. The conditions for recovery of the propagation parameters are given depending on properties of the measurement matrix, and on structural features such as the antenna geometry or the number of scatters to resolve. The proposed methodology is compared against several state-of-the-art parametric approaches.
\end{abstract}


\section{Introduction}
The ever-increasing demand for higher transmission data rates and the exponential growth in the number of simultaneously connected users/devices are the driving requirements for both the fifth generation (5G) and sixth generation (6G) of wireless communication systems \cite{Akyildiz20}. In order to meet these requirements, the use of massive \ac{MIMO} is mandatory \cite{Akyildiz20}. However, the spectral efficiency achieved by \ac{MIMO} alone is not enough to meet the current demand, surging the exploration of underused spectrum bands, such as the \ac{mmWave} band (in the 30-300 GHz range).

While \ac{mmWave} band systems can support a large number of users due to their significantly higher available bandwidth, channel estimation in these systems, tacitly done for a large number of antennas, is substantially more challenging. The primary characteristic of the \ac{mmWave} band is its sparse nature, as its impulse responses are dominated by a few key clusters of significant paths \cite{Lee16}. Consequently, the channel matrix is defined by a few dominant singular values that are considerably fewer than the number of the antennas. This matrix can be represented as a sparse model that contains relevant propagation information in a multidimensional space.  

Thus, the problem we aim to solve is a channel estimation problem in the multidimensional parametric domain that deals with the recovery of both \emph{i)} the channel matrix, defined as a sparse linear combination of steering vectors, and \emph{ii)} the multidimensional $\dd$-dimensional ($\dd$-D) frequency parameter that contains relevant channel propagation information embedded in the steering vectors, i.e., the \ac{AoD} and the \ac{AoA}. 

The problem of sparse parametric channel estimation, also known as sparse channel representation \cite{Bajwa10,Zhang24}, or beamspace channel estimation \cite{Gao19} is having a significant resonance given its improved complexity-reconstruction error trade-off \cite{Bajwa10} combined with the fact than it unveils relevant propagation parameters that can be used for precoder design \cite{Schniter14} or to reach multidimensional space awareness \cite{Akyildiz20}. Traditional methods for parametric channel estimation such as \ac{LS} \cite{Assalini09}, \ac{MUSIC} \cite{Liao15} or \ac{OMP} \cite{Swapna22} do not necessarily target the underlying sparse nature of the channel model. To enforce this sparse nature in the parameters, sparsity-inducing mixed-norm optimization approaches such as $\ell_0$ or $\ell_1$ \emph{atomic norms} are particularly suitable \cite{Chandrasekaran2012,Chi-2020}. The dictionary learning in these approaches typically relay on its discretization \cite{Gao19,Zhang24}, and therefore is always constrained by the margin of error inherent in the definition of the grid \cite{Heckel18}. In contrast, gridless approaches aim to solve the problem in the continuous domain of the parameter \cite{Chi-2020}, avoiding grid-induced limitations. State-of-the-art gridless approaches for multidimensional parametric models include \cite{Yang16, Sanchez-Fernandez21,Vega21}.

In this work, we propose a solution to the problem of gridless multidimensional spectral channel estimation in a \ac{mmWave} \ac{MIMO} scenario where the signal model builds up as a linear measurement in the complex domain of a sparse combination of steering vectors. Our approach does not restrict to uniform antenna deployments, enabling ubiquitous connectivity demands where the embedded antenna deployments can coexist with other uses of the same volumetric space \cite{Wu17b,Hu18}. The proposed approach leverages {\em an atomic norm cost function} in order to recover the channel matrix and extract the \ac{AoA} and \ac{AoD} propagation parameters. Compared to other gridless approaches to recover multidimensional parametric models \cite{Yang16, Sanchez-Fernandez21,Vega21}, in this work we generalize the measurement model to a structured matrix in the domain of the estimation pilot alphabet, whereas the measurement model showing up in \cite{Yang16, Sanchez-Fernandez21,Vega21} is a pure sampling scenario. In summary, the main contributions of these work are:
\begin{itemize}
\item We pose the sparse parametric channel estimation signal model as a linear measurement in the complex domain of a sparse combination of steering vectors. We reformulate the atomic norm cost function as a $\rank$ minimization or nuclear norm minimization problem, and provide the recovery techniques for both the channel model and the embedded multidimensional frequency parameters.
\item The measurement model arising from the transmission of a known pilot sequence is analyzed and, as recovery conditions, its structural features such as the number of antennas deployed in each of the array dimensions are linked to the number of scatters that we want to resolve.
\item A non-structured generalized linear measurement model is also analyzed and the properties of the measurement matrix needed to resolve the model are provided.
\item The proposed method is compared against several other state-of-the-art {\em compressed sensing} techniques, both parametric and non-parametric, in terms of both performance and computational complexity. 
\end{itemize}

\textit{Notation:} 
$\miI_N$ is the identity matrix of dimension $N$,
$[\cdot]^\top$ is the transpose, $[\cdot]^*$ is the conjugate, and $[\cdot]^{\H}$ is the Hermitian. $\miA^{-}$ is the weak inverse of the  $M \times N$ matrix  $\miA$ (i.e. any solution to $\miA\miA^{-}\miA=\miA$);  we refer to 
$\miA^{-}$ as left or right weak inverse  if $\miA^{-}\miA=\miI_N$ 
or $\miA\miA^{-}=\miI_M$ respectively. $\mathscr{C}(\cdot)$ represents the column row space generated by a matrix.  The operator $\diag(\mathbf{x})$ returns a diagonal matrix with diagonal given by $\mathbf{x}$. $\miX^{(\mathcal{I})}$ are the submatrix of $\miX$  and the subvector of $\mix$ given respectively by the rows and elements in the index set $\mathcal{I}$. For a given integer $K\in\Z$, $\left[K\right]=\left\{1,\dots,K\right\}$. $\T$ denotes the unit circle $\left[0, 1\right]$ by identifying the beginning and the ending points. The $y$--modulus of value $x$ is given by $\mod\left(x,y\right)$. $\mathbf{0}_{N\times M}$ is the $N\times M$ all-zeros matrix. 
The inner product of two vectors is represented with $\mix\cdot\miy$, the Kronecker product uses $\otimes$, and the Khatri--Rao $\odot$.

\section{Channel Model for sparse mmWave propagation}
\label{sec:II}

\subsection{Propagation sensing with $\dd$-D arrays}
\label{sec:sv}


{\arxiv Multidimensional antenna deployments enable different degrees of propagation sensing capabilities that depend on the dimensions $\dd=\{1,2,3\}$ where the antenna array is unfolded. For example, in a linear deployment, $\dd=1$,  only propagation direction in the azimuth $\theta$ can be perceived, unambiguously in $\theta\in\left[0,\pi\right]$. In $\dd=2$ arrays, hemispherical propagation is accounted for, providing unambiguous information in $\theta\in\left[0,\pi\right]$ and in the elevation $\phi$, in $\phi\in\left[-\frac{\pi}{2},\frac{\pi}{2}\right]$. Finally for $\dd=3$ deployments, we have full spherical angular propagation characterization with $\theta\in\left[0,2\pi\right]$ and $\phi\in\left[-\frac{\pi}{2},\frac{\pi}{2}\right]$.}

A $\dd$-D uniform array deployment is fully characterized with the \emph{dimension vector} $\NN=[\Nsf_1,\dots,\Nsf_\dd]$ containing the number of radiating elements $\Nsf_i$ uniformly spaced in each dimension $i=[\dd]$, and $\Nu=\prod_{i=1}^\dd \Nsf_i$ being the total number of antennas in the array. The normalized position of the $n$-th antenna in the uniform array is given by $\nn_n=[\n^1_n,\dots,\n^\dd_n]^\top$ with $\n^i_n\in\{0,1,\dots, \Nsf_{i}-1\}$ and $n\in[\Nu]$.

We define a \emph{normalized frequency parameter vector} $\mif=\left[\f^1,\dots,\f^\dd\right]^\top\in\T^\dd$ with $1\leq\dd\leq3$ containing the information on the propagation angular direction, where $\f^1=\mod\left(\delta_1\cos\theta,1\right)$, $\f^2=\mod\left(\delta_2\sin\theta\sin\phi,1\right)$ and $\f^3=\mod\left(\delta_3\sin\theta\cos\phi,1\right)$, and $\delta_i$ being the uniform spacing of the antennas in the $i$-dimension. 

The \emph{steering vector} of a $\dd$-D uniform array, identified by the \emph{dimension vector} $\NN$, provides information on the $\dd$-D propagation direction embed in $\mif\in\T^\dd$ as follows:  
\begin{eqnarray}
\miu_\NN\left(\mif\right)
=\frac{1}{\sqrt{\Nu}}\big[e^{\j 2\pi\mif\cdot \nn_1},\dots,e^{\j 2\pi\mif\cdot \nn_\Nu}\big]^\top=\Motimes_{i=1}^\dd\miu_{\Nsf_i}(\f^i)
\label{ec_sv}
\end{eqnarray}
where $\miu_{\Nsf_i}(\f^i)=\frac{1}{\sqrt{\Nsf_i}}\big[1,e^{\j 2\pi \f^i  },\dots,e^{\j 2\pi(\Nsf_i-1) \f^i  }\big]^\top$.

 \subsection{Heterogeneous array deployments}
 \label{sec:Nu_sv}
 
Heterogenous array deployments account for deployments emerging from an uniform structure where some antenna elements are missing or not active. These particular types of deployments are of interest, for example, when embedding the array in surfaces or volumetric spaces \cite{Wu17b,Hu18}.

More specifically, we address heterogenous antenna deployments whose steering vector $\miv_{\Na}(\mif)\in\C^\Na$ is obtained by removing some antennas from an uniform array characterized by $\miu_\NN(\mif)\in\C^\Nu$, by means of a so called \emph{sensing matrix} that we denote as $\miA\in\{0,1\}^{\Na\times \Nu}$ with $\Na\leq \Nu$:
\begin{equation}\miv_{\Na}(\mif)=\miA\miu_\NN(\mif)
\label{eq:nu_sv}
\end{equation}
The sensing matrix has elements $a_{ij}=1$  if the  $j\in[\Nu]$  antenna of the uniform array is active and it appears  as component $i\in[\Na]$ in $\miv_{\Na}(\mif)$, and $0$ otherwise. Furthermore, the sensing matrix $\miA\in\{0,1\}^{\Na\times \Nu}$ should ensure that antenna elements are effectively removed and also that one particular antenna of the uniform array is not mapped more than once in the sampled structure. This is equivalent to impose that the sensing matrix $\miA=\big[\miI_\Na | {\bf 0}_{\Na \times (\Nu-\Na)} \big] \miPi$ with $ \miPi\in\{0,1\}^{\Nu\times\Nu}$ a column permutation matrix. 

\begin{remark}
\label{rm1}

Note that for a given heterogeneous array deployment with steering vector $\miv_{\Na}(\mif)$, the \emph{underlying} uniform steering vector $\miu_\NN(\mif)$ and, correspondingly the sensing matrix $\miA$, whose product lead to $\miv_{\Na}(\mif)$ are not unique. 
In particular, the dimension $\Nu$ of $\miu_\NN(\mif)$ represents a trade-off between system complexity and performance, which will be further exploited in the following sections.

\end{remark}

In this work we specifically signal those heterogeneous arrays with a particular structure that we name \ntrivial:

\begin{definition}
\label{def:struct0}
Given  a $\dd$-D  heterogeneous array deployment whose steering vector is $\miv_{\Na}(\mif)\in\C^\Na$, we say that it has a \ntrivial~structure, if we can find a permutation matrix $\Pint$ such that 
\begin{equation}
\begin{split}
\Pint\miv_{\Na}(\mif)=\Pint\miA\miu_\NN(\mif)=\begin{bmatrix}
e^{\j2\pi {\bf\Delta}\cdot\mif}\miu_\RR({\mii}) \\
\mib
\end{bmatrix}
\end{split}
\label{struct}
\end{equation}
where  \emph{i)} $\miu_\RR({\mii})$ is a uniform steering vector with dimensions $\RR=[\Rsf_1,\dots \Rsf_\dd]\preceq\NN$ 
such that  $\sum_{i=1}^\dd \Rsf_{i} \geq (\dd+1)$, 
\emph{ii)} $\mii$ is any permutation of $\left[\alpha_1 \f^1,\dots,\alpha_\dd \f^\d\right]^\top$ with $\alpha_i$ a proper positive integer, and \emph{iii)} ${\bf\Delta}=[\Delta_1,\dots,\Delta_\dd]$ with $\Delta_i\in [\Nsf_i]$, and $i\in[\dd]$.
\end{definition}

We would like to highlight that the \ntrivial~structure, if it exists, does not impinge on the underlying uniform structure $\miu_\NN(\mif)$ whose product with the sensing matrix leads to $\miv_{\Na}(\mif)$. Note also that there is more that one permutation matrix $\Pint$ that allows to identify in $\Pint\miv_{\Na}(\mif)$ an \emph{embed uniform steering vector} $\miu_\RR({\mii})$. We are interested in the permutation matrix that provides the embed uniform steering vector with the largest value for $\sum_{i=1}^\dd \Rsf_{i}$. We define it as follows:

\begin{definition}
\label{kappa} 
Given an heterogeneous array deployment with a 
 \ntrivial~structure, as in Def.  
\ref{def:struct0}, we name 
the embed uniform steering vector with  largest value  for $\sum_{i=1}^\dd \Rsf_{i}$ 
as $\miu_\RRm({\mii})$, we have $\RRm=[\Rsfm{1},\dots \Rsfm{\dd}]$ and we define $\kappa=\sum_{i=1}^\dd \Rsfm{i}$ as the \rd~of the array deployment. 
\end{definition}

 In the following we denote with $\mathcal{A}_\NN(\Na,\Nu)$ the set of sensing matrices associated to heterogeneous array deployments with \ntrivial\, structure, whose \rd,  $\kappa$, is larger o equal than $(\dd+1)$.


{\arxiv
\begin{remark}
All discussions and definitions in Secs. \ref{sec:sv} and \ref{sec:Nu_sv} also apply to scenarios with $\dd\geq 3$.
\end{remark}

The motivation for this remark will be further clear in next sections. We will see that the channel propagation vector can be represented as the Kronecker product of the steering vectors at the transmitter and the receiver, both respectively following an structure as in \eqref{ec_sv}, therefore enabling an enlarged steering vector with a total \emph{dimension vector} with up to $6$ components.

}

\subsection{Parametrical MIMO Channel with heterogenous arrays}
\label{sec:IIc}

In a \ac{mmWave} propagation scenario, we assume $\K$ sources of scattering between the transmitter with $\Ma$ radiating elements deployed in $\dd_\t\leq3$ dimensions and the receiver with $\Na$ antennas deployed in $\dd_\r\leq3$ dimensions. The channel matrix $\miHa\in\C^{\Na\times\Ma}$, that is inherently modelled as a sparse linear combination of the product of the transmit and receive steering vectors $\miv_\Ma(\mig_{k})$ and $\miv_\Na(\mif_{k})$\footnote{Note that this structure is distinctly found in \ac{mmWave} propagation, where few propagation paths with very low scattering are present, leading to a sparse number of non-scattered propagation paths.}, is parametrized through the normalized frequencies $\mig_k\in\T^{\dd_\t}$ and $\mif_k\in\T^{\dd_\r}$ with $k\in[\K]$ containing, respectively, information on the \ac{AoD} and \ac{AoA}  of the $k$-th propagation path:
\begin{eqnarray}
\miHa(-\mig_{1:\K},\mif_{1:\K})&=&\sum_{k=1}^{\K}\gamma_k\miv_\Na(\mif_{k})\miv^\top_\Ma(-\mig_{k})\\
&=& \sum_{k=1}^{\K}\gamma_k\miA_{\r}\miu_\NN(\mif_{k})\miu_\MM^\top(-\mig_{k})\miA^\top_{\t}
\nonumber
\label{eq:channel}
\end{eqnarray}
where $\mig_{1:\K}=\left [\mig_{1}, \dots, \mig_{\K}\right ]\in\T^{\dd_\t\times\K}$ and $\mif_{1:\K}=\left [\mif_{1}, \dots, \mif_{\K}\right ]\in\T^{\dd_\r\times\K}$, $\gamma_k\in\C$ is the $k$-th multipath gain, $\miu_\NN(\mif_{k})\in\C^\Nu$ and $\miu_\MM(\mig_{k})\in\C^\Mu$ are respectively the chosen underlying uniform steering vectors at the receiver and transmitter, and $\Atx$, $\Arx$ are the sensing matrices that we assume belonging to $\mathcal{A}_\MM(\Ma,\Mu)$ and $\mathcal{A}_\NN(\Na,\Nu)$.

We define a vectorized version of the channel matrix where the uniform composite steering vectors follow a predetermined Kronecker ordering:
\begin{eqnarray}
\miha(-\mig_{1:\K},\mif_{1:\K})&=&\vc(\miHa(-\mig_{1:\K},\mif_{1:\K}))\nonumber\\
&=&\left(\Atx\otimes\Arx\right)\sum_{k=1}^K\gamma_k\miu_\MM(-\mig_k)\otimes\miu_\NN(\mif_k)\nonumber\\
&=&\left(\Atx\otimes\Arx\right)\sum_{k=1}^K\gamma_k\miPi_u\miu_\TT(\ff_k)\label{eq:channel_vec}\\
&=&\AL\mihu(\ff_{1:\K})=\AL\miU_\TT(\ff_{1:\K})\boldsymbol{\gamma}
\nonumber
\end{eqnarray}
where, if we define $\Tu=\Nu\Mu$,   $\miPi_u\in\{0,1\}^{\Tu\times\Tu}$ is a permutation matrix that allows a reordenation of the Knonecker product $\miu_\MM(-\mig_k)\otimes\miu_\NN(\mif_k)$ in terms of a composite steering vector $\miu_\TT(\ff_k)\in\C^\Tu$ such that the \emph{dimension vector} $\TT=[\Tsf_1,\dots,\Tsf_{\ddl}]$ with $\ddl=\dd_\t+\dd_\r$ follows the ordering $\Tsf_1 \leq\Tsf_2\leq\dots\leq\Tsf_{\ddl}$. Note that $\TT=\miPi_\dd[\MM,\NN]=\miPi_\dd[\Msf_1,\dots,\Msf_{\dd_\t},\Nsf_1,\dots,\Nsf_{\dd_\r}]$ where $\miPi_\dd\in\{0,1\}^{\ddl\times\ddl}$ is a permutation matrix that reorders the composite \emph{dimension vector} $[\MM,\NN]$ increasingly. Similarly, $\ff_k=[\ell^1_k,\dots,\ell^{\ddl}_k]=\miPi_\dd\left[-\mig_k,\mif_k\right]\in\T^{\ddl}$ is the frequency vector that contains information on the \ac{AoD} and \ac{AoA}. Finally in \eqref{eq:channel_vec} we also find the following definitions $\AL\triangleq(\Atx\otimes\Arx)\miPi_u$, $\mihu(\ff_{1:\K})\triangleq\sum_{k=1}^\K\gamma_k\miu_\TT(\ff_k)\in\C^\Tu$, $\miU_{\TT}(\ff_{1:\K}) \triangleq [\miu_\TT(\ff_{1}),\dots,\miu_\TT(\ff_{K})] \in\C^{\Tu\times K}$ and $\boldsymbol{\gamma}=[\gamma_1,\dots,\gamma_K]^\top$. 

{\arxiv Note that given the Kronecker structure defining each channel composite element $\miu_\TT(\ff)=\Motimes_{i=1}^{\ddl}\miu_{\Tsf_i}(\ell^i)$ (see \eqref{ec_sv}), the channel composite matrix $\miU_{\TT}(\ff_{1:\K})$ can also be represented in terms of a Khatri-Rao product within the frequency dimensions, i.e., $\miU_{\TT}(\ff_{1:\K})=\odot_{i=1}^{\ddl}\miU_{\Tsf_i}(\ell^i_{1:\K})$ where $\miU_{\Tsf_i}(\ell^i_{1:\K})=[\miu_{\Tsf_i}(\ell^i_{1}),\dots,\miu_{\Tsf_i}(\ell^i_{\K})]$ is a $\Tsf_i \times \K$ Vandermonde matrix with generating elements $\frac{1}{\sqrt{\Tsf_i}}e^{\j 2\pi\ell^i_{1}},\dots, \frac{1}{\sqrt{\Tsf_i}}e^{\j 2\pi\ell^i_{\K}}\in\C $.}

Finally, for the rest of this work, we follow:
\begin{assumption}{\bf (A.\ref{A1}):}
\label{A1}
The $\ddl$-D frequency vectors, $\ff_k$ with $ k\in[\K]$, associated to the  \ac{AoA}s and  \ac{AoD}s  of the $\K$ sources of scattering are modelled as independent and identically distributed random vectors whose components $\ell^{i}_k$ for $ k\in[\K]$, and $i\in[\ddl]$ are independent and uniformly distributed on $[0,1)$.
\end{assumption}
Note that in wireless propagation environments scattering sources are typically randomly distributed, therefore, the assumption is not restrictive. 

\section{Parametrical channel estimation using Atomic Norm}
\label{sec_system}

The proposed parametric channel estimation problem aims to identify, from a set of measurements $\miY\in\C^{\Na\times\P}$ received after transmission of a known pilot sequence $\miP\in\Pcal^{\Ma\times\P}$ from the symbol alphabet $\Pcal\subseteq\C$, the frequency vector $\ff_{1:\K}=\left[\ff_{1},\dots,\ff_{K}\right]\in\T^{\ddl\times K}$ that characterizes \ac{AoD} and \ac{AoA} and, alongside, the channel matrix $\miHa(-\mig_{1:\K},\mif_{1:\K})\triangleq\miH(\ff_{1:\K})\in\C^{\Na\times\Ma}$.
Precisely, the received signal $\miY\in\C^{\Na\times\P}$ follows the parametric linear model: 
\begin{equation}
\begin{split}
\miY&=\miHa(\ff_{1:\K})\miP+\miW
\end{split}
\label{ec_parametric}
\end{equation}
where $\miHa(\ff_{1:\K})$ matches the structure in \eqref{eq:channel} and $\miW\in\C^{\Na\times\P}$ is additive white Gaussian noise with component-wise variance $\sigma^2_w$. The received $\miY$ can be further vectorized:
\begin{eqnarray}
\miy&=&\vc(\miY)=\left(\miP^\top\otimes\miI_\Na\right)\mih(\ff_{1:\K})+\miw\label{ec_parametric_vec}\\
&=&\left(\miP^\top\otimes\miI_\Na\right)\AL\mihu(\ff_{1:\K})+\miw=\miQ\mihu(\ff_{1:\K})+\miw\nonumber\\
&=&\miQ\sum_{k=1}^K\gamma_k\miu_\TT(\ff_k)+\miw=\miQ\miU_\TT(\ff_{1:\K})\boldsymbol{\gamma}+\miw\nonumber 
\end{eqnarray}
where, $\miw=\vc(\miW)\in\C^{\P\Na}$ is the noise vector,  and, 
\begin{equation}
\miQ=\left(\miP^\top\otimes\miI_\Na\right)\AL
\label{Q_def}
\end{equation}
is a  linear mapping in $\{\Pcal\cup\{0\}\}^{\Ta\times\Tu}$, with $\Ta=\P\Na$,  built upon the pilot set and the sensing structures. 

\subsection{Structural features of the measurement vector}

Next we discuss the required assumptions and the relevant structural features of $\miQ\in\{\Pcal\cup\{0\}\}^{\Ta\times\Tu}$ and $\mih_u(\ff_{1:\K})$. 

To this end, 
since $\miQ$ is defined as 
the factorization of the matrices $\left(\miP^\top\otimes\miI_\Na\right)$ and $\AL=(\Atx\otimes\Arx)\miPi_u$, in the following 
we introduce  Proposition \ref{Atotal:}
and Assumption {\bf (A.\ref{A2})}.

\begin{proposition}
\label{Atotal:}
The $\Na\Ma \times \Tu$-matrix  $\AL=(\Atx\otimes\Arx)\miPi_u$, in \eqref{Q_def}, has 
\rd, $\kappaL = \kappa_\t + \kappa_\r$  where $\kappa_\t$ and $\kappa_\r$ are the \rds~of the heterogenous array deployments at the transmitter and receiver sides, and  belongs to the set $\mathcal{A}_\TT(\Na\Ma,\Tu)$ (i.e. $\kappaL \geq (\ddl+1)$).  
\end{proposition}
\begin{proof}
The proof of Prop. \ref{Atotal:} is provided in Appendix \ref{ProofAtotal:}
\end{proof}

Following with the pilot matrix $\miP$, we will throughout this work put forward next assumption:

\begin{assumption}{\bf (A.\ref{A2}):}
\label{A2}
$\miP\in\Pcal^{\Ma\times\P}$ is almost-sure full-rank.
\end{assumption}

From A.\ref{A2} and noting that the factorization of the measurement matrix $\miQ\in\{\Pcal\cup\{0\}\}^{\P\Na\times\Tu}$ can also be written as $\miQ=\left(\left(\miP^\top\Atx\right)\otimes\Arx\right)\miPi_u$, it is straightforward to show that $\rank\{\miQ\}=\Na\min\{P,\Ma\}$ which leads to a $\rank$-deficient scenario in those settings where $\P>\Ma$.

We finalize the discussion of the relevant structural features of $\miQ$, with the different measurement scenarios that can be found given the relation between $\Ta=\P\Na$ and $\Tu$. Scenario $\Ta\geq\Tu$ poses a condition on the number of pilots that need to be at least larger than the number of antenna elements in the uniform deployment of the transmitter, i.e. $\P\geq\frac{\Nu}{\Na}\Mu$, given that we always have that  $\frac{\Nu}{\Na}\geq 1$. Note that this measurement scenario, that lead to a \emph{tall} $\miQ$ matrix, is always a $\rank$-deficient scenario since $\P\geq\frac{\Nu}{\Na}\Mu\geq\Ma$ and therefore $\rank\{\miQ\}=\Na\Ma$. The counterpart scenario, i.e. when the length of the pilot sequence $\P<\frac{\Nu}{\Na}\Mu$, $\miQ$ matrix is \emph{fat} and would be full-$\rank$ when $\Ma=\Mu$ and $\Na=\Nu$.

Finally, heading back to the measurement vector in \eqref{ec_parametric_vec}, the we explicitly describe
\begin{eqnarray}\mih_u(\ff_{1:\K})=\sum_{k=1}^{\K}\gamma_k\miu_\TT(\ff_{k})=\miU_\TT(\ff_{1:\K})\boldsymbol{\gamma}
\label{ec_hu}
\end{eqnarray} 
as the parametric channel vector to be estimated, where we are targeting that both, the frequency set $\ff_{1:\K}=\left[\ff_1,\dots,\ff_K\right]$ that we assume that follow A.\ref{A1} and the channel fading coefficients $\boldsymbol{\gamma}=[\gamma_1,\dots,\gamma_\K]^\top$, are recovered. Note that with an estimate of $\mih_u(\ff_{1:\K})$, it is straightforward to obtain, using \eqref{eq:channel_vec}, the correspondent estimate of the  channel matrix $\miH(\ff_{1:\K})$ using the relation $\mih(\ff_{1:\K})=\vc(\miH(\ff_{1:\K}))=\AL\mih_u(\ff_{1:\K})$

\subsection{Sparse model recovery in the spectral domain}
\label{sec_sparse}

We identify a \emph{parametric sparse signal model}, for short \sparsem,  as a signal vector in $\C^\Tu$ that allows a parametric linear representation with few ($\K$) elements, also known as \atoms, compared to the dimension of the data available ($\Tu \gg \K$). The number of \atoms~$\K$ is also identified as the \sd. 

The parametric channel vector $\mihu(\ff_{1:\K})$ in \eqref{ec_hu} being a linear combination of $\K$ \atoms, $\miu_\TT(\ff_1),\dots\miu_\TT(\ff_\K)$ parametrized in a $\ddl$-D spectral domain, would follow a \sparsem~ if $\Tu \gg \K$, which would be generally the case in \ac{mmWave} propagation. Note that each \atom~$\miu_\TT(\ff_k)$ with $k\in[\K]$ belongs to the infinite-cardinality \atomset~ $\U^\F_\TT$ that has a known structure akin to \eqref{ec_sv} by means of the composite \emph{dimension vector} $\TT=\left[\Tsf_1,\dots,\Tsf_{\ddl}\right]$ and the continuous $\ddl$-D  spectral parameter $\ff\in\T^{\ddl}$: 
\begin{eqnarray}
\U^\F_\TT=\left\{\miu_{\TT}(\ff)=\frac{1}{\sqrt{\Tu}}[e^{\j2\pi\nn_1\cdot\ff},\dots,e^{\j2\pi\nn_{\Tu}\cdot\ff}]^\top:\ff\in\T^{\ddl}\right\}
\label{atom_set}
\end{eqnarray}
where $\nn_n=[\n^1_n,\dots,\n^{\ddl}_n]^\top$ with $n\in[\Tu]$ is now defined in $\n^i_n\in\{0,1,\dots, \Tsf_{i}-1\}$ with $i\in[\ddl]$.

Inferring a \sparsem~in $\U^\F_\TT$ following \eqref{ec_hu} from a linear measurement $\miQ\mihu(\ff_{1:\K})$ of the model $\mihu(\ff_{1:\K})$ with $\miQ\in\C^{\Ta\times\Tu}$ or from its noisy counterpart $\miQ\mihu(\ff_{1:\K})+\miw$ is a problem of high interest that matches, for example, the parametric estimation of sparse channels that we are addressing in this work which aim to unveil the composite atoms $\miu_\TT(\ff_1),\dots\miu_\TT(\ff_\K)$ and the embedded $\ddl$-D spectral parameters $\ff_{1:\K}$. Specifically, in this work, for the sparse channel estimation application, we address the following two problems, always under the assumption that the sparsity condition $\Tu \gg \K$ conveys to  $\Tu > 2\K$ and the measurement scenario to $\Ta>2\K$:

\begin{problem} \label{P_noiseless} Given a noiseless linear measurement of a \emph{sparse} channel model, $\miy=\miQ\mihu(\ff_{1:\K})$, and given the linear mapping that models the measurement process $\miQ$, the objective is to {\bf exactly recover} the composite \atoms~$\miu_\TT(\ff_k)$, specifically identifying the embed $\ddl$-D spectral parameters $\ff_k\in\T^{\ddl}$ gridlessly for $k\in[\K]$, and, along way, fully recover the \sparsem~$\mihu(\ff_{1:\K})$.
\end{problem}

\begin{problem} \label{P_noisy} Given a noisy linear measurement of a \emph{sparse} channel model, $\miy=\miQ\mihu(\ff_{1:\K})+\miw$, and given the linear mapping that models the measurement process $\miQ$, the objective is to propose and analize a recovery procedure for the composite \atoms~$\miu_\TT(\ff_k)$, specifically identifying the embed $\ddl$-D spectral  parameters $\ff_k\in\T^{\ddl}$ gridlessly for $k\in[\K]$, and along way recover the \sparsem~$\mihu(\ff_{1:\K})$.
\end{problem}

To enforce a \emph{sparse atom structure} on a model $\mihu$ where the \atoms~belong to a specific atom set, i.e. $\U^\F_\TT$, we rely on the definition of the following pseudo-norms \cite{Chandrasekaran2012},  which provide different measurements of the level of sparsity of a model.

\begin{definition}
	Given $\mihu\in\C^\Tu$, the $\ell_0$ \ac{AN} ($\ell_0$-\ac{AN}) of $\mih_u$ is defined as: 	
	\begin{equation*}
	\left\|\mih_u\right\|_{\U^\F_\TT,0}=
	\inf_{\ff\in \T^{\ddl},\alpha_k\in\C}\bigg\{ K : \, \,  \mih_u=\sum^\K_{k=1 }\alpha_k\miu_{\TT}\left(\ff_k\right)\bigg\}.
	\label{ec_4}
	\end{equation*}
\end{definition}

\begin{definition}
	Given $\mih_u\in\C^\Tu$, its $\ell_1$-\ac{AN} is defined as: 	
	\begin{equation*}
	\left\|\mih_u\right\|_{\U^\F_\TT,1}=
	\inf_{\ff\in \T^{\ddl},\alpha_k\in\C}\bigg\{\sum^\K_{k=1 } |\alpha_k| : \, \,  \mih_u=\sum^\K_{k=1 }\alpha_k\miu_{\TT}\left(\ff_k\right)\bigg\}.
	\label{ec_4}
	\end{equation*}
\end{definition}

Based on the definition of the $\ell_0$-\ac{AN}, the following optimization problem can be posed to enforce a \emph{sparse atom-based solution} of a measurement over a model, as a first step to address Problem \ref{P_noiseless}: 
	\begin{opteq}
		\underset{\mihu \in \C^{\Tu}}{\min} \left\|\mihu\right\|_{\U^\F_\TT,0} \quad {\rm s.t.}  \quad  \miy=\miQ\mihu
		\label{l0normoptAN}
	\end{opteq}
Note that the minimization problem \eqref{l0normoptAN}, if it had a solution, would first identify the \sd~$\K$ and also raise a model with the smallest number of atoms within the atom set $\U^\F_\TT$. {\arxiv 
Furthermore, according to \cite[ Th. 2.13]{CSbook}, a necessary  condition for the \sparsem~being uniquely recovered from \eqref{l0normoptAN} is that any subset $\U^\F_{\TT,2\K}\subseteq\U^\F_\TT$, such that $|\U^\F_{\TT,2\K}|=2\K$, defining a $\Tu\times2\K$ matrix $\miU_\TT(\ff_{1:2\K})$ trought the alignment of $2\K$ atoms at any order, generates a matrix $\miQ\miU_\TT(\ff_{1:2\K})$ that is injective as a map from $\C^{2\K}\to \C^{\Ta}$. Nevertheless despite this condition for unique recovery of a \sparsem, } 
solving \eqref{l0normoptAN} is a non-convex NP-hard problem \cite{CSbook} which moreover does not provide any insight on the procedure to recover the composite atoms and frequencies from the optimal solution. 
Alternative convex relaxation methods based on the $\ell_1$  have been widely proposed in the literature to solve $\ell_0$. Here we follow a similar approach and use the definition of $\ell_1$-\ac{AN} to search for a sparse solution in the atom set $\U^\F_\TT$: 
	\begin{opteq}
		\underset{\mihu \in \C^{\Tu}}{\min} \left\|\mihu\right\|_{\U^\F_\TT,1} \quad {\rm s.t.}  \quad  \miy=\miQ\mihu
		\label{l1normoptAN}
	\end{opteq}	
Finally, to address the noisy scenario in Problem \ref{P_noisy} targeting convex methods, we propose a generalization of the LASSO estimator \cite{Tibshirani96},  also known as atomic norm denoising approach \cite{Chi19}, to take into account the specific sparse structure of the model  $\mih_u$ in a continuous \emph{atom set}:
\begin{opteq}
	\begin{aligned}
		\min_{\mih_u \in \C^{\Tu}}   \|\mih_u\|_{{\U^\F_\TT,1}} \quad {\rm s.t.}  \quad \frac{1}{\P\Na}\|  \miQ \mih_u -  \miy \|^2_2\leq \sigma^2_w
	\end{aligned}
	\label{eq:SDPnoisy}
\end{opteq}

\section{Previous results for Atomic Norm-based optimization}

In the literature, there are different optimization approaches to solve $\ell_\beta$-\ac{AN} cost functions in the atom set $\U^\F_\TT$, with $\beta=\{0,1\}$  \cite{Yang16,Sanchez-Fernandez21}. A non-convex optimization approach shows up, based on the $\rank$ of a matrix with a certain structure (see Sec. \ref{sec_MLT}), when $\ell_0$-\ac{AN} wants to be minimized. Similarly, a convex  minimization problem of the nuclear norm of a matrix with the same structure as before appears when $\ell_1$-\ac{AN} wants to be enforced. Results on conditions for resolvability of either-both the rank-based or the nuclear norm-based approaches in the noiseless case can be found in \cite{Yang16,Sanchez-Fernandez21} for a  particular measurement scenario where $\Ta\leq\Tu$ samples of $\mihu(\ff_{1:\K})$ are observed without any further processing. This measurement process is a pure \emph{sampling} scenario where the linear mapping is built with a structure $\miQ=\big[\miI_\Ta|{\bf 0}_{\Ta\times (\Tu-\Ta)}\big]  \miPi_s\in\{0,1\}^{\Ta\times \Tu}$ with $\miPi_s\in\{0,1\}^{\Tu\times \Tu}$ being any permutation matrix. More general measurement scenarios, where the measurement matrix  $\miQ$ is defined on an arbitrary field with any relation between $\Ta$ and $\Tu$ have not been addressed, to the authors best knowledge, and will be addressed in this work.

Next we recall some known results that will be of help to use \ac{AN}-based optimization to recover a parametric \sparsem~following \eqref{ec_hu} from a linear measurement.

\subsection{Multi-level Toeplitz matrices and the Vandermonde decomposition}
\label{sec_MLT}

One of the key tools for solving the atomic norm in the $\U^\F_\TT$ domain is the matrix structure that, within the optimization process, is used to force the model to be in the span of a set of atoms in $\U^\F_\TT$ as in \eqref{atom_set}. This structure is termed \ac{MLT} and a matrix with this structure is defined next.

\begin{definition} \label{definitionMLT}
Consider the vector $\TT=[\Tsf_1,\dots,\Tsf_{\ddl}]$; also, let $\TT_{t:\ddl}=[\Tsf_t,\dots,\Tsf_{\ddl}]$  with $t\in[\ddl]$. A $\Tu \times \Tu$ matrix  $\miT_\TT$ is a  $\ddl$-\ac{MLT} matrix if it possess the following structure: $\miT_\TT$ is a block Hermitian Toeplitz matrix with $\Tsf_1\times\Tsf_1$ blocks:
	\begin{equation*}
	\miT_{\TT}=\begin{bmatrix}\miT_{0\TT_{2:\ddl}}& \miT_{1\TT_{2:\ddl}}&\dots &\miT_{(\Tsf_1-1)\TT_{2:\ddl}}\\\miT_{(-1)\TT_{2:\ddl}}&\miT_{0\TT_{2:\ddl}} &\dots&\miT_{(\Tsf_1-2)\TT_{2:\ddl}}\\\vdots & & &\vdots \\ \miT_{(-\Tsf_1+1)\TT_{2:\ddl}}& \miT_{(-\Tsf_1+2)\TT_{2:\ddl}}&\dots & \miT_{0\TT_{2:\ddl}}\\ \end{bmatrix}
	\label{eq:toeplitz}
	\end{equation*}
	where $\miT_{\a\TT_{2:\ddl}}= \miT_{-\a\TT_{2:\ddl}}^\H$; also, each block $\miT_{\a\TT_{2:\ddl}}$,  with  $-\Tsf_1+1 \leq \a  \leq \Tsf_1-1$, is in turn a block Hermitian Toeplitz matrix with $\Tsf_2\times\Tsf_2$ blocks defined as:
	{\small\begin{equation*}
	\miT_{\a\TT_{2:\ddl}}=\begin{bmatrix}\miT_{\a 0\TT_{3:\ddl}} & \miT_{\a 1\TT_{3:\ddl}} &\dots & \miT_{\a(\Tsf_2-1)\TT_{3:\ddl}}\\\miT_{\a(-1)\TT_{3:\ddl}} & \miT_{\a 0\TT_{3:\ddl}} &\dots&\miT_{\a(\Tsf_2-2)\TT_{3:\ddl}}\\\vdots & & &\vdots \\ \miT_{\a(-\Tsf_2+1)\TT_{3:\ddl}}&\miT_{\a(-\Tsf_2+2)\TT_{3:\ddl}} &\dots &\miT_{\a 0\TT_{3:\ddl}}\\ \end{bmatrix}\label{eq:toeplitz2}
	\end{equation*}}
	with $\miT_{\a\b\TT_{3:\ddl}}= \miT_{\a(-\b)\TT_{3:\ddl}}^\H$; each block $\miT_{\a\b\TT_{3:\ddl}}$, with  $-\Tsf_1+1 \leq \a  \leq \Tsf_1-1$ and  $-\Tsf_2+1 \leq \b  \leq \Tsf_2-1$ is again a $\Tsf_3\times\Tsf_3$-block Hermitian Toeplitz matrix. The above construction recursively repeats at each inner level until the last level is reached  containing a $\Tsf_{\ddl}\times\Tsf_{\ddl}$ Hermitian Toeplitz matrix.
\end{definition}

{\arxiv 
\begin{algorithm}[h!]
{\arxiv\caption{Vandermonde decomposition of a $\ddl$-\ac{MLT} matrix}
 \begin{algorithmic}\label{Alg:PrettyLemma}
 \STATE 
  \textbf{Definitions:} $\TT=[\Tsf_1,\dots,\Tsf_{\ddl}]$, $\TT_{t:{\ddl}}=[\Tsf_t,\dots,\Tsf_{\ddl}]$, $\miU_{\TT_{t:{\ddl}}}(\ff^{t:{\ddl}}_{1:r})=\odot_{s=t}^{\ddl}\miU_{\Tsf_s}(\ff^s_{1:r})$ with $t\in[{\ddl}]$, $\ff^{t:{\ddl}}_{1:r}=[\ff^{t:{\ddl}}_{1},\dots,\ff^{t:{\ddl}}_{r}]$, $\ff^{t:{\ddl}}_{k}=[\ell^t_k,\dots,\ell^{\ddl}_k]^\top$ and $\miU_{\Tsf_s}(\ff^{s}_{1:r})=[\miu_{\Tsf_s}(\ell^s_1),\dots,\miu_{\Tsf_s}(\ell^s_r)]$.
  \\
  \STATE  
 \textbf{Input:}  $\miT_{\TT}$ with $\rank\{\miT_{\TT}\}=r$.  \\
 \STATE 
 1) Obtain the Cholesky decomposition of $\miT_{\TT}=\miC_{1:{\ddl}}\miC_{1:{\ddl}}^\H$.
 
\textbf{Decomposing in the $i$ dimension}

 \FOR {$i=1,...,{\ddl}-1$}
 \STATE 
2) Define the sets $\mathcal{I}_{i}=\{  1 ,\dots,(\Tsf_i-1)\prod_{s=i+1}^{\ddl}\Tsf_s{ \} }$ and $\mathcal{I}^{+}_{i}=\{ 1+\prod_{s=i+1}^{\ddl}\Tsf_s,\dots,\prod_{s=i}^{\ddl}\Tsf_s\} $ and $\mathcal{I}_{i+1}=\{ 1,\dots,\prod_{s=i+1}^{\ddl}\Tsf_s\} $
\STATE 
3) Find the $r\times r$ $\miO_{i}$ unitary matrix such that $\miC^{(\mathcal{I}_{i})}_{i:{\ddl}}\miO_{i}=\miC^{(\mathcal{I}^{+}_{i})}_{i:{\ddl}}$.
\STATE 
4) Obtain the eigen-decomposition $\miO_{i}=\miK_{i}\miJ_i\miK_{i}^\dagger$ and we have that $\miJ_i=\diag([ e^{\j2\pi \ell'^i_1},\dots,e^{\j2\pi \ell'^i_r}] )$
\STATE 
5) Set the $\prod_{s=i+1}^{\ddl}\Tsf_s\times r$ matrix $\miC_{(i+1):{\ddl}}=\miC^{(\mathcal{I}_{i+1})}_{i:{\ddl}}$\\

\ENDFOR

 \STATE            
 6) Define the sets $\mathcal{I}_{{\ddl}}=\{ 1,\dots,\Tsf_{\ddl}-1\} $ and $\mathcal{I}^{+}_{{\ddl}}=\{ 2,\dots,\Tsf_{\ddl}\} $
 \STATE
 7) Find the $r\times r$ $\miO_{{\ddl}}$ unitary matrix such that $\miC^{(\mathcal{I}_{{\ddl}})}_{{\ddl}}\miO_{{\ddl}}=\miC^{(\mathcal{I}^{+}_{{\ddl}})}_{{\ddl}}$.
\STATE 
8) Obtain the eigen-decomposition $\miO_{{\ddl}}=\miK_{{\ddl}}\miJ_{\ddl}\miK_{{\ddl}}^\dagger$ and we have that $\miJ_{\ddl}=\diag([ e^{\j2\pi \ell'^{\ddl}_1},\dots,e^{\j2\pi \ell'^{\ddl}_r} ] )$

\textbf{Frequency pairing}
\STATE
9) Set $\ff^{{{\ddl}}}_{1:r}=\ff'^{{\ddl}}_{1:r}=[\ell'^{{\ddl}}_1,\dots,\ell'^{{\ddl}}_{r}]$

 \FOR {$i={\ddl}: {\ddl}-1:1$}
 \STATE
 10) Get $\miM_i=\big(\miU_{\TT_{i:{\ddl}}}(\ff^{i:{\ddl}}_{1:r})\big)^-\miC_{i:{\ddl}}\miC^\H_{i:{\ddl}}\big(\miU^\H_{\TT_{i:{\ddl}}}(\ff^{i:{\ddl}}_{1:r})\big)^-$
 \STATE
 11) Get $\miR_i=1/\sqrt{\miM_i}\big(\miU_{\TT_{i:{\ddl}}}(\ff^{i:{\ddl}}_{1:r})\big)^-\miC_{i:{\ddl}}$
 \STATE
 12) Define the paired set $$\ff^{{(i-1):{\ddl}}}_{1:r}=[(  \ff'^{i-1}_{1:r} )^\top ,(\ff^{{i:{\ddl}}}_{1:r}\miR_i)^\top]^\top$$

\ENDFOR
     
     \STATE \textbf{Output:} The recovered paired frequencies $\ff_{1:r}$, the atom set $\miU_\TT(\ff_{1:r})=[\miu_\TT(\ff_{1}),\dots, \miu_\TT(\ff_{r})]$ and the $r\times r$ diagonal matrix $\miD$ such that $\miT_{\TT}=\miU_\TT(\ff_{1:r})\miD\miU^\H_\TT(\ff_{1:r})$.
  \end{algorithmic}
  \label{Ag1}
  }
\end{algorithm}

}

A $\Tu \times \Tu$ matrix  $\miT_\TT$ is a canonical $\ddl$-\ac{MLT} matrix if its structure follows Definition \ref{definitionMLT} and we have that $\Tsf_1 \leq \Tsf_2\leq\dots \leq\Tsf_{\ddl}$\footnote{The ordering on the nesting in the \ac{MLT} matrix could be arbitrary. By enforcing this ordering, we allow the best resolvability conditions \cite{Sanchez-Fernandez21}.}.  We identify the set of all  \ac{PSD}  canonical $\ddl$-\ac{MLT} matrices on the \emph{dimension vector}  $\TT=[\Tsf_1,\dots,\Tsf_{\ddl}]$ with $\mathcal{T}^{\ddl}_\TT \subseteq \C^{\Tu \times \Tu}$.

Definition \ref{definitionMLT}, for $\ddl=1$, leads to a Hermitian Toeplitz matrix $\miT_{[\Tsf_1]}$. The classical result by Carath\'{e}odory and Fej\'{e}r \cite{Caratheodory11} states that a \ac{PSD} rank-deficient Toeplitz matrix with $\rank\{\miT_{[\Tsf_1]}\}=r<\Tsf_1$  can be uniquely decomposed as $\miT_{[\Tsf_1]}=\miU\miD\miU^\H$, where $\miU$ is a $\Tsf_1\times r$ Vandermonde matrix whose $r$ columns are atoms in $\U^\F_{[\Tsf_1]}$ with $1$-D frequencies $\{\ell_1,\dots\ell_r\}$ and $\miD=\diag\left([d_1,\dots,d_r]\right)$ is a diagonal matrix with real positive elements. {\arxiv Thus a mixture of the outer product of $r<\Tsf_1$ one dimensional complex atoms in $\U^\F_{[\Tsf_1]}$ is always a rank-deficient \ac{PSD} Toeplitz matrix. A similar result can be extended to $\ddl>1$ showing the conditions and method for which a \ac{PSD} rank-deficient \ac{MLT} matrix allows a Vandermonde-like decomposition parameterized by a $\ddl$-D frequency set  $\left[\ff_1,\dots\ff_r\right]$.  In that case, again, we could claim that a mixture of the outer product of atoms in $\U^\F_\TT$ is always a \ac{PSD} rank-deficient \ac{MLT} matrix and viceversa. }
The extension of Carath\'{e}odory and Fej\'{e}r result to $\ddl>1$ was initially addressed in \cite{Yang16} and further extended in \cite{Sanchez-Fernandez21} providing the following \ac{ML} Vandermonde decomposition result \cite[Lemma 1, Remark 3]{Sanchez-Fernandez21}, 

\begin{lemma}\label{PrettyLemma}
Let  $\miT_\TT$ be a canonical $\Tu \times \Tu$  \ac{PSD}  $\ddl$-\ac{MLT} matrix with $\rank$ $r<\Tu$. If the $\rank$ of the  $\Tsf_{\ddl} \times \Tsf_{\ddl}$ upper-left corner\footnote{The $\Tsf_{\ddl} \times \Tsf_{\ddl}$ upper-left corner of $\miT_\TT$, is the $\Tsf_{\ddl} \times \Tsf_{\ddl}$ sub block of $\miT_\TT$ obtained  considering the first $\Tsf_{\ddl}$ rows and the first $\Tsf_{\ddl}$ columns of $\miT_\TT$.} of $\miT_\TT$ is also equal to $r$ and $r<\Tsf_{\ddl}$ then $\miT_\TT$ can be uniquely decomposed, via \refAlg1~as $\miT_\TT=\miU_\TT\left(\ff_{1:r}\right)\miD \miU_\TT^\H\left(\ff_{1:r}\right)$, with $\ff_{1:r}=[\ff_{1},\dots,\ff_{r}]\in\T^{\ddl\times r}$ being a unique  set of frequencies, $\miU_\TT\left(\ff_{1:r}\right)=\left[ \miu_\TT(\ff_1),  \dots,  \miu_\TT(\ff_r)\right]\in\C^{\Tu \times r}$ with $\miu_\TT(\ff_k)\in\U^\F_\TT$, and $\miD=\diag\left([d_1,\dots,d_r]\right)$, $d_k\in \R^+$ with $k\in[r]$.
\end{lemma}
\begin{proof}  
The proof for $\ddl=3$ is given in \cite[Appendix A]{Sanchez-Fernandez21}. It is straightforward to extend it to $\ddl>3$.
\end{proof} 

Note that the aforementioned Vandermonde decomposition allows gridless recovery of parameters $\ff_{1:r}\in \T^{\ddl\times r}$.


\subsection{Enforcing a model $\mil_u$ to match a \sparsem~ in $\U_\TT^\F$}
\label{Sec_enforce}

The structure of a \ac{PSD} $\ddl$-\ac{MLT} matrix in $\TT=[\Tsf_1,\dots,\Tsf_{\ddl}]$  is of interest for enforcing that a generic model $\mil_u\in\C^\Tu$ is defined in the atom set $\U^\F_\TT$. We show this with next Lemma.

\begin{lemma}\label{Span_condition} Given a generic model $\mil_u\in\C^\Tu$ and a matrix $\miT_\TT\in\C^{\Tu\times\Tu}$ belonging to the set of \ac{PSD} canonical $\ddl$-\ac{MLT} matrices $\mathcal{T}^{\ddl}_\TT$ as in Definition \ref{definitionMLT} with \emph{dimension vector} $\TT=[\Tsf_1,\Tsf_2,\dots,\Tsf_{\ddl}]$ and an ordering such that $\Tsf_1 \leq \Tsf_2\leq\dots \leq\Tsf_{\ddl}$, if $\rank\{\miT_\TT\}<\Tsf_{\ddl}$  then condition:
\begin{equation}
\begin{bmatrix}\miT_{\TT} &\mil_u\\ \mil_u^\H & l\end{bmatrix}\succeq 0
\label{Span_eq}
\end{equation}
conveys the following structure to  $\mil_u$:
$$\mil_u=\sum_{k=1}^{\rank\{\miT_\TT\}}\beta_k\miu_\TT\left(\ll_{k}\right)=\miU_\TT(\ll_{1:\rank\{\miT_\TT\}})\boldsymbol{\beta}$$
with $\beta_k\in\C$, $\ll_k\in\T^{\ddl}$ for $k\in[\rank\{\miT_\TT\}]$ and $\boldsymbol{\beta}=[\beta_1,\dots,\beta_{\rank\{\miT_\TT\}}]^\top$.
\end{lemma}

\begin{proof}  
The proof is given in Appendix \ref{proofspan}. 
\end{proof} 

Lemma \ref{Span_condition} shows that the ordering of the canonical ${\ddl}$-\ac{MLT} matrix nesting following $\Tsf_1\leq\Tsf_2\leq\dots\leq\Tsf_{\ddl}$ is tranfered to the \emph{dimension vector} $\TT=[\Tsf_1,\dots,\Tsf_{\ddl}]$ of the model $\mil_u$ composite atoms $\miu_\TT(\ll_k)$ with $k\in [\rank\{\miT_\TT\}]$.

\subsection{Unique decomposition of a sparse model in $\U^\F_\TT$}
\label{Sec_unique}

We give next the conditions under which given a \sparsem~$\mihu$ as in \eqref{ec_hu} with frequencies following A.\ref{A1}, the decomposition of $\mihu$ in terms of a set of atoms in $\U^\F_\TT$ and a mixing vector is unique. This means that there does not exits two different representations in the atom set $\miu_{\TT}(\ff_{1}),\dots,\miu_{\TT}(\ff_{\K})$  and $\miu_{\TT}(\ll_{1}),\dots,\miu_{\TT}(\ll_{\K})$ with $\miu_{\TT}(\ff_{k})\neq\miu_{\TT}(\ll_{l})$ $\forall 
k,l\in[\K]$ such that $\mihu=\miU_{\TT}(\ff_{1:\K}){\boldsymbol\alpha}=\miU_{\TT}(\ll_{1:\K}){\boldsymbol\beta}$. This is, clearly true iff the matrix 
$\miU_{\TT}([\ff_{1:\K}\ll_{1:K}])$ has rank $2K$. The next proposition provide some sufficient condition for 
$ \miU_{\TT}([\ff_{1:\K},\ll_{1:K}])$ being injective as a map from $\C^{2K}\to\C^{\Ta}$.

\begin{proposition}\label{Prop2} \cite[Thm. 4]{Sidiropoulos01}: Given a sparse model $\mihu(\ff_{1:\K})$ as in \eqref{ec_hu} with frequencies $\ff_{1:\K}$ element-wise different, i.e. $\ell^{i}_k\neq\ell^{i}_p$ with $k,p\in[\K]$, $k\neq p$ and $i\in[\ddl]$, 
$\mihu(\ff_{1:\K})$ has a unique linear representation using $\K$ distinct \atoms, $\miu_\TT(\ff_1),\dots,\miu_\TT(\ff_\K)$, if $\sum_{i=1}^{\ddl} \Tsf_i\geq2\K+(\ddl-1)$.
\end{proposition}

\begin{corollary}\label{Prop3} From Proposition \ref{Prop2}, given $2\K$ frequencies $\ff_1,\dots\ff_{2\K}$ element-wise different, if $\sum_{i=1}^{\ddl} \Tsf_i\geq2\K+(\ddl-1)$, then 
$\rank\{\miU_\TT(\ff_{1:2\K})\}=2\K$.
\end{corollary}

\section{Atomic Norm-based model recovery for generalized linear measurements}
\label{Sec:AV}

In this section we establish the sufficient conditions under which Problem \ref{P_noiseless}, resorting to the optimizations in \eqref {l0normoptAN} or \eqref{l1normoptAN}, enable perfect recovery. To this end, we first identify, in Section \ref{sec:uniqueness}, the conditions under which a \sparsem~in $\U_\TT^\F$ can be uniquely recovered from an observation vector obtained through a generalized measurement matrix $\miQ\in\C^{\Ta\times\Tu}$. Different constructions for $\miQ$ will be addressed that include, but are not restricted to, the structure in  \eqref{Q_def}. Next, in Sec. \ref{Sec:recovery}, we pose two optimization problems based respectively on the minimization of a $\rank$ and of a nuclear norm, and provide the conditions under which they solve respectively \eqref {l0normoptAN} or \eqref{l1normoptAN}. Finally in Sec. \ref{Sec:noiseless} we state the optimization problem used to address Problem \ref{P_noisy}.

The recovery conditions we propose are fundamentally based on the \sd~$\K$, which corresponds to the number of scatters and, consequently, the richness of the propagation environment. These conditions focus on how $\K$ relates to structural features of the measurement scenario, such as array geometry and the pilot matrix. Given this interplay, we present two alternative scenarios for interpreting and applying our results.
In the first scenario, we assume $\K$ is known. With this information, we can design the array geometry and pilot matrix to meet the recovery conditions that depend on $\K$, ensuring accurate signal recovery. In the second scenario, instead of assuming a known $\K$, we start with a specific antenna deployment and pilot matrix. From this configuration, we evaluate the maximum $\K$ that can be identified while still ensuring reliable recovery.

\subsection{Unique recovery of a linear observation of a sparse model}
\label{sec:uniqueness}

We identify next the uniqueness conditions that will entail properties on $\miQ$ and on the atom set, under which, 
if the linear measurement  of two \sparsems, $\mihu(\ff_{1:\K})$ and  $\mil_u(\ll_{1:r})$, coincide then necessarily both, models and  frequencies are equal. 
We provide the result for \emph{any} relationship between $\K$ and $r$.
\begin{theorem}
\label{uniqueness}
Given an atom set $\U^\F_\TT$ as in \eqref{atom_set}, let $\mihu(\ff_{1:\K})= \miU_{\TT}(\ff_{1:\K})\boldsymbol{ \gamma}$ and $\mil_u(\ll_{1:r})= \miU_{\TT}(\ll_{1:r})\boldsymbol{\beta}$ be 
two  \sparsems~in $\U_\TT^\F$, 
with $\ff_{1:\K} \in \T^{\ddl\times K}$, $\ll_{1:r} \in \T^{\ddl\times r}$. 
Under the assumption that: 
\begin{enumerate}
\item[C.1a] $\miQ \miU_{\TT}(\ff_{1:K} )$ is injective as a map from $\C^{K}\to\C^{\Ta}$ under A.\ref{A1}, 
\item[C.1b] and $\rank\left\{\miQ \miU_{\TT}([\ff_{1:\K}\ll_{1:r}] ) \right\}= \K+\rank\left\{\miQ \miU_{\TT}(\ll_{1:r}) \right\}$ assuming that none of the frequency vectors among the sets $\ff_{1:\K}$ and $\ll_{1:r}$ coincide,
\end{enumerate}
the following two statements are equivalent: 
\begin{enumerate}
    \item[\emph{i)}] A linear measurement of both models coincide, i.e. $\miQ\mihu(\ff_{1:\K})=\miQ\mil_u(\ll_{1:r})$,
    \item[\emph{ii)}] $\K=r$, 
$\boldsymbol{ \gamma}=\boldsymbol{\beta}$ and  $\ff_{1:\K}=\ll_{1:r}$.
\end{enumerate}
\end{theorem}

\begin{proof}  The proof  is provided in Appendix \ref{proofuniqueness}.\end{proof}

Condition C.1 state the requisites needed in the $\rank$ of the measurement and \atom~matrices to ensure unique recovery of a sparse model given an observation. Nevertheless, in order to provide better insight on how to tailor system parameters to attain robust parametric channel estimation, it is of interest to link condition C.1 to the \emph{structure} of the problem, i.e. to the structure and properties of the pilot sequence $\miP$ or in a more general case to the measurement matrix $\miQ$, to the antenna geometry, and to the \sd~of the model $\K$. With this purpose, we provide next some uniqueness results, specifically addressing the  aforementioned structural features. 

\begin{theorem}
\label{Thm3}
Given a measurement matrix $\miQ\in\C^{\Ta\times\Tu}$ and given two \sparsems~in $\U_\TT^\F$, $\mihu(\ff_{1:\K})= \miU_{\TT}(\ff_{1:\K})\boldsymbol{ \gamma}$ and $\mil_u(\ll_{1:r})= \miU_{\TT}(\ll_{1:r})\boldsymbol{\beta}$  with $\ll_{1:r} \in \T^{\ddl\times r}$ and $\ff_{1:\K} \in \T^{\ddl\times K}$ satisfying A.\ref{A1}, then Condition C.1 in  Theorem \ref{uniqueness}  holds, if 

\emph{i)} $\sum_{i=1}^{\ddl} \Tsf_i{ \geq} K+r+(\ddl-1)$, and \emph{ii)} $\miQ$  admits a left pseudo inverse; 

or if

\emph{iii)}  $\sum_{i=1}^{\ddl}\Tsf_i{ \geq} K+r+(\ddl-1)$, 
  and \emph{iv)} $\miQ$ can be factorized  as $\miQ=\miQ_1\miQ_2$ where $\miQ_1$ admits a left pseudo inverse and $\miQ_2$ is a $\rankQ \times \Tu$ full-$\rank$ Kruskal-$\rank$ (\emph{k}-$\rank$) unitary-invariant matrix with $\rankQ=\rank\{\miQ\}\geq K+r$ and \emph{k}-$\rank$ equal to $\rho$. 
\end{theorem}

\begin{definition}
A matrix is \emph{k}-$\rank$ unitary-invariant if the right product with a unitary matrix does not change its \emph{k}-$\rank$. 
\end{definition}


\begin{theorem}
A matrix $\miM$ is a  Kruskal-rank unitary-invariant if any of the following assumption are met: 
\begin{enumerate}
    \item[1.] 
    $\miM$ has i.i.d. Gaussian entries \cite{Tulino04b};
    \item[2.] 
    $\miM= \miZ \miY$  where $\miZ$ is  either deterministic or random matrix and  $\miY$ is a classical Haar unitary matrix as in \cite{Tulino04b} and \cite{Tulino13};  
\item[3.]   $\miM= \miZ \miY$ where   $\miY$ an asymptotic liberating  matrix (as in \cite{Anderson14}), $\miZ$ is either deterministic or random.  
\end{enumerate} 
\end{theorem}
\begin{proof}  
The proof can be derived from the results in \cite{Tulino04b, Tulino13,Anderson14}.
\end{proof} 


Noting that Thm. \ref{Thm3} cannot be applied to the structure of $\miQ$ provided in \eqref{Q_def}, we provide next the following result.

\begin{theorem}
\label{Thm4}
Given two \sparsems,~$\mihu(\ff_{1:\K})= \miU_{\TT}(\ff_{1:\K})\boldsymbol{ \gamma}$ and $\mil_u(\ll_{1:r})= \miU_{\TT}(\ll_{1:r})\boldsymbol{\beta}$ as in  Thm.
\ref{Thm3} and a measurement matrix $\miQ\in\C^{\Ta\times\Tu}$ structured as in \eqref{Q_def}, then Condition C.1 in  Theorem \ref{uniqueness}  holds, if 

\emph{i)}  $\kappaL=\kappa_\t + \kappa_\r { \geq}  \K + r { +} (\ddl -1) $, and \emph{ii)} $\miP^\top$  admits a left pseudo inverse; 
    
  or if 

  \emph{iii)} $\kappaL=\kappa_\t + \kappa_\r  { \geq}  \K + r {+} (\ddl -1)$, \emph{iv)} and $\miP^\top$ is a full-rank \emph{k}-$\rank$ unitary-invariant matrix with \emph{k}-$\rank$ equal to $\P> \K+r$, where $\kappaL$,  $\kappa_\t$ and $\kappa_\r$ are defined  in Prop. \ref{Atotal:}. 
\end{theorem}

\begin{proof}  
The proof of Thm. \ref{Thm4} is provided in Appendix \ref{proofThm4}.
\end{proof} 

{\arxiv
}

\subsection{Noiseless sparse model recovery under generalized linear measurements}
\label{Sec:recovery}

We first address Problem \ref{P_noiseless}, which is to infer a \sparsem~following \eqref{ec_hu} from a linear measurement $\miQ\mihu(\ff_{1:\K})$ also unveiling its embedded $\ddl$-D spectral parameters $\ff_{1:\K}$. We use the results in Sec. \ref{sec_MLT}, Sec. \ref{Sec_enforce} and Thms. \ref{uniqueness}-\ref{Thm4}, to provide a non-convex $\rank$-based optimization and a convex nuclear norm-based optimization to solve the $\ell_{0}$-\ac{AN} and $\ell_{1}$-\ac{AN} cost functions in $\U^\F_\TT$ of respectively \eqref{l0normoptAN} and \eqref{l1normoptAN}. The conditions under which the $\rank$-based and the nuclear norm-based optimizations recover the sparse model $\mihu(\ff_{1:\K})$ and $\ff_{1:\K}$ are given in the next two Theorems.

\begin{theorem}
\label{Thm1}
Let $\miy=\miQ\mihu(\ff_{1:\K})$  be a noiseless linear measurement where $\mihu(\ff_{1:\K})$ is a \emph{sparse} channel model as in \eqref{ec_hu}  with frequencies $\ff_{1:\K}$ generated under A.\ref{A1}. Given the optimization problem:
	\begin{opteq}
	\begin{aligned}
		\min_{r,  \mil_u \in \C^{\Tu}, \miT_{\TT} \in {\mathcal{T}^{\ddl}_\TT} }  \, \,&  \rank \left \{ \miT_{\TT}\right\}\\\quad \quad   
		{\rm s.t.} 	\quad 	&\begin{bmatrix}\miT_{\TT}& \mil_u\\ \mil_u^\H & r\end{bmatrix}\succeq 0,\quad & \miy=\miQ \mil_u.
	\end{aligned}
	\label{ec_Thm1}
	\end{opteq}
 where  we set the {\em dimension vector} $\TT$ such that   $\K<  \Tsf_{\ddl}$, its  optimal solution $(r^\opt,\mil_u^\opt,\miT_{\TT}^\opt)$
uniquely identifies the sparse channel model, i.e.  $\mil_u^\opt=\mihu(\ff_{1:\K})$
iff the measurement matrix $\miQ$, and the atoms set, $\U^\F_\TT$, satisfies  Condition C.1 in Thm. \ref{uniqueness}  for all $r \leq K$. Furthermore,  via \refAlg1~, $\ff_{1:\K}\in \T^{\ddl\times K}$ can be uniquely and gridlessly recovered by Vandermonde Decomposition  of  $\miT_{\TT}^\opt$.
\end{theorem}

\begin{proof}  
The proof  is provided in Appendix \ref{proofThm1}.
\end{proof} 

{\arxiv \begin{remark}
Thm. \ref{Thm1} shows that $\TT$, and specifically $\Tsf_{\ddl}$, is the key structural parameter to regulate the trade-off between performance and complexity. The condition to ensure unique recovery of the \atoms, i.e. $\Tsf_{\ddl}>\K$ brings to the spotlight that larger values of $\Tsf_{\ddl}$ enable the recovery of richer propagation environments, at the expense of higher complexity of \refAlg1~ dictated by $\Tu= \prod_{i=1}^{\ddl}\Tsf_i$.
\end{remark}
}

From Thm. \ref{Thm1}, the following corollaries can be inferred.

\begin{corollary}
\label{cor+th1_1}
Given a noiseless linear measurement as in Thm. \ref{Thm1}, and  setting $\TT$ in \eqref{ec_Thm1}  such that 

\begin{eqnarray}
\Tsf_{\ddl}>\K, \qquad 
\sum_{i=1}^{\ddl} \Tsf_i{ \geq}2\K+(\ddl-1)
\label{sue2}
\end{eqnarray} 
under the assumption that $\miQ$ satisfies the conditions  ii) or iv) stated in Thm.  \ref{Thm3},
then the solution to the  optimization problem in \eqref{ec_Thm1} allows to uniquely reconstruct of the sparse model and uniquely recover the embed frequencies via \refAlg1. 
\end{corollary}

\begin{corollary}
\label{cor+th1_2}
Given a noiseless linear measurement as in Thm. \ref{Thm1} with $\miQ$ structured as in \eqref{Q_def},  and  setting $\TT$ in \eqref{ec_Thm1} such that 
$\Tsf_{\ddl}>\K$, under the assumption that  
$\miP^\top$ satisfies conditions  ii) or iv) stated in Thm.  \ref{Thm4}, and that $\AL$ in \eqref{Q_def} admits the associated \rd~satisfying condition  \emph{i)} stated in Thm.  \ref{Thm4}, i.e. 
 \begin{eqnarray}
     \kappaL \geq  2\K  + (\ddl -1)
     \label{sue5}
 \end{eqnarray}
then the solution to the optimization problem in \eqref{ec_Thm1} allows to uniquely reconstruct the  sparse model from its noiseless measurement  and uniquely recover the associated frequencies.   
\end{corollary}

{\arxiv\begin{remark}
{\arxiv
Corollaries \ref{cor+th1_1} and \ref{cor+th1_2} typically require  $\K$ to set 
$\TT$ in \eqref{ec_Thm1} according to \eqref{sue2} and \eqref{sue5}. However, this can be bypassed by checking if the rank of 
$\miT_{\TT}^\opt$  
  after solving \eqref{ec_Thm1} meets the conditions:
\begin{eqnarray} 
 \Tsf_{\ddl}> \rank \left \{ \miT_{\TT}^\opt\right\}, \quad 
\sum_{i=1}^{\ddl} \Tsf_i>2\rank \left \{ \miT_{\TT}^\opt\right\}+(\ddl-1).
\label{sue3} 
\end{eqnarray} 
and 
\begin{eqnarray} 
 \Tsf_{\ddl}> \rank \left \{ \miT_{\TT}^\opt\right\}, \quad 
 \kappaL >2\rank \left \{ \miT_{\TT}^\opt\right\}+(\ddl-1).
\label{sue6} 
\end{eqnarray}
If these hold, unique reconstruction and frequency recovery are ensured as shown in the proofs of Corollaries \ref{cor+th1_1} and \ref{cor+th1_2}.
}
\end{remark}}

The $\rank$-based optimization in Thm. \ref{Thm1}, addresses \eqref{l0normoptAN} and its corresponding $\ell_{0}$-\ac{AN} cost function in $\U^\F_\TT$. This non-convex approach can be relaxed by addressing the $\ell_{1}$-\ac{AN} cost function in \eqref{l1normoptAN} through the nuclear-norm based optimization posed in the next Theorem. 

\begin{theorem}
\label{Thm2}
Given a noiseless linear measurement $\miy=\miQ\mihu(\ff_{1:\K})$ with  $\mihu(\ff_{1:\K})$  a \emph{sparse} channel model as in \eqref{ec_hu}  whose frequencies $\ff_{1:\K}$ are generated under A.\ref{A1}, let us consider  the optimization problem:
\begin{opteq}
	\begin{aligned}
	\min_{t, \mil_u \in \C^{\Tu}, \miT_{\TT} \in \mathcal{T}^{\ddl}_\TT } &  \frac{1}{2}t+\frac{1}{2}\tr\left\{\miT_{\TT}\right\} 
	\\\quad {\rm s.t.} 	\quad 	&\begin{bmatrix}\miT_{\TT} &\mil_u\\ \mil_u^\H & t\end{bmatrix}\succeq 0,\quad & \miy=\miQ \mil_u.
	\end{aligned}
	\label{ec_Thm2}
\end{opteq}
 where  we set the {\em dimension vector} $\TT$ such that   $\K<  \Tsf_{\ddl}$. Then, the 
 optimal solution  $(t^\opt,\mil_u^\opt,\miT_{\TT}^\opt)$ to  \emph{\ref{ec_Thm2}}
uniquely identifies the sparse channel model and  the frequencies, if \emph{i)} the measurement matrix,  $\miQ$, and the \atom~set, $\U^\F_\TT$, satisfies  Condition C.1 in Thm. \ref{uniqueness}  for all $r \leq \rank \left \{ \miT_{\TT}^\opt\right\}$, and \emph{ii)}
 $\rank \left \{ \miT_{\TT}^\opt\right\}< \Tsf_{\ddl}$
The way of uniquely and gridlessly recovering the frequencies $\ff_{1:\K}\in \T^{\ddl\times K}$  is via \refAlg1, by Vandermonde Decomposition of  $\miT_{\TT}^\opt$.
\end{theorem}
\begin{proof}  
The proof  is provided in Appendix \ref{proofThm2}.
\end{proof}

{\arxiv \begin{remark}
{\arxiv Comparing Thm. \ref{Thm1} with Thm. \ref{Thm2}, it appears clear that solving for $\ell_{1}$-\ac{AN} requires stronger constraints on $\miQ$ and $\TT$. }
\end{remark}}

As done previously for Thm \ref{Thm1}, the following corollaries can be derived for Thm. \ref{Thm2}.

\begin{corollary}
\label{cor+th2_1}
Given a noiseless linear measurement as in Thm. \ref{Thm2}, and setting $\TT$ in \eqref{ec_Thm2} such that 
\begin{eqnarray}
\Tsf_{\ddl}>\K, \qquad 
\sum_{i=1}^{\ddl} \Tsf_i{ \geq}2\K+(\ddl-1)
\label{sue4}
\end{eqnarray}
if $\miQ$  satisfies the conditions  \emph{ii)} or \emph{iv)} stated in Thm.  \ref{Thm3}, then the solution to the optimization problem in \eqref{ec_Thm2} allows to uniquely reconstruct the  sparse model and uniquely recover the associated frequencies (i.e. \atoms) if $\rank\{\miT_{\TT}^\opt \} = \K$.
\end{corollary} 

\begin{corollary}
\label{cor+th2_2}
Given a noiseless linear measurement as in Thm. \ref{Thm2}, with  a measured matrix $\miQ$  structured as in \eqref{Q_def}, setting $\TT$ in \eqref{ec_Thm2} such that  $\Tsf_{\ddl}>\K$, under the assumption that $\miP^\top$ satisfies conditions \emph{ii)} or \emph{iv)} stated in Thm. \ref{Thm4}, 
and that $\AL=(\Atx\otimes\Arx)\miPi_u$ in \eqref{Q_def} admits an associated \rds~satisfying $\kappaL\geq2\K+(\ddl-1)$, then, the solution to the   optimization problem in \eqref{ec_Thm2} allows to uniquely reconstruct the  sparse model from its noiseless measurement  and uniquely recover the associated frequencies if $\rank\{\miT_{\TT}^\opt \} = \K$.   
\end{corollary} 

{\arxiv \begin{remark}
{\arxiv Similar to Corollaries \ref{cor+th1_1} and \ref{cor+th1_2}, Corollaries \ref{cor+th2_1} and \ref{cor+th2_2} ensure perfect reconstruction and recovery without prior knowledge of 
$\K$. This is achieved by verifying that the rank of $\miT_{\TT}^\opt$ satisfies all the conditions that 
$\K$ would need to meet.}
\end{remark}}

\subsection{Noisy sparse model recovery under generalized linear measurements}
\label{Sec:noiseless}

The atomic norm denoising approach given in \eqref{eq:SDPnoisy} addresses Problem \ref{P_noisy}, where now $\miy=\miQ\mihu(\ff_{1:\K})+\miw$ is a noisy observation of the sparse model $\mihu(\ff_{1:\K})$. In this approach, the $\ell_{1}$-\ac{AN} is recast as a nuclear norm minimization problem, similarly to \eqref{ec_Thm2}, where we modify the constraint on the noiseless observation to take into account the noise presence. The proposed optimization problem is given next:
\begin{opteq}
	\begin{aligned}
		\min_{t, \mil_u \in \C^{\Tu}, \miT_{\TT} \in \mathcal{T}^{\ddl}_\TT }  \!& \left( \frac{1}{2}t+\frac{1}{2}\tr\left\{\miT_{\TT}\right\}\right) \\\quad 
		{\rm s.t.} 	&\quad 	\begin{bmatrix}\miT_{\TT} &\mil_u\\ \mil_u^\H & t\end{bmatrix}\succeq 0, \quad  \frac{1}{\P\Na}\|  \miQ \mil_u -  \miy \|^2_2\leq \sigma^2_w
	\end{aligned}
	\label{eq:Tracenoisy}
\end{opteq}
The performance of this denoising approach is comparatively evaluated by simulation in next section.

\section{Benchmarks and Results}

In this section we provide numerical results to assess the performance of the proposed nuclear atomic norm approach for parametric channel estimation. We address both noiseless and noisy scenarios, i.e. we respectively solve \eqref{ec_Thm2} and \eqref{eq:Tracenoisy}. 

We choose the antenna deployment scenarios consisting of an uniform planar array array ($\dd_\r=2$) at the receiver side with $\NN=[\Nsf_1, \Nsf_2]=[4,6]$ and $\Na=\Nu=24$ antenna elements and an uniform linear array ($\dd_\t=1$) at the transmitter side with dimension vector $\MM=[\Msf_1]=[4]$ and $\Ma=\Mu=4$ antenna elements. Being both array deployments uniform, one option for the sensing matrices would be to set them $\Atx=\miI_\Mu$ and $\Arx=\miI_\Nu$ leading also to $\AL=(\Atx\otimes\Arx)\miPi_u=\miI_\Tu$. In this scenario, the underlying uniform composite array of $\ddl=3$, coincides with a simple concatenation of $\MM$ and $\NN$, i.e. $\TT=[\Tsf_1, \Tsf_2, \Tsf_3]=[\Msf_1, \Nsf_1, \Nsf_2]=[4,4,6]$ with $\Tu=\Mu\Nu=\Ma\Na=96$.  With this choice of underlying \emph{dimension vector} $\TT$ and sensing matrix $\AL=\miI_\Tu$, the \ntrivial~embed structure of Def. \ref{def:struct0} coincides with $\TT$ and the \rd~of Def. \ref{kappa}, in this case, is $\kappaL=\sum_{i=1}^3\Tsf_i=14$. Note that with this choice of $\kappaL$ the sparse channel model $\mihu$ is uniquely identified and recovered using \eqref{ec_Thm1} if the number of scatters $\K\leq \frac{\kappaL-2}{2}=6$, given that the conditions \emph{ii)} or \emph{iv)} in Thm. \ref{Thm4} would also hold. Note that, to in addition ensure the unique recovery of the \atoms~and frequency parameters, we have in this case a more restrictive condition since $\K<\Tsf_{\ddl}=6$.

To explore both potential recovery and non recovery scenarios,  we target in our simulations different sparse propagation scenarios with $\K\in\{1,2,3,4,5,6\}$ scatters. The scatters are located randomly at different \ac{AoD} and \ac{AoA} that translate to channel frequency parameters $\ff_k=[-\mig_k^\top, \mif_k^\top]^\top$ with $k\in[\K]$ generated with an uniform distribution in $\T^{3\times\K}$ to ensure that condition A.\ref{A1} holds. Finally, the $k$-th path gains $\gamma_k\in\C$ are randomly generated with a normal Gaussian i.i.d. distribution.

The measurement matrix $\miQ$ follows the structure in \eqref{Q_def}, i.e. $\miQ=\left(\miP^\top\otimes\miI_\Na\right)\AL=\left(\miP^\top\otimes\miI_\Na\right)\in\{\Pcal\cup\{0\}\}^{\Ta\times\Tu}$. Several pilot sequences $\miP$ have been evaluated using different pilot alphabets $\Pcal$ with symbols both, generated following A.\ref{A2} and not complying with this assumption, and different pilots sequence lengths $\P$. The pilot alphabets follow \emph{i)} a BPSK constellation with symbols generated as $\pm 1$ and identified as \bin~in the results, \emph{ii)} a QPSK constellation with symbols generated as $\pm 1\pm\j$ identified as \binc~in the results, and \emph{iii)} a benchmark constellation with pilots following a normal Gaussian distribution in the real domain, that we name \rand. All pilot sequences generated following the \bin~and \binc~constellations are uniform randomly generated. The pilot sequence lengths considered are $\P\in\{1,3,4,6\}$. These pilot length scenarios lead to different measurement scenarios given by the $\miQ\in\{\Pcal\cup\{0\}\}^{\Ta\times\Tu}$ measurement matrix, where $\Ta\in\{24,72,96,144\}$, i.e. the $\miQ$ evaluated consider both the $\Ta<\Tu$ and the $\Ta\geq\Tu$ scenarios. Finally, recalling that $\Ma=4$, $\miP\in\mathcal{P}^{4\times \P}$ comply almost surely with conditions \emph{ii)} of Thm. \ref{Thm4} when $\P=\{4,6\}$ with the exception of the BPSK constellation, where for a significant number of realizations the $\miP$ matrix is $\rank$-deficient, i.e. $\rank\{\miP\}<\min\{\P,\Ma\}$.

{\arxiv \begin{table*}[h!]
\begin{center}
{\arxiv \begin{tabular}{|p{3cm}| c|}
\hline
&\\
$\ell_1$-AN \eqref{ec_Thm2} / \eqref{eq:Tracenoisy} & $O\left(\Tu^{3.5}+\Tu^{2.5}+\sqrt{\Tu})\log(1/\epsilon)\right)$  \\&\\
\hline&\\
OMP \cite{Swapna22} & $O\big(\K\left(L_{\text{OMP}}^2\K + \Tu\K + L_{\text{OMP}}\right) + \sum_{k=1}^\K k^3 + 2\Tu k^2 + 2\Tu\K k\big)$   \\&\\
\hline&\\
MD-MUSIC \cite{Liao15} & $O\big(L_{\text{MUSIC}}\left(H^2 + H\ddl - H\K + 2H - \K + 1\right) + H(H-\K)^2\big)$ \\&\\\hline&\\
    IIC \cite{Grossi20} & $O\left(\K(L_{\text{IIC}} \Tu^2+\Tu^3+5\Tu^2+2L_{\text{IIC}})\right)$
  \\&\\
\hline
\end{tabular}
\caption{Complexity of benchmarks.} }
\label{resolvable}
\end{center}
\end{table*}}

\begin{figure*}
\centerline{
    \begin{subfigure}[\channelvec.]
    {
%
%
\pgfplotsset{
    legend image with text/.style={
        legend image code/.code={%
            \node[anchor=center] at (0.3cm,0cm) {#1};
        }
    },
}

\begin{tikzpicture}

\begin{axis}[%
width=0.6\figurewidth,
height=0.6\figureheight,
at={(0\figurewidth,0\figureheight)},
scale only axis,
xmin=1,
xmax=6,
xlabel style={font=\color{white!15!black}},
xlabel={$\K$},
ymode=log,
ymin=1e-25,
ymax=1.1e-1,
yminorticks=true,
ylabel={$\frac{1}{\Tu}\E\{\|\mihu - \widehat{\mih}_u\|_2^2\}$},
axis background/.style={fill=white},
axis x line*=bottom,
axis y line*=left,
xmajorgrids,
ymajorgrids,
legend columns=3, 
legend style={
    /tikz/column 5/.style={
        column sep=5pt,
    },
fill=none},
legend style={at={(0.2,0.65)}, anchor=north west, legend cell align=left, align=left, draw=none}
]

\addlegendimage{legend image with text={\tiny \rand{} }}
\addlegendentry{}
\addlegendimage{legend image with text={\tiny \binc{} }}
\addlegendentry{}
\addlegendimage{legend image with text={\tiny \bin{} }}
\addlegendentry{}


\addplot [color=red, dotted, line width=1.0pt, mark=square, mark options={solid, rotate=180, red}]
  table[row sep=crcr]{%
1	9.00625846605888e-05\\
2	8.21037015205213e-05\\
3	7.77435846585031e-05\\
4	8.53914733931776e-05\\
5	8.05614683274549e-05\\
6	8.07448685208916e-05\\
};
\addlegendentry{}

\addplot [color=blue, line width=1.0pt, mark=square, mark options={solid, rotate=180, blue}]
  table[row sep=crcr]{%
1	9.56461005830934e-05\\
2	8.66972354939276e-05\\
3	8.15560333914114e-05\\
4	8.78235536264738e-05\\
5	8.28905019808977e-05\\
6	8.40137796968066e-05\\
};
\addlegendentry{}

\addplot [color=black, dashdotted, line width=1.0pt, mark=square, mark options={solid, rotate=180, black}]
  table[row sep=crcr]{%
1	8.86538232206745e-05\\
2	8.17542905507572e-05\\
3	7.78218094379345e-05\\
4	8.51888146678123e-05\\
5	8.1092264568171e-05\\
6	8.15807328687055e-05\\
};
\addlegendentry{\tiny $\P=1$}


\addplot [color=red, dotted, line width=1.0pt, mark=star, mark options={solid, rotate=180, red}]
  table[row sep=crcr]{%
1	5.73671980293745e-08\\
2	1.34816085049858e-06\\
3	1.0973149664943e-06\\
4	2.05975596847467e-06\\
5	2.73617309163004e-06\\
6	3.86224604305412e-06\\
};
\addlegendentry{}

\addplot [color=blue, line width=1.0pt, mark=star, mark options={solid, rotate=180, blue}]
  table[row sep=crcr]{%
1	8.02265790032557e-06\\
2	6.22907757647443e-06\\
3	6.10598837858041e-06\\
4	7.72040795977499e-06\\
5	8.14697652896702e-06\\
6	9.78304600786135e-06\\
};
\addlegendentry{}

\addplot [color=black, dashdotted, line width=1.0pt, mark=star, mark options={solid, rotate=180, black}]
  table[row sep=crcr]{%
1	2.16561183221213e-05\\
2	1.82865585751679e-05\\
3	1.87499324598554e-05\\
4	2.18465885071144e-05\\
5	2.4182985991276e-05\\
6	2.68219895611781e-05\\
};
\addlegendentry{\tiny $\P=3$}


\addplot [color=red, dotted, line width=1.0pt, mark=triangle, mark options={solid, rotate=180, red}]
  table[row sep=crcr]{%
1	2.89195079949829e-23\\
2	2.74918445740074e-23\\
3	1.48167411926037e-23\\
4	5.2136153753185e-23\\
5	1.50777061461272e-22\\
6	5.02709056472672e-22\\
};
\addlegendentry{}

\addplot [color=blue, line width=1.0pt, mark=triangle, mark options={solid, rotate=180, blue}]
  table[row sep=crcr]{%
1	3.9571354848804e-20\\
2	4.18037167052584e-08\\
3	2.00477347975159e-08\\
4	3.90668784565146e-08\\
5	3.52976149682567e-07\\
6	4.42283174911447e-07\\
};
\addlegendentry{}

\addplot [color=black, dashdotted, line width=1.0pt, mark=triangle, mark options={solid, rotate=180, black}]
  table[row sep=crcr]{%
1	1.08105104996193e-05\\
2	8.5058708356764e-06\\
3	9.16276898451936e-06\\
4	1.21655731595093e-05\\
5	1.39821588816287e-05\\
6	1.65528977792083e-05\\
};
\addlegendentry{\tiny $\P=4$}


\addplot [color=red, dotted, line width=1.0pt, mark=o, mark options={solid, red}]
  table[row sep=crcr]{%
1	4.06385224915541e-23\\
2	3.14602338692174e-23\\
3	2.54959972661764e-23\\
4	1.02952787815346e-23\\
5	4.79402856949559e-24\\
6	1.16712866872579e-24\\
};
\addlegendentry{}

\addplot [color=blue, line width=1.0pt, mark=o, mark options={solid, blue}]
  table[row sep=crcr]{%
1	2.4835100055961e-23\\
2	2.59249968136535e-23\\
3	2.02574912613448e-23\\
4	9.54062006322288e-24\\
5	5.37860440299756e-24\\
6	6.06151647999978e-25\\
};
\addlegendentry{}

\addplot [color=black, dashdotted, line width=1.0pt, mark=o, mark options={solid, black}]
  table[row sep=crcr]{%
1	4.6135857796623e-06\\
2	3.28314290017203e-06\\
3	3.49179533796551e-06\\
4	4.71878069330314e-06\\
5	4.84715674236787e-06\\
6	5.82780748685492e-06\\
};
\addlegendentry{\tiny $\P=6$}

\end{axis}
\end{tikzpicture}%
}

    \end{subfigure}
    \begin{subfigure}[\freqvec.]
    {
%
%
\pgfplotsset{
    legend image with text/.style={
        legend image code/.code={%
            \node[anchor=center] at (0.3cm,0cm) {#1};
        }
    },
}

\begin{tikzpicture}

\begin{axis}[%
width=0.6\figurewidth,
height=0.6\figureheight,
at={(0\figurewidth,0\figureheight)},
scale only axis,
xmin=1,
xmax=6,
xlabel style={font=\color{white!15!black}},
xlabel={$\K$},
ymode=log,
ymin=1e-6,
ymax=1e-1,
yminorticks=true,
ylabel style={font=\color{white!15!black}},
ylabel={$\error$},
axis background/.style={fill=white},
axis x line*=bottom,
axis y line*=left,
xmajorgrids,
ymajorgrids,
legend columns=3, 
legend style={
    /tikz/column 5/.style={
        column sep=5pt,
    },
},
legend style={at={(0.0,1.35)}, anchor=north west, legend cell align=left, align=left, draw=none}
]




\addplot [color=red, dotted, line width=1.0pt, mark=square, mark options={solid, rotate=180, red}]
  table[row sep=crcr]{%
1	0.0367054181117952\\
2	0.0289572116742773\\
3	0.031211650247522\\
4	0.0308533649803278\\
5	0.0316255700561525\\
6	0.0349518840236854\\
};

\addplot [color=blue, line width=1.0pt, mark=square, mark options={solid, rotate=180, blue}]
  table[row sep=crcr]{%
1	0.0301183333258359\\
2	0.0279788030433877\\
3	0.0289899586769881\\
4	0.0284681546694885\\
5	0.0291392012316574\\
6	0.0339066445939259\\
};

\addplot [color=black, dashdotted, line width=1.0pt, mark=square, mark options={solid, rotate=180, black}]
  table[row sep=crcr]{%
1	0.0367052779933611\\
2	0.0271430838521819\\
3	0.0285558910340676\\
4	0.0283865815915517\\
5	0.0293810887644506\\
6	0.0324569202986861\\
};


\addplot [color=red, dotted, line width=1.0pt, mark=star, mark options={solid, rotate=180, red}]
  table[row sep=crcr]{%
1	4.39571745539806e-06\\
2	0.000406121939768133\\
3	0.000351525040787768\\
4	0.000721236062328038\\
5	0.0010358954483237\\
6	0.00579191364142636\\
};

\addplot [color=blue, line width=1.0pt, mark=star, mark options={solid, rotate=180, blue}]
  table[row sep=crcr]{%
1	0.00231208282512167\\
2	0.00221078616243823\\
3	0.00225870482165106\\
4	0.0026309289855142\\
5	0.00291287424749236\\
6	0.00816620497666183\\
};

\addplot [color=black, dashdotted, line width=1.0pt, mark=star, mark options={solid, rotate=180, black}]
  table[row sep=crcr]{%
1	0.00752813546398556\\
2	0.00734610085178527\\
3	0.00720180792089269\\
4	0.00738228814256245\\
5	0.00837881886621292\\
6	0.013297446678245\\
};


\addplot [color=red, dotted, line width=1.0pt, mark=triangle, mark options={solid, rotate=180, red}]
  table[row sep=crcr]{%
1	4.17074083760235e-21\\
2	2.44343686777832e-06\\
3	9.26734067090861e-06\\
4	9.80148243395939e-06\\
5	4.10210523521456e-05\\
6	0.0044398321121478\\
};

\addplot [color=blue, line width=1.0pt, mark=triangle, mark options={solid, rotate=180, blue}]
  table[row sep=crcr]{%
1	9.18630694155248e-20\\
2	6.79808195358659e-06\\
3	1.37005303436668e-05\\
4	1.37080762699233e-05\\
5	8.3777172885952e-05\\
6	0.00505560869375245\\
};

\addplot [color=black, dashdotted, line width=1.0pt, mark=triangle, mark options={solid, rotate=180, black}]
  table[row sep=crcr]{%
1	0.00403822581420873\\
2	0.00406272302735469\\
3	0.00410900565155575\\
4	0.00442263167953439\\
5	0.00524373429945414\\
6	0.0101710127707745\\
};


\addplot [color=red, dotted, line width=1.0pt, mark=o, mark options={solid, red}]
  table[row sep=crcr]{%
1	6.4235645751004e-21\\
2	2.44248572127208e-06\\
3	9.26705587159001e-06\\
4	9.80129562393312e-06\\
5	4.10216682759274e-05\\
6	0.00431512366255111\\
};

\addplot [color=blue, line width=1.0pt, mark=o, mark options={solid, blue}]
  table[row sep=crcr]{%
1	3.27235350883308e-21\\
2	2.44264581378741e-06\\
3	9.26702161855558e-06\\
4	9.80128544129624e-06\\
5	4.10222757312763e-05\\
6	0.00451636056227977\\
};

\addplot [color=black, dashdotted, line width=1.0pt, mark=o, mark options={solid, black}]
  table[row sep=crcr]{%
1	0.00192569100130568\\
2	0.00151324023545717\\
3	0.00155154766337676\\
4	0.00159178720282031\\
5	0.00193529420240062\\
6	0.00703421773711338\\
};

\end{axis}
\end{tikzpicture}%
}
    \end{subfigure}
    }
      
    \caption{\acs{MSE} performance in a noiseless scenario solving \eqref{ec_Thm2} for $\TT=[4,4,6]$ and different pilot alphabets with $\P=\{1,3,4,6\}$}.
    \label{fig:freq_error_noiseless}
\end{figure*}
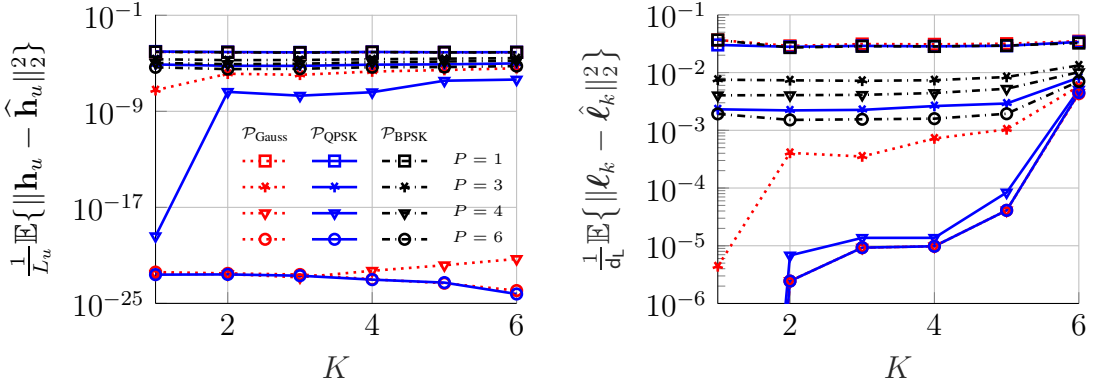
\begin{figure*}
    \centerline{
        \begin{subfigure}[\channelvec, $\P=3$]
            {\centering
%
%
\pgfplotsset{
    legend image with text/.style={
        legend image code/.code={%
            \node[anchor=center] at (0.3cm,0cm) {#1};
        }
    },
}

\begin{tikzpicture}

\begin{axis}[%
width=0.6\figurewidth,
height=0.6\figureheight,
at={(0\figurewidth,0\figureheight)},
scale only axis,
xmin=-10,
xmax=50,
xtick={-10, 0, 10,20,30,40,50},
xlabel style={font=\color{white!15!black}},
xlabel={SNR(dB)},
ymode=log,
ymin=1e-07,
ymax=0.0001,
ylabel style={font=\color{white!15!black}},
ylabel={$\frac{1}{\Tu}\E\{\|\mihu - \widehat{\mih}_u\|_2^2\}$},
axis background/.style={fill=white},
axis x line*=bottom,
axis y line*=left,
xmajorgrids,
ymajorgrids,
legend columns=3, 
legend style={
    /tikz/column 5/.style={
        column sep=5pt,
    },
fill=none},
legend style={at={(0.05,0.45)}, anchor=north west, legend cell align=left, align=left, draw=none}
]

\addlegendimage{legend image with text={\tiny \rand{} }}
\addlegendentry{}
\addlegendimage{legend image with text={\tiny \binc{} }}
\addlegendentry{}
\addlegendimage{legend image with text={\tiny \bin{} }}
\addlegendentry{}


\addplot [color=red, dotted, line width=1.0pt, mark=square, mark options={solid, rotate=180, red}]
  table[row sep=crcr]{%
-10	8.44701629813907e-05\\
-5	6.82448360628329e-05\\
-3	6.01983677888656e-05\\
0	4.7625186588006e-05\\
3	3.58229273993363e-05\\
5	2.90919914912832e-05\\
10	1.68134524117468e-05\\
20	5.73499551298119e-06\\
30	2.26111993271403e-06\\
50	1.10153045842248e-06\\
100	1.0968293133337e-06\\
};
\addlegendentry{}

\addplot [color=blue, line width=1.0pt, mark=square, mark options={solid, rotate=180, blue}]
  table[row sep=crcr]{%
-10	8.38982174300834e-05\\
-5	6.31809547109904e-05\\
-3	5.39693615805785e-05\\
0	4.15842564648189e-05\\
3	3.18713133916851e-05\\
5	2.68574533752615e-05\\
10	1.82527196885299e-05\\
20	1.13705089020953e-05\\
30	1.02619964362943e-05\\
50	9.83129076615377e-06\\
100	6.05711627472488e-06\\
};
\addlegendentry{}

\addplot [color=black, dashdotted, line width=1.0pt, mark=square, mark options={solid, rotate=180, black}]
  table[row sep=crcr]{%
-10	8.55328785740158e-05\\
-5	6.83710547892817e-05\\
-3	6.0820200476855e-05\\
0	5.0580819234934e-05\\
3	4.238251590392e-05\\
5	3.83891515979509e-05\\
10	3.22174089525163e-05\\
20	2.87733196518523e-05\\
30	2.77420239006881e-05\\
50	2.50058268038966e-05\\
100	1.88518473316577e-05\\
};
\addlegendentry{\tiny $\K=3$}


\addplot [color=red, dotted, line width=1.0pt, mark=o, mark options={solid, rotate=180, red}]
  table[row sep=crcr]{%
-10	9.08406815315324e-05\\
-5	7.429571735576e-05\\
-3	6.62418185524088e-05\\
0	5.3335725310172e-05\\
3	4.09885025868496e-05\\
5	3.36215523823331e-05\\
10	1.98981484403109e-05\\
20	7.61808512298534e-06\\
30	3.61496899456433e-06\\
50	2.11671141898549e-06\\
100	2.06002284673554e-06\\
};
\addlegendentry{}

\addplot [color=blue, line width=1.0pt, mark=o, mark options={solid, rotate=180, blue}]
  table[row sep=crcr]{%
-10	8.97251944736358e-05\\
-5	6.8776269336144e-05\\
-3	5.93779992354955e-05\\
0	4.63919439036032e-05\\
3	3.58265489217855e-05\\
5	3.02761054616286e-05\\
10	2.07029685906623e-05\\
20	1.29584438348687e-05\\
30	1.15483082000311e-05\\
50	1.0965509232767e-05\\
100	7.05989949348063e-06\\
};
\addlegendentry{}

\addplot [color=black, dashdotted, line width=1.0pt, mark=o, mark options={solid, rotate=180, black}]
  table[row sep=crcr]{%
-10	9.3690241673416e-05\\
-5	7.57727437762959e-05\\
-3	6.79138180228291e-05\\
0	5.69480741761102e-05\\
3	4.81105891544625e-05\\
5	4.34772720674437e-05\\
10	3.63730197546221e-05\\
20	3.20760004727468e-05\\
30	3.11376539936848e-05\\
50	2.7750561275731e-05\\
100	2.19962465553611e-05\\
};
\addlegendentry{\tiny $\K=4$}


\addplot [color=red, dotted, line width=1.0pt, mark=triangle, mark options={solid, rotate=180, red}]
  table[row sep=crcr]{%
-10	8.81865334990169e-05\\
-5	7.35060904898318e-05\\
-3	6.6207438820551e-05\\
0	5.43812658909409e-05\\
3	4.25548534752059e-05\\
5	3.52931401422274e-05\\
10	2.13572824111967e-05\\
20	8.68128914717479e-06\\
30	4.52369037807884e-06\\
50	2.80273964061309e-06\\
100	2.73620551417136e-06\\
};
\addlegendentry{}

\addplot [color=blue, line width=1.0pt, mark=triangle, mark options={solid, rotate=180, blue}]
  table[row sep=crcr]{%
-10	8.7636929440267e-05\\
-5	6.91886598940932e-05\\
-3	6.03467776671221e-05\\
0	4.75986215118144e-05\\
3	3.6927596618698e-05\\
5	3.1228735350904e-05\\
10	2.12674863889972e-05\\
20	1.27723368197697e-05\\
30	1.10128317881096e-05\\
50	1.0421839912929e-05\\
100	8.03244823836962e-06\\
};
\addlegendentry{}

\addplot [color=black, dashdotted, line width=1.0pt, mark=triangle, mark options={solid, rotate=180, black}]
  table[row sep=crcr]{%
-10	9.12520425528775e-05\\
-5	7.60489404218394e-05\\
-3	6.87665905026524e-05\\
0	5.79927733287731e-05\\
3	4.88769632565326e-05\\
5	4.38291038954015e-05\\
10	3.59664871915694e-05\\
20	3.1295649576622e-05\\
30	3.02061353302029e-05\\
50	2.81640494994094e-05\\
100	2.44181616293921e-05\\
};
\addlegendentry{\tiny $\K=5$}


\end{axis}
\end{tikzpicture}%
}
        \end{subfigure}
        \begin{subfigure}[\channelvec, $\P=4$]
            {\centering
%
%
\begin{tikzpicture}

\begin{axis}[%
width=0.6\figurewidth,
height=0.6\figureheight,
at={(0\figurewidth,0\figureheight)},
scale only axis,
xmin=-10,
xmax=50,
xtick={-10, 0, 10,20,30,40,50},
xlabel style={font=\color{white!15!black}},
xlabel={SNR(dB)},
ymode=log,
ymin=1e-7,
ymax=0.0001,
yminorticks=true,
ylabel style={font=\color{white!15!black}},
axis background/.style={fill=white},
axis x line*=bottom,
axis y line*=left,
xmajorgrids,
ymajorgrids,
legend style={at={(0.03,0.97)}, anchor=north west, legend cell align=left, align=left, draw=white!15!black}
]

\addplot [color=red, dotted, line width=1.0pt, mark=square, mark options={solid, rotate=180, red}]
  table[row sep=crcr]{%
-10	8.2844324650001e-05\\
-5	6.05725596850124e-05\\
-3	5.05700170429908e-05\\
0	3.67085247464913e-05\\
3	2.56159744704422e-05\\
5	1.98127195066445e-05\\
10	9.51716951434503e-06\\
20	1.84358673929529e-06\\
30	5.61076289479961e-07\\
50	6.1644418980475e-08\\
100	4.25623468665312e-12\\
};

\addplot [color=blue, line width=1.0pt, mark=square, mark options={solid, rotate=180, blue}]
  table[row sep=crcr]{%
-10	8.23717920113926e-05\\
-5	5.51812847029598e-05\\
-3	4.38081619392334e-05\\
0	2.93354100583893e-05\\
3	1.89040875798592e-05\\
5	1.39057721410565e-05\\
10	6.20432724224332e-06\\
20	1.29261355535382e-06\\
30	4.2058540493825e-07\\
50	3.83880240422361e-08\\
100	1.61897882529292e-08\\
};

\addplot [color=black, dashdotted, line width=1.0pt, mark=square, mark options={solid, rotate=180, black}]
  table[row sep=crcr]{%
-10	8.5168279304297e-05\\
-5	6.32833207432012e-05\\
-3	5.4052709203476e-05\\
0	4.20448343612485e-05\\
3	3.31194450541184e-05\\
5	2.8762210475363e-05\\
10	2.26950279349913e-05\\
20	1.97039720851596e-05\\
30	1.93016011958752e-05\\
50	1.63588952859926e-05\\
100	9.04802063985541e-06\\
};


\addplot [color=red, dotted, line width=1.0pt, mark=o, mark options={solid, rotate=180, red}]
  table[row sep=crcr]{%
-10	8.86847204975189e-05\\
-5	6.64834365808702e-05\\
-3	5.64750394726544e-05\\
0	4.2142267838793e-05\\
3	2.9947412064216e-05\\
5	2.33439348271332e-05\\
10	1.16027920012281e-05\\
20	2.56797528657724e-06\\
30	9.15724989326386e-07\\
50	9.26539078306499e-08\\
100	1.99079891111191e-11\\
};

\addplot [color=blue, line width=1.0pt, mark=o, mark options={solid, rotate=180, blue}]
  table[row sep=crcr]{%
-10	8.74208290718165e-05\\
-5	6.02157176237789e-05\\
-3	4.85777149866402e-05\\
0	3.33164442198822e-05\\
3	2.18245604008876e-05\\
5	1.6246334362595e-05\\
10	7.49391884491981e-06\\
20	1.61279709193638e-06\\
30	5.35480924792571e-07\\
50	9.78689804831878e-08\\
100	3.9034264348821e-08\\
};

\addplot [color=black, dashdotted, line width=1.0pt, mark=o, mark options={solid, rotate=180, black}]
  table[row sep=crcr]{%
-10	9.27176349254551e-05\\
-5	7.02174446893611e-05\\
-3	6.05338491694168e-05\\
0	4.75999493522954e-05\\
3	3.7663596873366e-05\\
5	3.28138442663675e-05\\
10	2.58178581097409e-05\\
20	2.21699878466911e-05\\
30	2.16270070576304e-05\\
50	1.83675131421866e-05\\
100	1.19311787106065e-05\\
};


\addplot [color=red, dotted, line width=1.0pt, mark=triangle, mark options={solid, rotate=180, red}]
  table[row sep=crcr]{%
-10	8.68617303246086e-05\\
-5	6.71781014480409e-05\\
-3	5.79191130019438e-05\\
0	4.42730724029723e-05\\
3	3.2214504129586e-05\\
5	2.54898265263956e-05\\
10	1.33930541668826e-05\\
20	3.59925086339795e-06\\
30	1.6735503640201e-06\\
50	1.39732566461827e-07\\
100	7.668708046872e-12\\
};

\addplot [color=blue, line width=1.0pt, mark=triangle, mark options={solid, rotate=180, blue}]
  table[row sep=crcr]{%
-10	8.60471946571829e-05\\
-5	6.19238516831459e-05\\
-3	5.08519039157331e-05\\
0	3.5922574411804e-05\\
3	2.4168208606882e-05\\
5	1.82914005755917e-05\\
10	8.82087052787501e-06\\
20	2.07632322684413e-06\\
30	7.51922534740466e-07\\
50	3.86956712332931e-07\\
100	3.50082101878105e-07\\
};

\addplot [color=black, dashdotted, line width=1.0pt, mark=triangle, mark options={solid, rotate=180, black}]
  table[row sep=crcr]{%
-10	9.05869964752742e-05\\
-5	7.12619631476147e-05\\
-3	6.21450968995779e-05\\
0	4.9327133706183e-05\\
3	3.8855059575162e-05\\
5	3.3565648588637e-05\\
10	2.54893588780985e-05\\
20	2.13322892454885e-05\\
30	2.06492952947984e-05\\
50	1.91017589286286e-05\\
100	1.37924468821556e-05\\
};

\addplot [color=red, dotted, line width=1.0pt, mark=triangle, mark options={solid, rotate=180, black}]
  table[row sep=crcr]{%
-10	1.48167411926037e-23\\
-5	1.48167411926037e-23\\
-3	1.48167411926037e-23\\
0	1.48167411926037e-23\\
3	1.48167411926037e-23\\
5	1.48167411926037e-23\\
10	1.48167411926037e-23\\
20	1.48167411926037e-23\\
30	1.48167411926037e-23\\
50	1.48167411926037e-23\\
100	1.48167411926037e-23\\
};

\end{axis}
\end{tikzpicture}%
}
        \end{subfigure}
        \begin{subfigure}[\channelvec, $\P=6$]
            {\centering
%
%
\begin{tikzpicture}

\begin{axis}[%
width=0.6\figurewidth,
height=0.6\figureheight,
at={(0\figurewidth,0\figureheight)},
scale only axis,
xmin=-10,
xmax=50,
xtick={-10, 0, 10,20,30,40,50},
xlabel style={font=\color{white!15!black}},
xlabel={SNR(dB)},
ymode=log,
ymin=1e-7,
ymax=0.0001,
yminorticks=true,
ylabel style={font=\color{white!15!black}},
axis background/.style={fill=white},
axis x line*=bottom,
axis y line*=left,
xmajorgrids,
ymajorgrids,
legend style={at={(0.03,0.97)}, anchor=north west, legend cell align=left, align=left, draw=white!15!black}
]


\addplot [color=red, dotted, line width=1.0pt, mark=square, mark options={solid, rotate=180, red}]
  table[row sep=crcr]{%
-10	7.35982836859203e-05\\
-5	4.5947044386103e-05\\
-3	3.5888721156708e-05\\
0	2.37992255564654e-05\\
3	1.5186643174622e-05\\
5	1.09694396472561e-05\\
10	4.46012943224563e-06\\
20	6.00273877470399e-07\\
30	7.37289044587067e-08\\
50	1.07064012150485e-09\\
100	1.42918783301562e-14\\
};

\addplot [color=blue, line width=1.0pt, mark=square, mark options={solid, rotate=180, blue}]
  table[row sep=crcr]{%
-10	7.34681878724528e-05\\
-5	4.3445564036697e-05\\
-3	3.28309077042855e-05\\
0	2.03818947234549e-05\\
3	1.20547010487187e-05\\
5	8.31066219413166e-06\\
10	3.09835557167661e-06\\
20	3.77172053642257e-07\\
30	4.32135920380449e-08\\
50	5.68696229214635e-10\\
100	6.6491765306133e-15\\
};

\addplot [color=black, dashdotted, line width=1.0pt, mark=square, mark options={solid, rotate=180, black}]
  table[row sep=crcr]{%
-10	7.51700189355009e-05\\
-5	4.73800735276891e-05\\
-3	3.74243864860296e-05\\
0	2.56181891653025e-05\\
3	1.75743301606263e-05\\
5	1.4009169816737e-05\\
10	9.27055615676321e-06\\
20	6.88714804236854e-06\\
30	6.72823396062326e-06\\
50	6.41076460111646e-06\\
100	3.39362961761822e-06\\
};


\addplot [color=red, dotted, line width=1.0pt, mark=o, mark options={solid, rotate=180, red}]
  table[row sep=crcr]{%
-10	7.93833263197443e-05\\
-5	5.16881755492975e-05\\
-3	4.10890333866381e-05\\
0	2.77610145744359e-05\\
3	1.79210644037015e-05\\
5	1.30563900683964e-05\\
10	5.40977223181877e-06\\
20	7.63024376695786e-07\\
30	9.68988352338392e-08\\
50	1.47521277464575e-09\\
100	2.64067556440578e-14\\
};

\addplot [color=blue, line width=1.0pt, mark=o, mark options={solid, rotate=180, blue}]
  table[row sep=crcr]{%
-10	7.85050325165538e-05\\
-5	4.88018859236183e-05\\
-3	3.76691208558362e-05\\
0	2.40437121296178e-05\\
3	1.44879893168932e-05\\
5	1.00753371200231e-05\\
10	3.80943052277113e-06\\
20	4.72555973134693e-07\\
30	5.47637167950479e-08\\
50	7.3799354076725e-10\\
100	9.20020006944827e-15\\
};

\addplot [color=black, dashdotted, line width=1.0pt, mark=o, mark options={solid, rotate=180, black}]
  table[row sep=crcr]{%
-10	8.20479025258284e-05\\
-5	5.37946388804317e-05\\
-3	4.30824842154928e-05\\
0	3.00317300470443e-05\\
3	2.07119927405013e-05\\
5	1.65295886659088e-05\\
10	1.08055435302659e-05\\
20	7.92801544318032e-06\\
30	7.63378522081817e-06\\
50	7.27576205407869e-06\\
100	4.56342141825561e-06\\
};


\addplot [color=red, dotted, line width=1.0pt, mark=triangle, mark options={solid, rotate=180, red}]
  table[row sep=crcr]{%
-10	7.92455652468991e-05\\
-5	5.40345068882737e-05\\
-3	4.3717661796608e-05\\
0	3.01692811657142e-05\\
3	1.9756852592254e-05\\
5	1.45567217780023e-05\\
10	6.21913783803927e-06\\
20	9.15179431353425e-07\\
30	1.21800834627904e-07\\
50	2.06338241376753e-09\\
100	3.36772027677913e-14\\
};

\addplot [color=blue, line width=1.0pt, mark=triangle, mark options={solid, rotate=180, blue}]
  table[row sep=crcr]{%
-10	7.86730557500415e-05\\
-5	5.1583837272356e-05\\
-3	4.06027772970041e-05\\
0	2.66863112433269e-05\\
3	1.65167329235832e-05\\
5	1.16847628033189e-05\\
10	4.58018502757582e-06\\
20	5.92176700379197e-07\\
30	7.01608107195672e-08\\
50	9.68480619814922e-10\\
100	1.3347387665179e-14\\
};

\addplot [color=black, dashdotted, line width=1.0pt, mark=triangle, mark options={solid, rotate=180, black}]
  table[row sep=crcr]{%
-10	8.14730662226169e-05\\
-5	5.59166810918678e-05\\
-3	4.54233625109671e-05\\
0	3.20145185558407e-05\\
3	2.20803916041983e-05\\
5	1.74739123508602e-05\\
10	1.09342515440957e-05\\
20	7.54918279964095e-06\\
30	7.10475226213788e-06\\
50	6.83903421186234e-06\\
100	4.72352157663526e-06\\
};


\end{axis}
\end{tikzpicture}%
}
        \end{subfigure}
    }
    \centerline{
        \begin{subfigure}[\freqvec, $\P=3$]
            {\centering
%
%
\pgfplotsset{
    legend image with text/.style={
        legend image code/.code={%
            \node[anchor=center] at (0.3cm,0cm) {#1};
        }
    },
}

\begin{tikzpicture}

\begin{axis}[%
width=0.6\figurewidth,
height=0.6\figureheight,
at={(0\figurewidth,0\figureheight)},
scale only axis,
xmin=-10,
xmax=50,
xtick={-10, 0, 10,20,30,40,50},
xlabel style={font=\color{white!15!black}},
xlabel={SNR(dB)},
ymode=log,
ymin=0.00001,
ymax=0.1,
yminorticks=true,
ylabel style={font=\color{white!15!black}},
ylabel={$\error$},
axis background/.style={fill=white},
axis x line*=bottom,
axis y line*=left,
xmajorgrids,
ymajorgrids,
legend columns=3, 
legend style={
    /tikz/column 5/.style={
        column sep=5pt,
    },
fill=none},
legend style={at={(0.05,0.39)}, anchor=north west, legend cell align=left, align=left, draw=none}
]

\addlegendimage{legend image with text={\tiny \rand{} }}
\addlegendentry{}
\addlegendimage{legend image with text={\tiny \binc{} }}
\addlegendentry{}
\addlegendimage{legend image with text={\tiny \bin{} }}
\addlegendentry{}


\addplot [color=red, dotted, line width=1.0pt, mark=square, mark options={solid, rotate=180, red}]
  table[row sep=crcr]{%
-10	0.0524346798775834\\
-5	0.0438763787301861\\
-3	0.039264002748531\\
0	0.0306264217025095\\
3	0.0222676435965926\\
5	0.0175125383872317\\
10	0.00902098230889555\\
20	0.00255170792928227\\
30	0.00099426389481413\\
50	0.000382865969073868\\
100	0.000351907918216356\\
};
\addlegendentry{}

\addplot [color=blue, line width=1.0pt, mark=square, mark options={solid, rotate=180, blue}]
  table[row sep=crcr]{%
-10	0.0509236452637736\\
-5	0.0387208886256563\\
-3	0.0330441184525578\\
0	0.0241735770166666\\
3	0.0174975873279624\\
5	0.0141704622891839\\
10	0.00880063757549084\\
20	0.00509116191531304\\
30	0.0037051666644602\\
50	0.00308827071058169\\
100	0.00228828477340041\\
};
\addlegendentry{}

\addplot [color=black, dashdotted, line width=1.0pt, mark=square, mark options={solid, rotate=180, black}]
  table[row sep=crcr]{%
-10	0.0514964420509715\\
-5	0.0415391426161334\\
-3	0.0367177976806059\\
0	0.029385656254484\\
3	0.0227387955471782\\
5	0.0196873300498229\\
10	0.014859696174029\\
20	0.0116472874065071\\
30	0.0105035186643644\\
50	0.00922242607895546\\
100	0.00759054771281284\\
};
\addlegendentry{\tiny $\K=3$}


\addplot [color=red, dotted, line width=1.0pt, mark=o, mark options={solid, rotate=180, red}]
  table[row sep=crcr]{%
-10	0.04814743652107\\
-5	0.0419149719384955\\
-3	0.0377852018551435\\
0	0.0308825568265353\\
3	0.0235471755341672\\
5	0.0188284180436533\\
10	0.0104811099194679\\
20	0.00316753292771715\\
30	0.00141417043503072\\
50	0.000691532655288415\\
100	0.000741760347775901\\
};
\addlegendentry{}

\addplot [color=blue, line width=1.0pt, mark=o, mark options={solid, rotate=180, blue}]
  table[row sep=crcr]{%
-10	0.0455104327603826\\
-5	0.0371663945384212\\
-3	0.032603146097088\\
0	0.0254136599662879\\
3	0.0193142076755281\\
5	0.0160050296207974\\
10	0.0104032119807432\\
20	0.00586169669186094\\
30	0.0040708159263262\\
50	0.00340222511138201\\
100	0.00252484307575692\\
};
\addlegendentry{}

\addplot [color=black, dashdotted, line width=1.0pt, mark=o, mark options={solid, rotate=180, black}]
  table[row sep=crcr]{%
-10	0.0465971019896252\\
-5	0.0396907431511802\\
-3	0.035676991935973\\
0	0.0298075618750076\\
3	0.0241387568046557\\
5	0.0210492101722904\\
10	0.0162887276416466\\
20	0.0123022734632905\\
30	0.0107260384155245\\
50	0.00911977924133866\\
100	0.00759241133909195\\
};
\addlegendentry{\tiny $\K=4$}


\addplot [color=red, dotted, line width=1.0pt, mark=triangle, mark options={solid, rotate=180, red}]
  table[row sep=crcr]{%
-10	0.045244924402773\\
-5	0.0404715590995562\\
-3	0.0373636140399516\\
0	0.0319711825464104\\
3	0.025335025424836\\
5	0.0213045595130673\\
10	0.0131748444080086\\
20	0.00551868344229396\\
30	0.00293865328044518\\
50	0.00136309785956019\\
100	0.00103643342597198\\
};
\addlegendentry{}

\addplot [color=blue, line width=1.0pt, mark=triangle, mark options={solid, rotate=180, blue}]
  table[row sep=crcr]{%
-10	0.0429731359290111\\
-5	0.0364967847555686\\
-3	0.0330970048748768\\
0	0.0275995406454647\\
3	0.0219981865361335\\
5	0.0188369575325212\\
10	0.0131365447322886\\
20	0.00774422942157833\\
30	0.00519393457003891\\
50	0.00360188083254224\\
100	0.00290957083570429\\
};
\addlegendentry{}

\addplot [color=black, dashdotted, line width=1.0pt, mark=triangle, mark options={solid, rotate=180, black}]
  table[row sep=crcr]{%
-10	0.0438946331057556\\
-5	0.0385199917994581\\
-3	0.035459534112558\\
0	0.0304547328170074\\
3	0.0258340244325431\\
5	0.022878397854208\\
10	0.0180570885344514\\
20	0.0135262750830035\\
30	0.0115988750159421\\
50	0.00982002213696115\\
100	0.00850705013696744\\
};
\addlegendentry{\tiny $\K=5$}

\addplot [color=gray, dashdotted, line width=1.0pt, mark=star, mark options={solid, rotate=180, gray}]
  table[row sep=crcr]{%
-10	0.000351525040787768\\
-5	0.000351525040787768\\
-3	0.000351525040787768\\
0	0.000351525040787768\\
3	0.000351525040787768\\
5	0.000351525040787768\\
10	0.000351525040787768\\
20	0.000351525040787768\\
30	0.000351525040787768\\
50	0.000351525040787768\\
100	0.000351525040787768\\
};

\end{axis}
\end{tikzpicture}%
}
        \end{subfigure}
        \begin{subfigure}[\freqvec, $\P=4$]
            {\centering
%
%
\begin{tikzpicture}

\begin{axis}[%
width=0.6\figurewidth,
height=0.6\figureheight,
at={(0\figurewidth,0\figureheight)},
scale only axis,
xmin=-10,
xmax=50,
xtick={-10, 0, 10,20,30,40,50},
xlabel style={font=\color{white!15!black}},
xlabel={SNR(dB)},
ymode=log,
ymin=9.22514864264197e-06,
ymax=0.1,
yminorticks=true,
ylabel style={font=\color{white!15!black}},
axis background/.style={fill=white},
axis x line*=bottom,
axis y line*=left,
xmajorgrids,
ymajorgrids,
legend style={at={(0.03,0.97)}, anchor=north west, legend cell align=left, align=left, draw=white!15!black}
]

\addplot [color=red, dotted, line width=1.0pt, mark=square, mark options={solid, rotate=180, red}]
  table[row sep=crcr]{%
-10	0.047419869873702\\
-5	0.0360629762169071\\
-3	0.030703961060906\\
0	0.0225442387141767\\
3	0.0156430171175327\\
5	0.0119060902253104\\
10	0.00549323810269527\\
20	0.000968849560206262\\
30	0.000417257026650633\\
50	3.83115036201127e-05\\
100	9.22514864264197e-06\\
};

\addplot [color=blue, line width=1.0pt, mark=square, mark options={solid, rotate=180, blue}]
  table[row sep=crcr]{%
-10	0.0468023488296889\\
-5	0.0332419548457618\\
-3	0.0269654251706598\\
0	0.018231296238066\\
3	0.0118475231778716\\
5	0.00892364870195328\\
10	0.0039571901216182\\
20	0.000801998340131726\\
30	0.000216811359844797\\
50	6.75078107610139e-05\\
100	8.49504312398372e-05\\
};

\addplot [color=black, dashdotted, line width=1.0pt, mark=square, mark options={solid, rotate=180, black}]
  table[row sep=crcr]{%
-10	0.0485512343420669\\
-5	0.0375747504081986\\
-3	0.0326041966972602\\
0	0.0244932259487153\\
3	0.0182377400159629\\
5	0.015334777049879\\
10	0.0109025893594077\\
20	0.00802245231664747\\
30	0.00719755699696994\\
50	0.00610160805166792\\
100	0.00428182519120463\\
};


\addplot [color=red, dotted, line width=1.0pt, mark=o, mark options={solid, rotate=180, red}]
  table[row sep=crcr]{%
-10	0.0435378323142876\\
-5	0.0353072267003372\\
-3	0.0308716178840802\\
0	0.0238017797721193\\
3	0.0171472231576952\\
5	0.0136455738536359\\
10	0.00674379479261047\\
20	0.00149872752779119\\
30	0.000595635946217047\\
50	8.20758986831649e-05\\
100	9.90911378140924e-06\\
};

\addplot [color=blue, line width=1.0pt, mark=o, mark options={solid, rotate=180, blue}]
  table[row sep=crcr]{%
-10	0.0427838550304674\\
-5	0.0327877981636405\\
-3	0.0279370566514917\\
0	0.0202147967407955\\
3	0.0138339343927947\\
5	0.0105593112492464\\
10	0.00489220804269097\\
20	0.00115725153854061\\
30	0.000356465592250281\\
50	0.00011479312204536\\
100	6.99618350846932e-05\\
};

\addplot [color=black, dashdotted, line width=1.0pt, mark=o, mark options={solid, rotate=180, black}]
  table[row sep=crcr]{%
-10	0.0446924334725191\\
-5	0.0363859734877409\\
-3	0.0321069021481963\\
0	0.0256477662627098\\
3	0.0195989847783366\\
5	0.0168547559676582\\
10	0.0120630725213266\\
20	0.00838585575359892\\
30	0.0072505802163416\\
50	0.00624875406411672\\
100	0.0044924771185318\\
};


\addplot [color=red, dotted, line width=1.0pt, mark=triangle, mark options={solid, rotate=180, red}]
  table[row sep=crcr]{%
-10	0.0412095153751419\\
-5	0.0349934403799903\\
-3	0.0312507199517619\\
0	0.0254854319841639\\
3	0.019647372693292\\
5	0.0163111866407563\\
10	0.00967449784761677\\
20	0.00337940382949727\\
30	0.00150183135535429\\
50	0.000269367151890129\\
100	4.06677117710851e-05\\
};

\addplot [color=blue, line width=1.0pt, mark=triangle, mark options={solid, rotate=180, blue}]
  table[row sep=crcr]{%
-10	0.0407059593372196\\
-5	0.0335277363942468\\
-3	0.0293323781409428\\
0	0.02304981955184\\
3	0.0170942095151773\\
5	0.0138181675467888\\
10	0.00782647185590791\\
20	0.00261522915709347\\
30	0.000954037279114973\\
50	0.000271887784319253\\
100	9.49760596149603e-05\\
};

\addplot [color=black, dashdotted, line width=1.0pt, mark=triangle, mark options={solid, rotate=180, black}]
  table[row sep=crcr]{%
-10	0.0425282993276905\\
-5	0.0359900944978731\\
-3	0.0326111202912927\\
0	0.027035983114634\\
3	0.0219245768143716\\
5	0.0189518781715039\\
10	0.0139779505181501\\
20	0.00938642733831296\\
30	0.00809432183251905\\
50	0.00663014813491942\\
100	0.00524843258747836\\
};

\addplot [color=gray, dashdotted, line width=1.0pt, mark=star, mark options={solid, rotate=180, gray}]
  table[row sep=crcr]{%
-10	9.26734067090861e-06\\
-5	9.26734067090861e-06\\
-3	9.26734067090861e-06\\
0	9.26734067090861e-06\\
3	9.26734067090861e-06\\
5	9.26734067090861e-06\\
10	9.26734067090861e-06\\
20	9.26734067090861e-06\\
30	9.26734067090861e-06\\
50	9.26734067090861e-06\\
100	9.26734067090861e-06\\
};

\end{axis}
\end{tikzpicture}%
}
        \end{subfigure}
        \begin{subfigure}[\freqvec, $\P=6$]
            {\centering
%
%
\begin{tikzpicture}

\begin{axis}[%
width=0.6\figurewidth,
height=0.6\figureheight,
at={(0\figurewidth,0\figureheight)},
scale only axis,
xmin=-10,
xmax=50,
xtick={-10, 0, 10,20,30,40,50},
xlabel style={font=\color{white!15!black}},
xlabel={SNR(dB)},
ymode=log,
ymin=9.26379360385285e-06,
ymax=0.1,
yminorticks=true,
ylabel style={font=\color{white!15!black}},
axis background/.style={fill=white},
axis x line*=bottom,
axis y line*=left,
xmajorgrids,
ymajorgrids,
legend style={at={(0.03,0.97)}, anchor=north west, legend cell align=left, align=left, draw=white!15!black}
]

\addplot [color=red, dotted, line width=1.0pt, mark=square, mark options={solid, rotate=180, red}]
  table[row sep=crcr]{%
-10	0.0426264439659384\\
-5	0.0284293431494398\\
-3	0.0225739585324497\\
0	0.0154109851089269\\
3	0.00979758668397732\\
5	0.00688543192125888\\
10	0.00287967969420975\\
20	0.000405161183009712\\
30	7.0307557744611e-05\\
50	1.17235348892689e-05\\
100	9.26609801753874e-06\\
};

\addplot [color=blue, line width=1.0pt, mark=square, mark options={solid, rotate=180, blue}]
  table[row sep=crcr]{%
-10	0.0440192823496135\\
-5	0.0281003940496555\\
-3	0.0216210591476908\\
0	0.0137734501124582\\
3	0.00864405794746264\\
5	0.00592364063100553\\
10	0.00270220772377507\\
20	0.000263807177932048\\
30	6.18426108559295e-05\\
50	1.07722184819919e-05\\
100	9.26379360385285e-06\\
};

\addplot [color=black, dashdotted, line width=1.0pt, mark=square, mark options={solid, rotate=180, black}]
  table[row sep=crcr]{%
-10	0.0454496927974985\\
-5	0.0309762705692818\\
-3	0.0253067139974729\\
0	0.0174282659851246\\
3	0.0119326911092283\\
5	0.00918293932077781\\
10	0.00527768831042493\\
20	0.00286012443162783\\
30	0.00248917876784624\\
50	0.00222782082154959\\
100	0.00157247653752262\\
};


\addplot [color=red, dotted, line width=1.0pt, mark=o, mark options={solid, rotate=180, red}]
  table[row sep=crcr]{%
-10	0.04018785999608\\
-5	0.0288538205730584\\
-3	0.0236230663525156\\
0	0.016627153075343\\
3	0.0111368748578634\\
5	0.00813725319594816\\
10	0.00351521927607889\\
20	0.000557451039941526\\
30	0.000160079564987761\\
50	1.65386856611492e-05\\
100	9.79130749912759e-06\\
};

\addplot [color=blue, line width=1.0pt, mark=o, mark options={solid, rotate=180, blue}]
  table[row sep=crcr]{%
-10	0.0416902483462065\\
-5	0.0290779263908391\\
-3	0.0232930311868669\\
0	0.015638220190642\\
3	0.00995084200529718\\
5	0.00709093687808449\\
10	0.00293701033540402\\
20	0.00043353887586831\\
30	0.000113715086371185\\
50	2.18382338314643e-05\\
100	9.79339549444241e-06\\
};

\addplot [color=black, dashdotted, line width=1.0pt, mark=o, mark options={solid, rotate=180, black}]
  table[row sep=crcr]{%
-10	0.042614720838741\\
-5	0.0309985750650956\\
-3	0.0258573378967375\\
0	0.0187730039813728\\
3	0.0133111541655543\\
5	0.0105104602155997\\
10	0.00607744088481997\\
20	0.00322089809449209\\
30	0.00257377584071991\\
50	0.00228174867468584\\
100	0.00163654329054821\\
};


\addplot [color=red, dotted, line width=1.0pt, mark=triangle, mark options={solid, rotate=180, red}]
  table[row sep=crcr]{%
-10	0.0385756702881886\\
-5	0.0294812178925722\\
-3	0.0251814127544682\\
0	0.0191586135535119\\
3	0.01392708905518\\
5	0.0110247307007634\\
10	0.00599224657109569\\
20	0.00165024361590412\\
30	0.000695496826234274\\
50	0.000110225377548808\\
100	4.0669652769947e-05\\
};

\addplot [color=blue, line width=1.0pt, mark=triangle, mark options={solid, rotate=180, blue}]
  table[row sep=crcr]{%
-10	0.040163929207555\\
-5	0.0303923810591224\\
-3	0.0255904586891298\\
0	0.0187693012265503\\
3	0.0130437636483931\\
5	0.01020627296946\\
10	0.00531047677364605\\
20	0.00150219477505748\\
30	0.000640240973296027\\
50	0.000104823638527785\\
100	4.07512176970787e-05\\
};

\addplot [color=black, dashdotted, line width=1.0pt, mark=triangle, mark options={solid, rotate=180, black}]
  table[row sep=crcr]{%
-10	0.0408991533290374\\
-5	0.0318560025252071\\
-3	0.0275021967483093\\
0	0.0213440967617228\\
3	0.0157396162745025\\
5	0.0129790121551812\\
10	0.00812538184398874\\
20	0.00428437120672224\\
30	0.00321645794115028\\
50	0.00232814947736287\\
100	0.00187600626755537\\
};

\addplot [color=gray, dashdotted, line width=1.0pt, mark=star, mark options={solid, rotate=180, gray}]
  table[row sep=crcr]{%
-10	9.26705587159001e-06\\
-5	9.26705587159001e-06\\
-3	9.26705587159001e-06\\
0	9.26705587159001e-06\\
3	9.26705587159001e-06\\
5	9.26705587159001e-06\\
10	9.26705587159001e-06\\
20	9.26705587159001e-06\\
30	9.26705587159001e-06\\
50	9.26705587159001e-06\\
100	9.26705587159001e-06\\
};

\end{axis}
\end{tikzpicture}%
}
        \end{subfigure}
    }
    \caption{ \acs{MSE} performance in a noisy scenario solving \ref{eq:Tracenoisy} for $\TT=[4,4,6]$, and different pilot alphabets with (a)-(d) $\P=3$ and $\Ta=72$ ($\Ta$ < $\Tu$), (b)-(e) $\P=4$ and $\Ta=96$ ($\Ta$ = $\Tu$), (c)-(f) $\P=6$ and $\Ta=144$ ($\Ta$ > $\Tu$). The noiseless baseline performance, when visible, is shown in gray with $\color{gray}{\star}$ marker.}
    \label{fig:freqs_error_pilots}
\end{figure*}

\begin{figure*}
    \centerline{
        \begin{subfigure}[\channelvec, $\P=3$]
            {\centering
%
%
\definecolor{mycolor1}{rgb}{0.00000,1.00000,1.00000}%
\definecolor{mycolor2}{rgb}{1.00000,0.00000,1.00000}%

\pgfplotsset{
    legend image with text/.style={
        legend image code/.code={%
            \node[anchor=center] at (0.3cm,0cm) {#1};
        }
    },
}
\begin{tikzpicture}

\begin{axis}[%
width=0.6\figurewidth,
height=0.6\figureheight,
at={(0\figurewidth,0\figureheight)},
scale only axis,
unbounded coords=jump,
xmin=-10,
xmax=50,
xtick={-10, 0, 10,20,30,40,50},
xlabel style={font=\color{white!15!black}},
xlabel={SNR(dB)},
ymode=log,
ymin=1e-07,
ymax=0.001,
yminorticks=true,
ylabel style={font=\color{white!15!black}},
ylabel={$\frac{1}{\Tu}\E\{\|\mihu - \widehat{\mih}_u\|_2^2\}$},
axis background/.style={fill=white},
axis x line*=bottom,
axis y line*=left,
xmajorgrids,
ymajorgrids,
legend columns=5, 
legend style={
    /tikz/column 5/.style={
        column sep=0pt,
    },
fill=none},
legend style={at={(-0.03,0.53)}, anchor=north west, legend cell align=left, align=left, draw=none}
]

\addlegendimage{legend image with text={\tiny \ANnoisy\\}}
\addlegendentry{}
\addlegendimage{legend image with text={\tiny \music{}}}
\addlegendentry{}
\addlegendimage{legend image with text={\tiny \omp{}}}
\addlegendentry{}
\addlegendimage{legend image with text={\tiny \iic{}}}
\addlegendentry{}
\addlegendimage{legend image with text={\tiny \mmse{}}}
\addlegendentry{}

\addplot [color=blue, line width=1.0pt, mark=square, mark options={solid, blue}]
  table[row sep=crcr]{%
-10	8.38982174300834e-05\\
-5	6.31809547109904e-05\\
-3	5.39693615805785e-05\\
0	4.15842564648189e-05\\
3	3.18713133916851e-05\\
5	2.68574533752615e-05\\
10	1.82527196885299e-05\\
20	1.13705089020953e-05\\
30	1.02619964362943e-05\\
50	9.83129076615377e-06\\
100	6.05711627472488e-06\\
};
\addlegendentry{}

\addplot [color=red, dashed, line width=1.0pt, mark=square, mark options={solid, red}]
  table[row sep=crcr]{%
-10	0.00600740570394894\\
-5	0.00227050887801994\\
-3	0.00142500666190799\\
0	0.000730153988085677\\
3	0.000237161704301499\\
5	0.00014883094417946\\
10	5.10671682081006e-05\\
20	4.60876152010933e-05\\
30	4.74120509963125e-05\\
50	4.69543404215608e-05\\
100	4.70410144076081e-05\\
};
\addlegendentry{}

\addplot [color=green, dotted, line width=1.0pt, mark=square, mark options={solid, green}]
  table[row sep=crcr]{%
-10	0.000312312522606568\\
-5	0.000128245856758789\\
-3	9.1528573556593e-05\\
0	5.83700827365054e-05\\
3	4.1531818699024e-05\\
5	3.5045105819361e-05\\
10	2.73640778577155e-05\\
20	2.46050807646462e-05\\
30	2.45341703234829e-05\\
50	2.451537088769e-05\\
100	2.4573842281356e-05\\
};
\addlegendentry{}

\addplot [color=black, dashdotdotted, line width=1.0pt, mark=square, mark options={solid, black}]
  table[row sep=crcr]{%
-10	0.000235410534068197\\
-5	8.793137800102e-05\\
-3	6.74508337337494e-05\\
0	5.2440586291494e-05\\
3	4.39311776835277e-05\\
5	3.91006712070676e-05\\
10	3.04763839690351e-05\\
20	2.08370381873279e-05\\
30	1.92222990032533e-05\\
50	1.97211997418274e-05\\
100	1.98520170294421e-05\\
};
\addlegendentry{}

\addplot [color=mycolor1, densely dashdotdotted, line width=1.0pt, mark=square, mark options={solid, mycolor1}]
  table[row sep=crcr]{%
-10	8.88378287683461e-05\\
-5	8.01900900702337e-05\\
-3	7.4700741639182e-05\\
0	6.48314936715141e-05\\
3	5.44525327444432e-05\\
5	4.80417095160852e-05\\
10	3.55076838360025e-05\\
20	2.47160728955884e-05\\
30	2.29555897232955e-05\\
50	2.2745751622192e-05\\
100	2.27436158360647e-05\\
};
\addlegendentry{\tiny $\K=3$}

\addplot [color=mycolor2, line width=1.0pt, mark=o, mark options={solid, mycolor2}]
  table[row sep=crcr]{%
-10	nan\\
-5	nan\\
-3	nan\\
0	nan\\
3	nan\\
5	nan\\
10	nan\\
20	nan\\
30	nan\\
50	nan\\
100	nan\\
};
\addlegendentry{}

\addplot [color=blue, line width=1.0pt, mark=o, mark options={solid, blue}]
  table[row sep=crcr]{%
-10	8.97251944736358e-05\\
-5	6.8776269336144e-05\\
-3	5.93779992354955e-05\\
0	4.63919439036032e-05\\
3	3.58265489217855e-05\\
5	3.02761054616286e-05\\
10	2.07029685906623e-05\\
20	1.29584438348687e-05\\
30	1.15483082000311e-05\\
50	1.0965509232767e-05\\
100	7.05989949348063e-06\\
};
\addlegendentry{}

\addplot [color=red, dashed, line width=1.0pt, mark=o, mark options={solid, red}]
  table[row sep=crcr]{%
-10	0.00285601582536392\\
-5	0.00119096259094696\\
-3	0.000529463206832424\\
0	0.000298743105683465\\
3	0.000158440598015292\\
5	0.000100486490893267\\
10	6.68167141512823e-05\\
20	6.66364960431035e-05\\
30	7.83570793024282e-05\\
50	5.26759874596388e-05\\
100	5.29324162273636e-05\\
};
\addlegendentry{}

\addplot [color=green, dotted, line width=1.0pt, mark=o, mark options={solid, green}]
  table[row sep=crcr]{%
-10	0.000399768411587914\\
-5	0.000159321574408729\\
-3	0.000113871225174661\\
0	7.16268997218442e-05\\
3	4.93134401129791e-05\\
5	4.06342804289701e-05\\
10	3.0676596382914e-05\\
20	2.65864239401316e-05\\
30	2.63885588619897e-05\\
50	2.62727922236398e-05\\
100	2.62767217098587e-05\\
};
\addlegendentry{}

\addplot [color=black, dashdotdotted, line width=1.0pt, mark=o, mark options={solid, black}]
  table[row sep=crcr]{%
-10	0.000267035200528039\\
-5	0.000101400015773918\\
-3	7.82100699820251e-05\\
0	6.15289826008183e-05\\
3	5.2134779328275e-05\\
5	4.7163124410943e-05\\
10	3.53495083332345e-05\\
20	2.47446599476389e-05\\
30	2.24838012920673e-05\\
50	2.30154060426783e-05\\
100	2.2754934295726e-05\\
};
\addlegendentry{\tiny $\K=4$}

\addplot [color=mycolor1, densely dashdotdotted, line width=1.0pt, mark=o, mark options={solid, mycolor1}]
  table[row sep=crcr]{%
-10	9.51325444994079e-05\\
-5	8.49162500108639e-05\\
-3	7.87451479286169e-05\\
0	6.79964980663126e-05\\
3	5.70000032602119e-05\\
5	5.03168545652116e-05\\
10	3.74444330588573e-05\\
20	2.65523058342687e-05\\
30	2.4790508393045e-05\\
50	2.45806922714012e-05\\
100	2.45785572217598e-05\\
};
\addlegendentry{}

\addplot [color=mycolor2, dashed, line width=1.0pt, mark=o, mark options={solid, mycolor2}]
  table[row sep=crcr]{%
-10	nan\\
-5	nan\\
-3	nan\\
0	nan\\
3	nan\\
5	nan\\
10	nan\\
20	nan\\
30	nan\\
50	nan\\
100	nan\\
};
\addlegendentry{}

\addplot [color=blue, line width=1.0pt, mark=triangle, mark options={solid, blue}]
  table[row sep=crcr]{%
-10	8.7636929440267e-05\\
-5	6.91886598940932e-05\\
-3	6.03467776671221e-05\\
0	4.75986215118144e-05\\
3	3.6927596618698e-05\\
5	3.1228735350904e-05\\
10	2.12674863889972e-05\\
20	1.27723368197697e-05\\
30	1.10128317881096e-05\\
50	1.0421839912929e-05\\
100	8.03244823836962e-06\\
};
\addlegendentry{}

\addplot [color=red, dashed, line width=1.0pt, mark=triangle, mark options={solid, red}]
  table[row sep=crcr]{%
-10	0.00274589412855848\\
-5	0.000805662728688651\\
-3	0.0006411521986798\\
0	0.000298400988045733\\
3	0.000174743028887093\\
5	0.000150646913533428\\
10	6.13914039396982e-05\\
20	5.38010536001154e-05\\
30	4.64114847089164e-05\\
50	4.91624230423516e-05\\
100	4.68163010306291e-05\\
};
\addlegendentry{}

\addplot [color=green, dotted, line width=1.0pt, mark=triangle, mark options={solid, green}]
  table[row sep=crcr]{%
-10	0.000494687390002449\\
-5	0.000188701507207557\\
-3	0.000131885036216287\\
0	8.06625184593669e-05\\
3	5.39021292871833e-05\\
5	4.31503724578056e-05\\
10	3.18739632994896e-05\\
20	2.69169118174635e-05\\
30	2.6628104336484e-05\\
50	2.65179033248401e-05\\
100	2.65230232903637e-05\\
};
\addlegendentry{\tiny $\K=5$}

\addplot [color=black, dashdotdotted, line width=1.0pt, mark=triangle, mark options={solid, black}]
  table[row sep=crcr]{%
-10	0.000289764165367648\\
-5	0.000113239024127944\\
-3	8.52260164077359e-05\\
0	6.57453209716972e-05\\
3	5.54282236534865e-05\\
5	5.00918127851502e-05\\
10	3.91327992282827e-05\\
20	2.61134245953913e-05\\
30	2.28721007076201e-05\\
50	2.35102288266481e-05\\
100	2.47697741132981e-05\\
};

\addplot [color=mycolor1, densely dashdotdotted, line width=1.0pt, mark=triangle, mark options={solid, mycolor1}]
  table[row sep=crcr]{%
-10	9.06632772106615e-05\\
-5	8.14108280506683e-05\\
-3	7.57133560144392e-05\\
0	6.56535511878273e-05\\
3	5.52253861776958e-05\\
5	4.88239454728742e-05\\
10	3.63078873449503e-05\\
20	2.54862052528062e-05\\
30	2.37244543804694e-05\\
50	2.35146368926478e-05\\
100	2.35125017894743e-05\\
};


\end{axis}
\end{tikzpicture}%
}
        \end{subfigure}
        \begin{subfigure}[\channelvec, $\P=4$]
            {\centering
%
%
\definecolor{mycolor1}{rgb}{0.00000,1.00000,1.00000}%
\definecolor{mycolor2}{rgb}{1.00000,0.00000,1.00000}%
\begin{tikzpicture}

\begin{axis}[%
width=0.6\figurewidth,
height=0.6\figureheight,
at={(0\figurewidth,0\figureheight)},
scale only axis,
unbounded coords=jump,
xmin=-10,
xmax=50,
xtick={-10, 0, 10,20,30,40,50},
xlabel style={font=\color{white!15!black}},
xlabel={SNR(dB)},
ymode=log,
ymin=1e-07,
ymax=0.001,
yminorticks=true,
ylabel style={font=\color{white!15!black}},
axis background/.style={fill=white},
axis x line*=bottom,
axis y line*=left,
xmajorgrids,
ymajorgrids,
legend style={
    /tikz/column 5/.style={
        column sep=5pt,
    },
},
legend style={at={(0.0,1.32)}, anchor=north west, legend cell align=left, align=left, draw=none}
]
\addplot [color=blue, line width=1.0pt, mark=square, mark options={solid, blue}]
  table[row sep=crcr]{%
-10	8.23717920113926e-05\\
-5	5.51812847029598e-05\\
-3	4.38081619392334e-05\\
0	2.93354100583893e-05\\
3	1.89040875798592e-05\\
5	1.39057721410565e-05\\
10	6.20432724224332e-06\\
20	1.29261355535382e-06\\
30	4.2058540493825e-07\\
50	3.83880240422361e-08\\
100	1.61897882529292e-08\\
};

\addplot [color=red, dashed, line width=1.0pt, mark=square, mark options={solid, red}]
  table[row sep=crcr]{%
-10	0.000254689917597019\\
-5	0.000134719457263939\\
-3	0.000107788279477396\\
0	7.48997640481953e-05\\
3	5.52061602330058e-05\\
5	4.86450881885484e-05\\
10	3.72855302496818e-05\\
20	2.49099112723084e-05\\
30	2.07179754431608e-05\\
50	2.53511509442973e-05\\
100	2.57528326837152e-05\\
};

\addplot [color=green, dotted, line width=1.0pt, mark=square, mark options={solid, green}]
  table[row sep=crcr]{%
-10	0.00029984608759842\\
-5	0.000144057810037244\\
-3	0.000106088136359478\\
0	6.42291287422256e-05\\
3	3.79048839086391e-05\\
5	2.75490224945302e-05\\
10	1.53128301687123e-05\\
20	1.12350560122865e-05\\
30	1.0779422414646e-05\\
50	1.07463792915162e-05\\
100	1.07655540731616e-05\\
};

\addplot [color=black, dashdotdotted, line width=1.0pt, mark=square, mark options={solid, black}]
  table[row sep=crcr]{%
-10	0.000150879366380373\\
-5	6.61354410087111e-05\\
-3	5.5048240881817e-05\\
0	4.61693886708015e-05\\
3	3.9642574819298e-05\\
5	3.52163545752511e-05\\
10	2.61534654206709e-05\\
20	1.73582082509748e-05\\
30	1.65167305426748e-05\\
50	1.67109614932995e-05\\
100	1.65900320874821e-05\\
};

\addplot [color=mycolor1, densely dashdotdotted, line width=1.0pt, mark=square, mark options={solid, mycolor1}]
  table[row sep=crcr]{%
-10	8.60312960050387e-05\\
-5	7.4407379172745e-05\\
-3	6.76271182920191e-05\\
0	5.60307508961368e-05\\
3	4.42907078725255e-05\\
5	3.71519434178683e-05\\
10	2.28861650818891e-05\\
20	7.1012164525266e-06\\
30	3.37229525496062e-06\\
50	2.88294496837326e-06\\
100	2.87790807443074e-06\\
};

\addplot [color=mycolor2, line width=1.0pt, mark=o, mark options={solid, mycolor2}]
  table[row sep=crcr]{%
-10	nan\\
-5	nan\\
-3	nan\\
0	nan\\
3	nan\\
5	nan\\
10	nan\\
20	nan\\
30	nan\\
50	nan\\
100	nan\\
};

\addplot [color=blue, line width=1.0pt, mark=o, mark options={solid, blue}]
  table[row sep=crcr]{%
-10	8.74208290718165e-05\\
-5	6.02157176237789e-05\\
-3	4.85777149866402e-05\\
0	3.33164442198822e-05\\
3	2.18245604008876e-05\\
5	1.6246334362595e-05\\
10	7.49391884491981e-06\\
20	1.61279709193638e-06\\
30	5.35480924792571e-07\\
50	9.78689804831878e-08\\
100	3.9034264348821e-08\\
};

\addplot [color=red, dashed, line width=1.0pt, mark=o, mark options={solid, red}]
  table[row sep=crcr]{%
-10	0.000580984867750975\\
-5	0.00023349725151431\\
-3	0.000169312709006162\\
0	0.000106664148204947\\
3	7.16557178358809e-05\\
5	6.00562786349321e-05\\
10	4.62665946003832e-05\\
20	4.96920701816435e-05\\
30	4.27239353366084e-05\\
50	3.28699311668732e-05\\
100	3.40307072174767e-05\\
};

\addplot [color=green, dotted, line width=1.0pt, mark=o, mark options={solid, green}]
  table[row sep=crcr]{%
-10	0.000388039823556173\\
-5	0.000173100354497101\\
-3	0.000127477739860825\\
0	7.81198801779921e-05\\
3	4.59726827198977e-05\\
5	3.34355055224173e-05\\
10	1.77559399079194e-05\\
20	1.21368106998184e-05\\
30	1.18021487440244e-05\\
50	1.17364484679607e-05\\
100	1.17458936016257e-05\\
};

\addplot [color=black, dashdotdotted, line width=1.0pt, mark=o, mark options={solid, black}]
  table[row sep=crcr]{%
-10	0.000167595468192199\\
-5	7.73088044245808e-05\\
-3	6.50417477520045e-05\\
0	5.5212343122604e-05\\
3	4.73802180731442e-05\\
5	4.23433257633413e-05\\
10	3.08690404223296e-05\\
20	1.98336339133628e-05\\
30	1.94596215406476e-05\\
50	1.87493985470528e-05\\
100	1.90712776227121e-05\\
};

\addplot [color=mycolor1, densely dashdotdotted, line width=1.0pt, mark=o, mark options={solid, mycolor1}]
  table[row sep=crcr]{%
-10	9.23454214604542e-05\\
-5	7.90174555087769e-05\\
-3	7.14309706186763e-05\\
0	5.87194451548004e-05\\
3	4.61406193422831e-05\\
5	3.86121446403073e-05\\
10	2.37592844920266e-05\\
20	7.52823529895682e-06\\
30	3.76786905392193e-06\\
50	3.2781357251072e-06\\
100	3.27309920312315e-06\\
};

\addplot [color=mycolor2, dashed, line width=1.0pt, mark=o, mark options={solid, mycolor2}]
  table[row sep=crcr]{%
-10	nan\\
-5	nan\\
-3	nan\\
0	nan\\
3	nan\\
5	nan\\
10	nan\\
20	nan\\
30	nan\\
50	nan\\
100	nan\\
};

\addplot [color=blue, line width=1.0pt, mark=triangle, mark options={solid, blue}]
  table[row sep=crcr]{%
-10	8.60471946571829e-05\\
-5	6.19238516831459e-05\\
-3	5.08519039157331e-05\\
0	3.5922574411804e-05\\
3	2.4168208606882e-05\\
5	1.82914005755917e-05\\
10	8.82087052787501e-06\\
20	2.07632322684413e-06\\
30	7.51922534740466e-07\\
50	3.86956712332931e-07\\
100	3.50082101878105e-07\\
};

\addplot [color=red, dashed, line width=1.0pt, mark=triangle, mark options={solid, red}]
  table[row sep=crcr]{%
-10	0.000681513496210722\\
-5	0.000259085465791996\\
-3	0.000184009057434676\\
0	0.000114565397876197\\
3	0.000103039650967648\\
5	5.79657603492819e-05\\
10	4.05837045581161e-05\\
20	2.7920615478462e-05\\
30	2.23435072639952e-05\\
50	2.88408101760581e-05\\
100	3.27404397159203e-05\\
};

\addplot [color=green, dotted, line width=1.0pt, mark=triangle, mark options={solid, green}]
  table[row sep=crcr]{%
-10	0.000454650165438711\\
-5	0.000193028822785093\\
-3	0.000140813485512375\\
0	8.68003711264522e-05\\
3	5.19391335134327e-05\\
5	3.74273466821672e-05\\
10	1.89742577832115e-05\\
20	1.17105094230191e-05\\
30	1.11693651451423e-05\\
50	1.11331126291404e-05\\
100	1.11142061894834e-05\\
};

\addplot [color=black, dashdotdotted, line width=1.0pt, mark=triangle, mark options={solid, black}]
  table[row sep=crcr]{%
-10	0.000178549623270587\\
-5	8.19589019031662e-05\\
-3	6.75064721106205e-05\\
0	5.59192910435233e-05\\
3	4.75088767254547e-05\\
5	4.33449002871597e-05\\
10	3.20359361656745e-05\\
20	2.08595457245843e-05\\
30	1.87911329457353e-05\\
50	1.79773878952926e-05\\
100	1.8677200217987e-05\\
};

\addplot [color=mycolor1, densely dashdotdotted, line width=1.0pt, mark=triangle, mark options={solid, mycolor1}]
  table[row sep=crcr]{%
-10	8.79750308600499e-05\\
-5	7.57448671451821e-05\\
-3	6.87014242868347e-05\\
0	5.67798708640847e-05\\
3	4.48518750210164e-05\\
5	3.76567831063428e-05\\
10	2.33432295398111e-05\\
20	7.47477932920579e-06\\
30	3.74004536676646e-06\\
50	3.25092871187135e-06\\
100	3.24589228980287e-06\\
};

\addplot [color=mycolor2, dotted, line width=1.0pt, mark=o, mark options={solid, mycolor2}]
  table[row sep=crcr]{%
-10	nan\\
-5	nan\\
-3	nan\\
0	nan\\
3	nan\\
5	nan\\
10	nan\\
20	nan\\
30	nan\\
50	nan\\
100	nan\\
};

\end{axis}
\end{tikzpicture}%
}
        \end{subfigure}
        \begin{subfigure}[\channelvec, $\P=6$]
            {\centering
%
%
\definecolor{mycolor1}{rgb}{0.00000,1.00000,1.00000}%
\definecolor{mycolor2}{rgb}{1.00000,0.00000,1.00000}%
\begin{tikzpicture}

\begin{axis}[%
width=0.6\figurewidth,
height=0.6\figureheight,
at={(0\figurewidth,0\figureheight)},
scale only axis,
xmin=-10,
xmax=50,
xtick={-10, 0, 10,20,30,40,50},
xlabel style={font=\color{white!15!black}},
xlabel={SNR(dB)},
ymode=log,
ymin=1e-7,
ymax=0.001,
yminorticks=true,
ylabel style={font=\color{white!15!black}},
axis background/.style={fill=white},
axis x line*=bottom,
axis y line*=left,
xmajorgrids,
ymajorgrids,
legend style={
    /tikz/column 5/.style={
        column sep=5pt,
    },
},
legend style={at={(0.0,1.32)}, anchor=north west, legend cell align=left, align=left, draw=none}
]
\addplot [color=blue, line width=1.0pt, mark=square, mark options={solid, blue}]
  table[row sep=crcr]{%
-10	7.34681878724528e-05\\
-5	4.3445564036697e-05\\
-3	3.28309077042855e-05\\
0	2.03818947234549e-05\\
3	1.20547010487187e-05\\
5	8.31066219413166e-06\\
10	3.09835557167661e-06\\
20	3.77172053642257e-07\\
30	4.32135920380449e-08\\
50	5.68696229214635e-10\\
100	6.6491765306133e-15\\
};

\addplot [color=red, dashed, line width=1.0pt, mark=square, mark options={solid, red}]
  table[row sep=crcr]{%
-10	0.000166273343929639\\
-5	7.731333794294e-05\\
-3	5.8931479339904e-05\\
0	4.44311573965944e-05\\
3	3.75724301311357e-05\\
5	3.47326802433254e-05\\
10	2.77291581736863e-05\\
20	1.53734193479881e-05\\
30	1.77977332764355e-05\\
50	2.18549669112181e-05\\
100	2.20220474185819e-05\\
};

\addplot [color=green, dotted, line width=1.0pt, mark=square, mark options={solid, green}]
  table[row sep=crcr]{%
-10	0.000201567408283253\\
-5	7.5773344502009e-05\\
-3	4.97513086327112e-05\\
0	2.78152557899182e-05\\
3	1.68256494826257e-05\\
5	1.29651635786307e-05\\
10	9.25978678301404e-06\\
20	7.88182988185314e-06\\
30	7.66022025152549e-06\\
50	7.64302463719144e-06\\
100	7.64265235223769e-06\\
};

\addplot [color=black, dashdotdotted, line width=1.0pt, mark=square, mark options={solid, black}]
  table[row sep=crcr]{%
-10	0.000101443375706925\\
-5	5.53209310779086e-05\\
-3	4.85750630184084e-05\\
0	4.22029388323964e-05\\
3	3.61012343131871e-05\\
5	3.18220652236269e-05\\
10	2.29996094245043e-05\\
20	1.37657120172191e-05\\
30	1.24865086938119e-05\\
50	1.24403350049775e-05\\
100	1.23949958121147e-05\\
};

\addplot [color=mycolor1, densely dashdotdotted, line width=1.0pt, mark=square, mark options={solid, mycolor1}]
  table[row sep=crcr]{%
-10	8.3713706569741e-05\\
-5	6.95167143178384e-05\\
-3	6.15975280833898e-05\\
0	4.84689313678936e-05\\
3	3.54684895751178e-05\\
5	2.76234380133595e-05\\
10	1.26801673529948e-05\\
20	1.61825714514168e-06\\
30	1.66568599708166e-07\\
50	1.67116171314277e-09\\
100	1.67123218880513e-14\\
};


\addplot [color=blue, line width=1.0pt, mark=o, mark options={solid, blue}]
  table[row sep=crcr]{%
-10	7.85050325165538e-05\\
-5	4.88018859236183e-05\\
-3	3.76691208558362e-05\\
0	2.40437121296178e-05\\
3	1.44879893168932e-05\\
5	1.00753371200231e-05\\
10	3.80943052277113e-06\\
20	4.72555973134693e-07\\
30	5.47637167950479e-08\\
50	7.3799354076725e-10\\
100	9.20020006944827e-15\\
};

\addplot [color=red, dashed, line width=1.0pt, mark=o, mark options={solid, red}]
  table[row sep=crcr]{%
-10	0.000212462054244596\\
-5	9.58276377517078e-05\\
-3	7.35806106881665e-05\\
0	5.41459725298109e-05\\
3	4.36977992497663e-05\\
5	3.84378256532165e-05\\
10	2.80548845253903e-05\\
20	1.81516480342399e-05\\
30	1.74831663251088e-05\\
50	2.50628292958954e-05\\
100	2.56870645969507e-05\\
};

\addplot [color=green, dotted, line width=1.0pt, mark=o, mark options={solid, green}]
  table[row sep=crcr]{%
-10	0.000246329949038641\\
-5	9.48186182562279e-05\\
-3	6.26093936490567e-05\\
0	3.48038511658066e-05\\
3	2.04375254057745e-05\\
5	1.53582393142131e-05\\
10	1.01187872914165e-05\\
20	8.39445289057415e-06\\
30	8.30505493481881e-06\\
50	8.28005754570362e-06\\
100	8.2900328509812e-06\\
};

\addplot [color=black, dashdotdotted, line width=1.0pt, mark=o, mark options={solid, black}]
  table[row sep=crcr]{%
-10	0.00011547231877734\\
-5	6.51328223560822e-05\\
-3	5.78534712709739e-05\\
0	5.00944378134921e-05\\
3	4.32278651454995e-05\\
5	3.85660956431266e-05\\
10	2.73918934445605e-05\\
20	1.56480420083036e-05\\
30	1.3786004836898e-05\\
50	1.36100009108653e-05\\
100	1.35951082550346e-05\\
};

\addplot [color=mycolor1, densely dashdotdotted, line width=1.0pt, mark=o, mark options={solid, mycolor1}]
  table[row sep=crcr]{%
-10	9.01903223163069e-05\\
-5	7.43773226350895e-05\\
-3	6.56493914770048e-05\\
0	5.13269587961186e-05\\
3	3.73127704820587e-05\\
5	2.89252884974215e-05\\
10	1.30891589321778e-05\\
20	1.62837721741018e-06\\
30	1.66675875439469e-07\\
50	1.67116180024763e-09\\
100	1.67123215094239e-14\\
};


\addplot [color=blue, line width=1.0pt, mark=triangle, mark options={solid, blue}]
  table[row sep=crcr]{%
-10	7.86730557500415e-05\\
-5	5.1583837272356e-05\\
-3	4.06027772970041e-05\\
0	2.66863112433269e-05\\
3	1.65167329235832e-05\\
5	1.16847628033189e-05\\
10	4.58018502757582e-06\\
20	5.92176700379197e-07\\
30	7.01608107195672e-08\\
50	9.68480619814922e-10\\
100	1.3347387665179e-14\\
};

\addplot [color=red, dashed, line width=1.0pt, mark=triangle, mark options={solid, red}]
  table[row sep=crcr]{%
-10	0.000239675118143537\\
-5	0.000104509765824287\\
-3	7.80067963075131e-05\\
0	5.51589851747043e-05\\
3	4.22907527993545e-05\\
5	3.68470171230006e-05\\
10	2.86506310334872e-05\\
20	1.83313680804697e-05\\
30	1.88257364667874e-05\\
50	2.71235554655316e-05\\
100	2.94563861581337e-05\\
};

\addplot [color=green, dotted, line width=1.0pt, mark=triangle, mark options={solid, green}]
  table[row sep=crcr]{%
-10	0.00028836998081199\\
-5	0.00010971243580665\\
-3	7.34373100453845e-05\\
0	4.02310948260513e-05\\
3	2.28616215806663e-05\\
5	1.66588083144059e-05\\
10	9.88209991829911e-06\\
20	7.64838678698112e-06\\
30	7.47355955740575e-06\\
50	7.43927708663367e-06\\
100	7.41803954219229e-06\\
};

\addplot [color=black, dashdotdotted, line width=1.0pt, mark=triangle, mark options={solid, black}]
  table[row sep=crcr]{%
-10	0.000122852120113281\\
-5	6.83815852728331e-05\\
-3	6.01481262937523e-05\\
0	5.10043986403753e-05\\
3	4.39469698394716e-05\\
5	3.93715505719623e-05\\
10	2.86607986893851e-05\\
20	1.5839838138026e-05\\
30	1.32410007462355e-05\\
50	1.30665440355424e-05\\
100	1.31038733069811e-05\\
};

\addplot [color=mycolor1, densely dashdotdotted, line width=1.0pt, mark=triangle, mark options={solid, mycolor1}]
  table[row sep=crcr]{%
-10	8.58121493368738e-05\\
-5	7.11398691705458e-05\\
-3	6.29960204207074e-05\\
0	4.95409568942456e-05\\
3	3.62445513513351e-05\\
5	2.82162821921081e-05\\
10	1.28951377518736e-05\\
20	1.62356505780196e-06\\
30	1.66601425312342e-07\\
50	1.67112673880835e-09\\
100	1.6712320542117e-14\\
};


\end{axis}
\end{tikzpicture}%
}
        \end{subfigure}
    }        
    \centerline{
        \begin{subfigure}[\freqvec, $\P=3$]
            {\centering
%
%
\pgfplotsset{
    legend image with text/.style={
        legend image code/.code={%
            \node[anchor=center] at (0.3cm,0cm) {#1};
        }
    },
}

\begin{tikzpicture}

\begin{axis}[%
width=0.6\figurewidth,
height=0.6\figureheight,
at={(0\figurewidth,0\figureheight)},
scale only axis,
xmin=-10,
xmax=50,
xtick={-10, 0, 10,20,30,40,50},
xlabel style={font=\color{white!15!black}},
xlabel={SNR(dB)},
ymode=log,
ymin=0.00001,
ymax=0.1,
yminorticks=true,
ylabel style={font=\color{white!15!black}},
ylabel={$\error$},
axis background/.style={fill=white},
axis x line*=bottom,
axis y line*=left,
xmajorgrids,
ymajorgrids,
legend columns=4, 
legend style={
    /tikz/column 5/.style={
        column sep=5pt,
    },
fill=none},
legend style={at={(0.0,0.55)}, anchor=north west, legend cell align=left, align=left, draw=none}
]

\addlegendimage{legend image with text={\tiny \ANnoisy \\}}
\addlegendentry{}
\addlegendimage{legend image with text={\tiny \music{}}}
\addlegendentry{}
\addlegendimage{legend image with text={\tiny \omp{}}}
\addlegendentry{}
\addlegendimage{legend image with text={\tiny \iic{}}}
\addlegendentry{}

\addplot [color=blue, line width=1.0pt, mark=square, mark options={solid, blue}]
  table[row sep=crcr]{%
-10	0.0509236452637736\\
-5	0.0387208886256563\\
-3	0.0330441184525578\\
0	0.0241735770166666\\
3	0.0174975873279624\\
5	0.0141704622891839\\
10	0.00880063757549084\\
20	0.00509116191531304\\
30	0.0037051666644602\\
50	0.00308827071058169\\
100	0.00228828477340041\\
};
\addlegendentry{}

\addplot [color=red, dashed, line width=1.0pt, mark=square, mark options={solid, red}]
  table[row sep=crcr]{%
-10	0.0747742786817343\\
-5	0.0668186742065502\\
-3	0.062632605614433\\
0	0.0564276722983039\\
3	0.0517767297457411\\
5	0.0496165116589699\\
10	0.0468667210445626\\
20	0.0448354162058949\\
30	0.0437278526909644\\
50	0.043225874660041\\
100	0.0431776618589844\\
};
\addlegendentry{}

\addplot [color=green, dotted, line width=1.0pt, mark=square, mark options={solid, green}]
  table[row sep=crcr]{%
-10	0.0570336356764928\\
-5	0.0483973922902494\\
-3	0.0435442176870748\\
0	0.0353089569160997\\
3	0.027125850340136\\
5	0.0227847694633409\\
10	0.0167613378684807\\
20	0.0145209750566893\\
30	0.0143968253968253\\
50	0.0144969765684051\\
100	0.0145672713529856\\
};
\addlegendentry{}

\addplot [color=black, dashdotdotted, line width=1.0pt, mark=square, mark options={solid, black}]
  table[row sep=crcr]{%
-10	0.065456289946308\\
-5	0.0534168482808347\\
-3	0.0498449558855001\\
0	0.045897022632159\\
3	0.0417914514395728\\
5	0.0387969625500531\\
10	0.031370860471837\\
20	0.016636325823\\
30	0.0121990754129296\\
50	0.0115010439569376\\
100	0.0115793098631449\\
};
\addlegendentry{\tiny $\K=3$}

\addplot [color=blue, line width=1.0pt, mark=o, mark options={solid, blue}]
  table[row sep=crcr]{%
-10	0.0455104327603826\\
-5	0.0371663945384212\\
-3	0.032603146097088\\
0	0.0254136599662879\\
3	0.0193142076755281\\
5	0.0160050296207974\\
10	0.0104032119807432\\
20	0.00586169669186094\\
30	0.0040708159263262\\
50	0.00340222511138201\\
100	0.00252484307575692\\
};
\addlegendentry{}

\addplot [color=red, dashed, line width=1.0pt, mark=o, mark options={solid, red}]
  table[row sep=crcr]{%
-10	0.0724860336479352\\
-5	0.0674387056591007\\
-3	0.0653803377382372\\
0	0.0607340726486379\\
3	0.0571860067827696\\
5	0.0550218570427794\\
10	0.0515259844796558\\
20	0.0482676170330998\\
30	0.0474471393904859\\
50	0.0474072436914026\\
100	0.0473588572823112\\
};
\addlegendentry{}

\addplot [color=green, dotted, line width=1.0pt, mark=o, mark options={solid, green}]
  table[row sep=crcr]{%
-10	0.0506309523809524\\
-5	0.0447414965986395\\
-3	0.0409736394557823\\
0	0.0344931972789115\\
3	0.0278462301587301\\
5	0.0240238095238095\\
10	0.0180999149659863\\
20	0.0156825396825396\\
30	0.0156822562358276\\
50	0.0156560374149659\\
100	0.015688775510204\\
};
\addlegendentry{}

\addplot [color=black, dashdotdotted, line width=1.0pt, mark=o, mark options={solid, black}]
  table[row sep=crcr]{%
-10	0.0676188514680747\\
-5	0.0591235836042529\\
-3	0.0561941233356892\\
0	0.0525801413974836\\
3	0.0477377476529642\\
5	0.0439607930664576\\
10	0.0333910807023568\\
20	0.019089051250133\\
30	0.0137879594853595\\
50	0.0126554194456301\\
100	0.0123859211408716\\
};
\addlegendentry{\tiny $\K=4$}

\addplot [color=blue, line width=1.0pt, mark=triangle, mark options={solid, blue}]
  table[row sep=crcr]{%
-10	0.0429731359290111\\
-5	0.0364967847555686\\
-3	0.0330970048748768\\
0	0.0275995406454647\\
3	0.0219981865361335\\
5	0.0188369575325212\\
10	0.0131365447322886\\
20	0.00774422942157833\\
30	0.00519393457003891\\
50	0.00360188083254224\\
100	0.00290957083570429\\
};
\addlegendentry{}

\addplot [color=red, dashed, line width=1.0pt, mark=triangle, mark options={solid, red}]
  table[row sep=crcr]{%
-10	0.0697063750195967\\
-5	0.0661799570947924\\
-3	0.0635095180888209\\
0	0.0593724606572445\\
3	0.0552360085227973\\
5	0.0527431247182903\\
10	0.0481125886091259\\
20	0.0442570041988536\\
30	0.0444856806201002\\
50	0.0447591271702881\\
100	0.0447637184501191\\
};
\addlegendentry{}

\addplot [color=green, dotted, line width=1.0pt, mark=triangle, mark options={solid, green}]
  table[row sep=crcr]{%
-10	0.0464815192743764\\
-5	0.042068820861678\\
-3	0.0395241496598639\\
0	0.0341278911564626\\
3	0.0283748299319728\\
5	0.0248750566893424\\
10	0.018909977324263\\
20	0.0160993197278912\\
30	0.0159125850340136\\
50	0.0160478458049887\\
100	0.0161269841269842\\
};
\addlegendentry{}

\addplot [color=black, dashdotdotted, line width=1.0pt, mark=triangle, mark options={solid, black}]
  table[row sep=crcr]{%
-10	0.0675737150166801\\
-5	0.0603287295885735\\
-3	0.0576735289150205\\
0	0.0531666330634006\\
3	0.0479957169198904\\
5	0.0439167249609065\\
10	0.0342071354842768\\
20	0.0205063461498765\\
30	0.0147608300955802\\
50	0.013802712855277\\
100	0.0138565690111683\\
};
\addlegendentry{\tiny $\K=5$}

\end{axis}
\end{tikzpicture}%
}
        \end{subfigure}
        \begin{subfigure}[\freqvec, $\P=4$]
            {\centering
%
%
\begin{tikzpicture}

\begin{axis}[%
width=0.6\figurewidth,
height=0.6\figureheight,
at={(0\figurewidth,0\figureheight)},
scale only axis,
xmin=-10,
xmax=50,
xtick={-10, 0, 10,20,30,40,50},
xlabel style={font=\color{white!15!black}},
xlabel={SNR(dB)},
ymode=log,
ymin=1e-05,
ymax=0.1,
yminorticks=true,
ylabel style={font=\color{white!15!black}},
axis background/.style={fill=white},
axis x line*=bottom,
axis y line*=left,
xmajorgrids,
ymajorgrids,
legend style={
    /tikz/column 5/.style={
        column sep=5pt,
    },
},
legend style={at={(0.0,1.32)}, anchor=north west, legend cell align=left, align=left, draw=none}
]
\addplot [color=blue, line width=1.0pt, mark=square, mark options={solid, blue}]
  table[row sep=crcr]{%
-10	0.0468023488296889\\
-5	0.0332419548457618\\
-3	0.0269654251706598\\
0	0.018231296238066\\
3	0.0118475231778716\\
5	0.00892364870195328\\
10	0.0039571901216182\\
20	0.000801998340131726\\
30	0.000216811359844797\\
50	6.75078107610139e-05\\
100	8.49504312398372e-05\\
};

\addplot [color=red, dashed, line width=1.0pt, mark=square, mark options={solid, red}]
  table[row sep=crcr]{%
-10	0.0766398376387572\\
-5	0.0728662514402043\\
-3	0.0695788278183374\\
0	0.0620264008496162\\
3	0.0547206732995674\\
5	0.0513193964628608\\
10	0.046361495398034\\
20	0.0308112437939283\\
30	0.017944063669773\\
50	0.0170955841984441\\
100	0.0173849313048969\\
};

\addplot [color=green, dotted, line width=1.0pt, mark=square, mark options={solid, green}]
  table[row sep=crcr]{%
-10	0.0586857520786092\\
-5	0.0541409674981103\\
-3	0.0498832199546486\\
0	0.04159410430839\\
3	0.0321298185941043\\
5	0.0259608843537415\\
10	0.0134686318972033\\
20	0.00663076341647769\\
30	0.00622297808012092\\
50	0.00621447467876038\\
100	0.00623771730914587\\
};

\addplot [color=black, dashdotdotted, line width=1.0pt, mark=square, mark options={solid, black}]
  table[row sep=crcr]{%
-10	0.0623564187067566\\
-5	0.0518168864256549\\
-3	0.0489056840443108\\
0	0.0448494549190772\\
3	0.0408215404064856\\
5	0.037834519545289\\
10	0.0298106425790802\\
20	0.0150051121762376\\
30	0.0111748555767363\\
50	0.0102610534639439\\
100	0.0101261913687524\\
};

\addplot [color=blue, line width=1.0pt, mark=o, mark options={solid, blue}]
  table[row sep=crcr]{%
-10	0.0427838550304674\\
-5	0.0327877981636405\\
-3	0.0279370566514917\\
0	0.0202147967407955\\
3	0.0138339343927947\\
5	0.0105593112492464\\
10	0.00489220804269097\\
20	0.00115725153854061\\
30	0.000356465592250281\\
50	0.00011479312204536\\
100	6.99618350846932e-05\\
};

\addplot [color=red, dashed, line width=1.0pt, mark=o, mark options={solid, red}]
  table[row sep=crcr]{%
-10	0.074837977893781\\
-5	0.0721235721856161\\
-3	0.0700601067803467\\
0	0.0651522511388156\\
3	0.0586823689599814\\
5	0.0552988642214184\\
10	0.0492618110551429\\
20	0.0340343992809858\\
30	0.0222778829689291\\
50	0.0180264475543839\\
100	0.0185746182233399\\
};

\addplot [color=green, dotted, line width=1.0pt, mark=o, mark options={solid, green}]
  table[row sep=crcr]{%
-10	0.0523272392290249\\
-5	0.048953939909297\\
-3	0.0461449829931973\\
0	0.0399411848072563\\
3	0.031640589569161\\
5	0.0263801020408163\\
10	0.0145428004535147\\
20	0.00673582766439907\\
30	0.00650878684807254\\
50	0.00644912131519272\\
100	0.00645266439909295\\
};

\addplot [color=black, dashdotdotted, line width=1.0pt, mark=o, mark options={solid, black}]
  table[row sep=crcr]{%
-10	0.0662310047194337\\
-5	0.0582757461295034\\
-3	0.0560912949655922\\
0	0.0521954500116126\\
3	0.0470722948668373\\
5	0.0432024575360548\\
10	0.0322242082729655\\
20	0.0173815057245693\\
30	0.0129304412929507\\
50	0.0113112210575247\\
100	0.0112636390390891\\
};

\addplot [color=blue, line width=1.0pt, mark=triangle, mark options={solid, blue}]
  table[row sep=crcr]{%
-10	0.0407059593372196\\
-5	0.0335277363942468\\
-3	0.0293323781409428\\
0	0.02304981955184\\
3	0.0170942095151773\\
5	0.0138181675467888\\
10	0.00782647185590791\\
20	0.00261522915709347\\
30	0.000954037279114973\\
50	0.000271887784319253\\
100	9.49760596149603e-05\\
};

\addplot [color=red, dashed, line width=1.0pt, mark=triangle, mark options={solid, red}]
  table[row sep=crcr]{%
-10	0.071499695816148\\
-5	0.0694422048915289\\
-3	0.0670647204816571\\
0	0.06225492507603\\
3	0.0563265778768395\\
5	0.0525997990831995\\
10	0.0455845597411357\\
20	0.0339409484273245\\
30	0.0236354123674311\\
50	0.0197335548625904\\
100	0.0204256331211952\\
};

\addplot [color=green, dotted, line width=1.0pt, mark=triangle, mark options={solid, green}]
  table[row sep=crcr]{%
-10	0.0477774376417234\\
-5	0.0451592970521542\\
-3	0.0429206349206349\\
0	0.0378492063492063\\
3	0.0314326530612245\\
5	0.0268518140589569\\
10	0.0167250566893424\\
20	0.00865056689342404\\
30	0.00831156462585032\\
50	0.00837097505668934\\
100	0.00835918367346938\\
};

\addplot [color=black, dashdotdotted, line width=1.0pt, mark=triangle, mark options={solid, black}]
  table[row sep=crcr]{%
-10	0.0669607965674837\\
-5	0.0595006164400904\\
-3	0.056655925570757\\
0	0.0522677713718142\\
3	0.0465174660413321\\
5	0.042893822891537\\
10	0.0325279685747854\\
20	0.0193074824038727\\
30	0.0139610312683192\\
50	0.0129802372399022\\
100	0.0128557185926877\\
};

\end{axis}
\end{tikzpicture}%
}
        \end{subfigure}
        \begin{subfigure}[\freqvec, $\P=6$]
            {\centering
%
%
\begin{tikzpicture}

\begin{axis}[%
width=0.6\figurewidth,
height=0.6\figureheight,
at={(0\figurewidth,0\figureheight)},
scale only axis,
xmin=-10,
xmax=50,
xtick={-10, 0, 10,20,30,40,50},
xlabel style={font=\color{white!15!black}},
xlabel={SNR(dB)},
ymode=log,
ymin=1e-05,
ymax=0.1,
yminorticks=true,
ylabel style={font=\color{white!15!black}},
axis background/.style={fill=white},
axis x line*=bottom,
axis y line*=left,
xmajorgrids,
ymajorgrids,
legend style={
    /tikz/column 5/.style={
        column sep=5pt,
    },
},
legend style={at={(0.0,1.32)}, anchor=north west, legend cell align=left, align=left, draw=none}
]
\addplot [color=blue, line width=1.0pt, mark=square, mark options={solid, blue}]
  table[row sep=crcr]{%
-10	0.0440192823496135\\
-5	0.0281003940496555\\
-3	0.0216210591476908\\
0	0.0137734501124582\\
3	0.00864405794746264\\
5	0.00592364063100553\\
10	0.00270220772377507\\
20	0.000263807177932048\\
30	6.18426108559295e-05\\
50	1.07722184819919e-05\\
100	9.26379360385285e-06\\
};

\addplot [color=red, dashed, line width=1.0pt, mark=square, mark options={solid, red}]
  table[row sep=crcr]{%
-10	0.0706972705492471\\
-5	0.0585614645573905\\
-3	0.0547273551719516\\
0	0.0499897147879901\\
3	0.0468869115974526\\
5	0.0452661521201248\\
10	0.0389403049850269\\
20	0.0185334648589329\\
30	0.0143445308421425\\
50	0.0144619807712792\\
100	0.0145843108755847\\
};

\addplot [color=green, dotted, line width=1.0pt, mark=square, mark options={solid, green}]
  table[row sep=crcr]{%
-10	0.0536649659863946\\
-5	0.040687641723356\\
-3	0.0338059334845049\\
0	0.0239368858654573\\
3	0.0151477702191987\\
5	0.0109234693877551\\
10	0.00632520786092212\\
20	0.00475850340136053\\
30	0.00447978080120937\\
50	0.00442309145880574\\
100	0.00442176870748299\\
};

\addplot [color=black, dashdotdotted, line width=1.0pt, mark=square, mark options={solid, black}]
  table[row sep=crcr]{%
-10	0.0591778883521736\\
-5	0.0495104133354222\\
-3	0.0469497577261634\\
0	0.043527483977738\\
3	0.039544445311322\\
5	0.0367293788420433\\
10	0.0284833727949787\\
20	0.0130776231185689\\
30	0.00975416519839397\\
50	0.00889187685965834\\
100	0.0088110181076089\\
};

\addplot [color=blue, line width=1.0pt, mark=o, mark options={solid, blue}]
  table[row sep=crcr]{%
-10	0.0416902483462065\\
-5	0.0290779263908391\\
-3	0.0232930311868669\\
0	0.015638220190642\\
3	0.00995084200529718\\
5	0.00709093687808449\\
10	0.00293701033540402\\
20	0.00043353887586831\\
30	0.000113715086371185\\
50	2.18382338314643e-05\\
100	9.79339549444241e-06\\
};

\addplot [color=red, dashed, line width=1.0pt, mark=o, mark options={solid, red}]
  table[row sep=crcr]{%
-10	0.0691772235415548\\
-5	0.0614786160889472\\
-3	0.0583193208097673\\
0	0.054131860240306\\
3	0.0511525320540023\\
5	0.048627069000152\\
10	0.0403387576928965\\
20	0.024055824787382\\
30	0.0148468616594124\\
50	0.0141099755878729\\
100	0.0145072333074718\\
};

\addplot [color=green, dotted, line width=1.0pt, mark=o, mark options={solid, green}]
  table[row sep=crcr]{%
-10	0.047719387755102\\
-5	0.038780753968254\\
-3	0.0330695861678005\\
0	0.0247439058956916\\
3	0.0164410430839002\\
5	0.0119024943310657\\
10	0.00631405895691607\\
20	0.00466269841269839\\
30	0.00469940476190475\\
50	0.00464087301587301\\
100	0.00463435374149659\\
};

\addplot [color=black, dashdotdotted, line width=1.0pt, mark=o, mark options={solid, black}]
  table[row sep=crcr]{%
-10	0.0645498229674536\\
-5	0.0573748320745814\\
-3	0.055065415893027\\
0	0.0510338415704109\\
3	0.0462666801457941\\
5	0.0424294451744727\\
10	0.0318004169852354\\
20	0.0158444933761086\\
30	0.011247350844932\\
50	0.010007944467701\\
100	0.00985724156841313\\
};

\addplot [color=blue, line width=1.0pt, mark=triangle, mark options={solid, blue}]
  table[row sep=crcr]{%
-10	0.040163929207555\\
-5	0.0303923810591224\\
-3	0.0255904586891298\\
0	0.0187693012265503\\
3	0.0130437636483931\\
5	0.01020627296946\\
10	0.00531047677364605\\
20	0.00150219477505748\\
30	0.000640240973296027\\
50	0.000104823638527785\\
100	4.07512176970787e-05\\
};

\addplot [color=red, dashed, line width=1.0pt, mark=triangle, mark options={solid, red}]
  table[row sep=crcr]{%
-10	0.0663487687249089\\
-5	0.0597891250479339\\
-3	0.0559051143571071\\
0	0.0515095796220147\\
3	0.0473141132763958\\
5	0.0447038366474497\\
10	0.0386283104458064\\
20	0.0252170396260286\\
30	0.0181504977143689\\
50	0.0172812490843719\\
100	0.0177822737892772\\
};

\addplot [color=green, dotted, line width=1.0pt, mark=triangle, mark options={solid, green}]
  table[row sep=crcr]{%
-10	0.0443743764172336\\
-5	0.0369831065759637\\
-3	0.032858843537415\\
0	0.025822335600907\\
3	0.0188768707482993\\
5	0.0146505668934241\\
10	0.00870158730158731\\
20	0.00662936507936504\\
30	0.00643628117913827\\
50	0.00664149659863941\\
100	0.00662131519274371\\
};

\addplot [color=black, dashdotdotted, line width=1.0pt, mark=triangle, mark options={solid, black}]
  table[row sep=crcr]{%
-10	0.0662093070703294\\
-5	0.0588354122924651\\
-3	0.0557725441810306\\
0	0.0507266255639864\\
3	0.0457040377241967\\
5	0.0417475222231975\\
10	0.0318690197592685\\
20	0.017489519406396\\
30	0.01277339249449\\
50	0.0116872899612518\\
100	0.0115536692739323\\
};

\end{axis}
\end{tikzpicture}%
}
        \end{subfigure}
    }        
    \caption{\acs{MSE} performance in a noisy scenario solving \ref{eq:Tracenoisy} for $\TT=[4,4,6]$, and \binc{} pilot alphabet with (a)-(d) $\P=3$ and $\Ta=72$ ($\Ta$ < $\Tu$), (b)-(e) $\P=4$ and $\Ta=96$ ($\Ta$ = $\Tu$), (c)-(f) $\P=6$ and $\Ta=144$ ($\Ta$ > $\Tu$).}
    \label{fig:freqs_error_bin_complex}
\end{figure*}
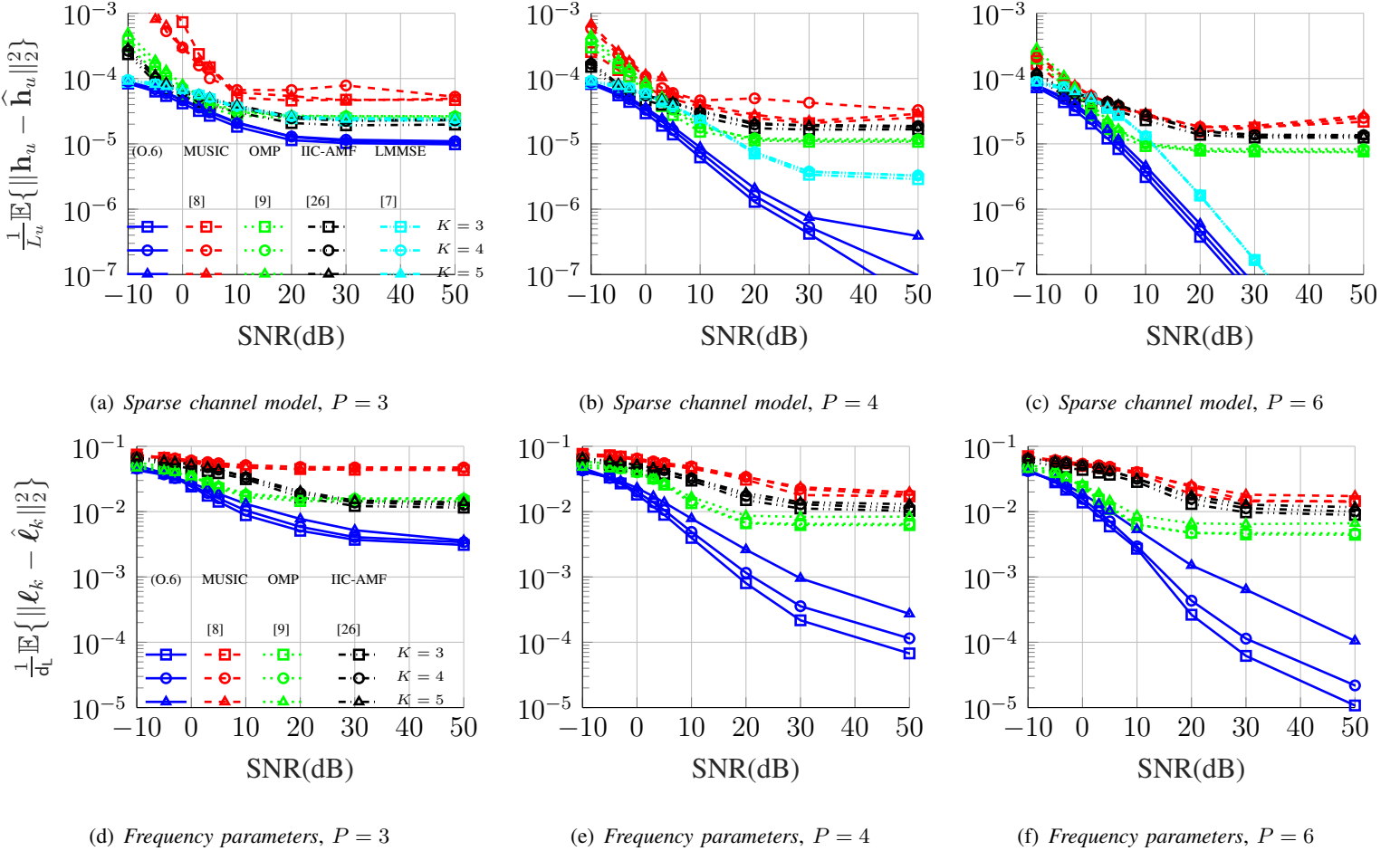

We evaluate by simulation the recovery performance of the sparse channel model $\mihu$ and of the $\ddl$-D channel frequency parameters $\ff_k$ with $k\in[\K]$ solving respectively \eqref{ec_Thm2} and \eqref{eq:Tracenoisy} for the noiseless and noisy observation scenarios, where the measurement matrix $\miQ$ and the noise variance $\sigma^2_w$ are known. Furthermore, in the implementation of all optimization algorithms, the number $\K$ of scatters is also assumed known. The estimate of the sparse channel model $\widehat{\mih}_u$ is the optimal solution $\mil^\opt_u$ of  \eqref{ec_Thm2} or \eqref{eq:Tracenoisy}. Similarly, the estimated channel frequency parameters containing information on the \ac{AoD} and \ac{AoA}, named $\hat{\ff}_k$ with $k\in[\K]$, are the result of applying \refAlg1~to the optimal \ac{MLT}-matrix $\miT_{\TT}^\opt$ obtained from either \eqref{ec_Thm2} and \eqref{eq:Tracenoisy}. In both cases, it is evaluated the \ac{MSE}, i.e. $\frac{1}{\Tu}\E\{\|\mih_u - \widehat{\mih}_u\|_2^2\}$ or $\error$ where the average is taken with respect the frequency samples, the fading coefficients, the pilot sequence realizations and, where corresponds, the noise samples. In the noisy observation scenarios the $\SNR$ is defined as $\SNR=\frac{\E\{\|\mihu\|^2\}}{\sigma^2_w}$.

To show the validity of our nuclear atomic norm approach, we compare with other state-of-the-art approaches. For the recovery performance of the the sparse channel model $\mihu$ and frequency parameters we explore three parametric solutions fundamentally relaying on a finite dictionary exploration, namely \ac{OMP} \cite{Swapna22}, \ac{MD-MUSIC} \cite{Liao15} and \ac{IIC-AMF} \cite{Grossi20}. Additionally, for the recovery of the sparse channel model $\mihu$, we include \ac{LMMSE} \cite{Assalini09} as an additional non-parametric state-of-the-art method. In order to make a fair comparison between our proposed method and the on-grid parametric state-of-the-art approaches, we parametrize all algorithms so that they have the same complexity. The complexity expressions 
can be seen in Table \ref{resolvable}, 
where $L_{\text{OMP}}$, $L_{\text{MUSIC}}$ and $L_{\text{IIC}}$ denotes the respective lengths of the dictionaries of each method and, for \ac{MUSIC}, $H<\Tu$ is an additional tunable parameter related to the dimension of a Hankel matrix that defines a stacking of the observation vector. In all cases, we choose the dictionaries lengths matching the complexity of our atomic norm based optimization problem  and we set $H=\Tu-1$. For the non-parametric approach \ac{LMMSE}, we recall that since this method depend on the inversion of a matrix, its complexity is $O\left(\Tu^{3}\right)$.

Fig. \ref{fig:freq_error_noiseless}.a shows the recovery performance of the sparse channel model for three different pilot constellations with respect to several values of $\K$ in a noiseless scenario using \eqref{ec_Thm2}. All $\K$ values explored comply with condition $\K\leq \frac{\kappaL-2}{2}=6$. Also, $\miP^\top$ admits a left pseudo-inverse for $\P=4$ and $\P=6$ for the \binc~and \rand~constellations. The recovery performance is notoriously better with $P=6$ for those constellations where $\miP^\top$ admits a left pseudo-inverse reflecting that the optimal solution is also complying with $\rank\{\miT_{\TT}^\opt \} = \K$ in those scenarios. For the rest of scenarios, as $\P$ is reduced, the recovery error increases, being the worst overall performance when ~\bin~pilots are used. In this specific scenario recall that the $\miP$ does not comply with A.\ref{A2}, i.e. is not almost sure full-$\rank$. Regarding the different constellation alphabets explored, the practical \binc~performs significantly well, close in many cases to the unfeasible benchmark based on a continuous Gaussian distribution \rand. Fig. \ref{fig:freq_error_noiseless}.b explores the recovery performance of the frequency parameter. In this result, not all $\K$ values comply with $\K<\Tsf_{\ddl}=6$. Indeed, it is noticeable that the estimation error increases with respect to $\K$ and it gets significantly high when $\K = 6$.
Fig. \ref{fig:freqs_error_pilots} explore the analogous scenarios to Fig. \ref{fig:freq_error_noiseless} for the noisy scenario solving the optimization problem in \eqref{eq:Tracenoisy}. All $\K$ values explored comply in this case with the conditions $\K\leq \frac{\kappaL-2}{2}=6$ and $\K<\Tsf_{\ddl}=6$. The combination of the different pilot alphabets and the different values of $\P$ once more examine scenarios where $\miP^\top$ does not have a left pseudo-inverse, for example when $\P=3$ or when the pilot constellation is drawn from \bin. These are the worst performance scenarios, both in the sparse channel model and in the frequency recovery. In addition to the noisy recovery performance of the frequency parameters, the result for $\K=3$ in the noiseless scenario \eqref{ec_Thm2} is added as a baseline showing the convergence for large $\SNR$ values. 
Finally our proposed method is compared with  other state-of-the-art parametric and non-parametric methods. We can see this comparison, carried out using pilots in the \binc~alphabet in a noisy scenario solving \eqref{eq:Tracenoisy} in Fig. \ref{fig:freqs_error_bin_complex} for the sparse channel model and frequency parameter recoveries. For any given value of $\P$, $\K$ or $\SNR$, the atomic norm based optimization method outperforms the others. It is specially noticeable the $\miP=3$ case, since this method is able to resolve with much less error an underdetermined system, which means that a shorter pilot sequence is enough for the method proposed in this work to achieve a better performance than the other solutions.

 \section{Conclusions} 
 \label{sec:VI}
 
This work presents a parametric channel estimation approach that enables joint recovery of the channel matrix and gridless parameter recovery of both \ac{AoA} and \ac{AoD}, each of them be characterized in up to $3$-D. We provide the recovery conditions in the noiseless case relating key structural features of the measurement scenario with the \sd~of the problem, unveiling the trade-off between recovery performance of richer propagation environments with complexity. The proposed \ac{AN} minimization technique is compared with state-of-the-art on--grid techniques, outperforming all of them for equal complexity. 

\appendices
\section{Proof of Proposition \ref{Atotal:} }
\label{ProofAtotal:}

The proof of Proposition   \ref{Atotal:} follows easily from the fact that $\Atx\in\mathcal{A}_\MM(\Ma,\Mu)$ and $\Arx\in\mathcal{A}_\NN(\Na,\Nu)$ respectively. Let us first prove that $\AL$ is a sensing matrix:     
\begin{eqnarray}
\AL \!\!\! \!\!&=& \!\! \!\! \!(\Atx\otimes\Arx)\miPi_u  
 = 
\left (\big[\miI_\Ma | {\bf O}_\t \big] \miPi_\t 
\right)  \otimes
\left (\big[\miI_\Na | {\bf O}_\r   \big] \miPi_\r \right)  \miPi_u 
\nonumber
\\ 
 \!\!\! \!\!&=& \!\! \!\! \!
\left (\big[\miI_\Ma | {\bf O}_\t \big] \otimes \big[\miI_\Na |  {\bf O}_\r \big] \right)  
(\miPi_\t \otimes \miPi_\r) \miPi_u \\
 \!\!\! \!\!&=& \!\! \!\! \!
 \big[\miI_{\Ma\Na}|{\bf 0}_{\Ma\Na\times (\Tu-\Ma\Na)}\big]\miPi_s
 \nonumber
\end{eqnarray}

where ${\bf O}_\t \triangleq {\bf 0}_{\Ma \times (\Mu-\Ma)}$, 
${\bf O}_\r \triangleq {\bf 0}_{\Na \times (\Nu-\Na)}$, $ \miPi_\t\in\{0,1\}^{\Mu\times\Mu}$ and 
$ \miPi_\r\in\{0,1\}^{\Nu\times\Nu}$  are a column permutation matrices and finally   $\miPi_s = (\miPi_\t \otimes \miPi_\r) \miPi_u $. 
Furthermore, given the steering vector $\miu_\TT(\ff)$, in \eqref{struct2} we show that we can find, embed in $\AL  \miu_\TT(\ff)$, a uniform steering vector $\miu_{\SS}({\mii})$ with dimensions $\SS=[\LL,\RR]=[\Ssf_1,\dots \Ssf_{\ddl}]\preceq\TT$ such that  $\sum_{i=1}^{\ddl} \Ssf_{i} \geq (\ddl+1)$. Given this relationship, it is straightforward to show that the \rd~$\kappaL$ of $\AL  \miu_\TT(\ff)$ is the sum of the \rd~of $\Atx  \miu_\MM(\mif)$ and the \rd~of $\Arx  \miu_\NN(\mif)$, i.e. $\kappaL=\kappa_\t+\kappa_\r$.

\begin{figure*}[ht]
\begin{eqnarray}
\label{struct2}
\AL  \miu_\TT(\ff)&=&
(\Atx\otimes\Arx) \miPi_u   \miu_\TT(\ff)
=
(\Atx\otimes\Arx) (
\miu_\MM^*(\mig)\otimes\miu_\NN(\mif))
=
(\Atx \miu_\MM^*(\mig) )\otimes( \Arx \miu_\NN(\mif))
\nonumber
\\
&=&
(\Atx \miu_\MM^*(\mig) )\otimes( \Arx \miu_\NN(\mif)) 
= 
\left(  \Pint^{\t\H}\begin{bmatrix}
e^{\j2\pi {\bf\Delta}\cdot\mig}\miu_{\LL}({\mii_\t}) \\
\mib_\t
\end{bmatrix} \right) \otimes
\Pint^{\r\H}\left (\begin{bmatrix}
e^{\j2\pi {\bf\Delta}\cdot\mif}\miu_{\RR}({\mii_\r}) \\
\mib_\r
\end{bmatrix} \right) \nonumber\\
&=&
\left( \Pint^{\t\H}
\otimes
 \Pint^{\r\H} \right) \left( \begin{bmatrix}
e^{\j2\pi {\bf\Delta}_\t\cdot\mif}\miu_{\LL}({\mii_\t}) \\
\mib_\t
\end{bmatrix} \otimes
\begin{bmatrix}
e^{\j2\pi {\bf\Delta}_\r\cdot\mif}\miu_{\RR}({\mii_\r}) \\
\mib_\r
\end{bmatrix}
\right) 
\label{struct2}
\\&=&
\Pint^{\mathsf{L}\H} \begin{bmatrix}
e^{\j2\pi {\bf\Delta}_\t\cdot\mig+{\bf\Delta}_\r\cdot\mif}\miu_{\SS}({\mii}) \\
\mib_\t\otimes\mib_\r
\end{bmatrix} 
\nonumber
\end{eqnarray}
\end{figure*}

\section{Proof of Lemma \ref{Span_condition}} 
\label{proofspan}


Matrix $\miT_\TT\in\mathcal{T}^{\ddl}_\TT$ with $r_1\triangleq\rank\{\miT_\TT\}<\Tsf_{\ddl}$ has a \ac{MLT} structure as in Def. \ref{definitionMLT} and the \emph{dimension vector} $\TT=[\Tsf_1,\Tsf_2,\dots,\Tsf_{\ddl}]$ has an ordering such that $\Tsf_1 \leq \Tsf_2\leq\dots \leq\Tsf_{\ddl}$ with $\Tu=\prod_{i=1}^{\ddl}\Tsf_i$. Furthermore, for $1\leq t\leq\ddl$, we recall that the (sub)matrix nested along the left upper block diagonal\footnote{$\miT_{\mathbf{0}\TT_{t:\ddl}}$ is the $\prod_{i=t}^{\ddl}\Tsf_i\times\prod_{i=t}^{\ddl}\Tsf_i$ subblock of $\miT_\TT$ obtained considering the first $\prod_{i=t}^{\ddl}\Tsf_i$ rows and the first $\prod_{i=t}^{\ddl}\Tsf_i$ columns of $\miT_\TT$. } $\miT_{\mathbf{0}\TT_{t:\ddl}}\in\C^{\prod_{i=t}^{\ddl}\Tsf_i\times\prod_{i=t}^{\ddl}\Tsf_i}$ is also a  $(\ddl-t-1)$-\ac{MLT} matrix with  $r_t\triangleq\rank\left\{\miT_{\mathbf{0}\TT_{t:\ddl}}\right\}$. 
From \cite[Thm. 4.1]{Meyer73} we have that $r_{\ddl}\leq r_{\ddl-1}\leq\dots\leq r_{1} $ and therefore $r_{\ddl}\leq r_{\ddl-1}\leq\dots\leq r_{1} <\Tsf_{\ddl}$.
By \cite[Lemma in Proposition 1]{Gurvits_prueba} we have that, if $r_t<\prod_{i=t+1}^{\ddl}\Tsf_i$,  a \ac{PSD} block matrix $\miT_{\mathbf{0}\TT_{t:\ddl}}$ is decomposed as:
$$
\miT_{\mathbf{0}\TT_{t:\ddl}}=\sum_{j=1}^{r_t} 
(\miu_{\Tsf_t}(\nu^t_j)\otimes  \mig^{t+1}_{j})(\miu_{\Tsf_t}(\nu^t_j)\otimes  \mig^{t+1}_{j})^\H=\miC_{t:\ddl} \miC_{t:\ddl}^\H
$$
where $\nu^t_j\in \T$, with $j \in [r_t]$ and $\miC_{t:\ddl}\in\C^{\prod_{i=t}^{\ddl}\Tsf_i\times r_t}$ can be also written as:
\begin{eqnarray}
\miC_{t:\ddl}&=&[\mic^t_1,..,\mic^t_{r_t} ]=[\miu_{\Tsf_t}(\nu^t_1)\otimes  \mig^{t+1}_{1},..,\miu_{\Tsf_t}(\nu^t_{r_t})\otimes  \mig^{t+1}_{r_t}] \nonumber
\\&=&\miU_{\Tsf_t}(\ll^{t}_{1:r_t}) \odot \miG_{t+1:\ddl}.\nonumber
\label{machec0}
\end{eqnarray}

Denoting $\miT_{\mathbf{0}\a\TT_{t+1:\ddl}}$ as the $\a$-th block of $\miT_{\mathbf{0}\TT_{t:\ddl}}$ for  $-\Tsf_t+1\leq \a\leq\Tsf_t-1$ we have that
\begin{eqnarray}
\miT_{\mathbf{0}\a\TT_{t+1:\ddl}}&=&\sum_{j=1}^{r_t} e^{\j2\pi \a \nu^{t}_j}  \mig^{t+1}_{j} \mig^{t+1\H}_{j}\label{incredibile00}
=\miG_{t+1:\ddl}\miD_t^\a\miG_{t+1:\ddl}^\H
\end{eqnarray}
with $\miD_t= \diag(e^{\j2\pi \nu^{t}_1}, \ldots, e^{\j2\pi \nu^{t}_{r_t}})$. Given that we also have that $\miT_{\mathbf{0}\TT_{t+1:\ddl}}=\miC_{t+1:\ddl} \miC_{t+1:\ddl}^\H$, setting $\a=0$ in \eqref{incredibile00}, we necessarily have that $\miG_{t+1:\ddl}\miG_{t+1:\ddl}^\H=\miC_{t+1:\ddl} \miC_{t+1:\ddl}^\H$ and therefore, we can always find  a $r_{t+1}\times r_{t}$ unitary matrix, $\miO_t$, such that $\miG_{t+1:\ddl}=\miC_{t+1:\ddl}\miO_{t}$. Given this, for $j\in[r_t]$ and $t\leq\ddl-2$ we have: 
\begin{equation}
\begin{split}
\mic_j^{t}&=\miu_{\Tsf_t}(\nu^{t}_j) \otimes\sum_{i=1}^{r_{t+1}}o^{t}_{ji}\miu_{\Tsf_{t+1}}(\nu^{t+1}_i) \otimes\mig^{t+2}_{i}\end{split}
\label{cxyz}
\end{equation}
where $o^{t}_{ji}$ is the element in row $j$ and column $i$ of matrix $\miO_{t}$.

Applying recursively the relationship in \eqref{cxyz}, for $t=1$, i.e. for $\miT_\TT=\miC_{1:\ddl} \miC_{1:\ddl}^\H$, and $i_1=[r_1]$ we have:
\begin{eqnarray}
\mic_{i_1}^{1}&=&\miu_{\Tsf_1}(\nu^{1}_{i_1}) \otimes\sum_{i_2=1}^{r_{2}}o^{2}_{i_2i_1}\miu_{\Tsf_{2}}(\nu^{2}_{i_2}) \otimes\dots\nonumber\\&&\dots\otimes\sum_{i_{\ddl}=1}^{r_{\ddl}}o^{\ddl}_{i_{\ddl}i_{\ddl-1}}p_{i_{\ddl}}\miu_{\Tsf_{\ddl}}(\nu^{\ddl}_{i_{\ddl}})\nonumber\\
&=&\sum_{i_2=1}^{r_{2}}\dots\sum_{i_{\ddl}=1}^{r_{\ddl}}p_{i_{\ddl}}\Motimes_{s=1}^{\ddl}o^{s}_{i_{s}i_{s-1}}\miu_{\Tsf_s}(\nu^{s}_{i_s})
\label{Cxyz}
\end{eqnarray}
where $p_{i_{\ddl}}$ with $i_{\ddl}\in[r_{\ddl}]$ are the elements of the diagonal of the Vandermonde decomposition of the Toeplitz matrix $\miT_{\mathbf{0}\Tsf_{\ddl}}$ found in the very last nesting and we define $o^{1}_{i_{1}i_{0}}=1$.

The column elements of $\miC_{1:\ddl}=[\mic_{1}^{1},\dots,\mic_{r_1}^{1}]$ described in \eqref{Cxyz} are generated by $\prod_{s=1}^{\ddl}r_s$ vectors $\Motimes_{s=1}^{\ddl}\miu_{\Tsf_s}(\nu^{s}_{i_s}) $ with $i_s\in[r_s]$, which necessarily need to be linearly dependent given that $\rank\{\miT_\TT\}=\rank\{\miC_{1:\ddl}\miC_{1:\ddl}^\H\}=r_1$. We define $\mathcal{M}$ as the set of indexes $(i_{1,m},\dots,i_{\ddl,m})$ such that the vectors  $\Motimes_{s=1}^{\ddl}\miu_{\Tsf_s}(\nu^{s}_{i_{s,m}})$ are linearly independent, note that $|\mathcal{M}|=r_1$ and $m\in[r_1]$. Then, the linearly independent generating vectors of matrix $\miC_{1:\ddl}$, now indexed in $m\in[r_1]$, are $\Motimes_{s=1}^{\ddl}\miu_{\Tsf_s}(\nu^{s}_{i_{s,m}})=\miu_\TT(\ll_m)$ where $\ll_m=[\nu^{1}_{i_{1,m}},\dots,\nu^{\ddl}_{i_{\ddl,m}}]^\top$.

Due to Schur complement lemma, the model $\mil_u$ that satisfies constraint $\begin{bmatrix}\miT_{\TT} &\mil_u\\ \mil_u^\H & l\end{bmatrix}\succeq 0$ also satisfies $\miT_{\TT} -( l)^{-1}\mil_u\mil_u^\H\succeq 0$ which implies that $\mil_u\in\spn\{\miT_{\TT}\}=\spn\{\miC_{1:\ddl}\miC_{1:\ddl}^\H\}$. Therefore there exist two set of coefficients $[\alpha_1, \ldots, \alpha_{r_1}]$ and $[\beta_1, \ldots, \beta_{r_1}]$ such that 
\begin{eqnarray}
\mil_u&=&\sum_{k=1}^{r_1}\alpha_k\mic_{k}^{1}
=\sum_{i_1=1}^{r_1}\dots\sum_{i_{\ddl}=1}^{r_{\ddl}}\alpha_{i_1}p_{i_{\ddl}}\Motimes_{s=1}^{\ddl}o^{s}_{i_{s}i_{s-1}}\miu_{\Tsf_s}(\nu^{s}_{i_s})\nonumber\\
&=&\sum_{k=1}^{r_1}\beta_k\miu_\TT\left(\ll_{k}\right)
\label{decomp}
\end{eqnarray}

\vspace*{-0.5cm}
\section{Proof of Thm. \ref{uniqueness}} 
 \label{proofuniqueness}

Given the two \sparsems~ $\mihu(\ff_{1:\K})= \miU_{\TT}(\ff_{1:\K})\boldsymbol{ \gamma}$ and $\mil_u(\ll_{1:r})= \miU_{\TT}(\ll_{1:r})\boldsymbol{\beta}$, we need to show that, under C1.a and C1.b, it is equivalent to state that the sparse models are equal, also in the composite frequencies, i.e. that $\K=r$, $\ff_{1:\K}=\ll_{1:r}$, $\boldsymbol{ \gamma}=\boldsymbol{\beta}$,  and that 
$\miQ\mihu(\ff_{1:\K})=\miQ\mil_u(\ll_{1:r})$.
 
From the condition that $\miQ \miU_{\TT}(\ff_{1\K} )$ is injective as a map from $\C^{\K}\to\C^{\Ta}$, we have that  $\rank\left\{\miQ \miU_{\TT}(\ff_{1:\K} ) \right\}= \K $. Also, let $\rank\left\{\miQ \miU_{\TT}(\ll_{1:r}) \right\}=r'$ where $r'\leq r$. Without loss of generality, let us identify the $r'$ independent columns of $\miQ \miU_{\TT}(\ll_{1:r})$ with its first $r'$ columns. Then 
$\miQ \miU_{\TT}(\ll_{1:r})\boldsymbol{\beta}=\sum_{k=1}^{r}\miQ\miu_{\TT}(\ll_{k})\beta_k=\sum_{k=1}^{r'}\miQ\miu_{\TT}(\ll_{k})\beta'_k=\miQ \miU_{\TT}(\ll_{1:r'})\boldsymbol{\beta}'$.

Given the sets $\ff_{1:\K}$ and $\ll_{1:r'}$, we denote by $\e$, with $0\leq \e\leq \min\{\K,r'\}$, the number of $\ll_{j_\e}$ with $j_\e\in [\min\{\K,r\}]$ such that it exits a $k \in [\K]$ for which $\ll_{j_\e}=\ff_k$. Furthermore, with no loss of generality, we assume that $j_\e \in [\e]$, i.e., the first $\e$ frequency in $\ll_{1:r'}$ are already included in the first $\e$ elements of $\ff_{1:\K}$. Then,  we have that $\miQ\mil_u(\ll_{1:r})=\miQ \miU_{\TT}(\ll_{1:r'})\boldsymbol{\beta}'=\miQ \miU_{\TT}([\ff_{1:\e},\ll_{\e+1:r'}])\boldsymbol{\beta}'$ where $\boldsymbol{\beta}'$ is partitioned as $\boldsymbol{\beta}'=[\boldsymbol{\beta}^{'\top}_\e, \boldsymbol{\beta}^{'\top}_{\n\e}]^\top$. Similarly $\miQ\mihu(\ff_{1:\K})= \miQ\miU_{\TT}(\ff_{1:\K})\boldsymbol{\gamma}= \miQ\miU_{\TT}([\ff_{1:\e},\ff_{\e+1:\K}])\boldsymbol{\gamma}$ where again we  partition $\boldsymbol{\gamma}=[\boldsymbol{\gamma}^{\top}_\e, \boldsymbol{\gamma}^{\top}_{\n\e}]^\top$. We are now ready to state the equivalence, starting from $\miQ\mihu(\ff_{1:\K})=\miQ\mil_u(\ll_{1:r})$:

\begin{eqnarray}
&&\miQ\miU_{\TT}([\ff_{1:\e},\ff_{\e+1:\K}])\begin{bmatrix}\boldsymbol{\gamma}_\e\\\boldsymbol{\gamma}_{\n\e}\end{bmatrix}-\miQ \miU_{\TT}([\ff_{1:\e},\ll_{\e+1:r'}])\begin{bmatrix}\boldsymbol{\beta}'_\e\\\boldsymbol{\beta}'_{\n\e}\end{bmatrix}=\nonumber\\
&&\miQ\miU_{\TT}([\ff_{1:\e},\ff_{\e+1:\K},\ll_{\e+1:r'}])[\boldsymbol{(\gamma}_\e-\boldsymbol{\beta}'_\e)^\top \boldsymbol{\gamma}_{\n\e}^\top \boldsymbol{\beta}'^\top_{\n\e}]^\top=\mathbf{0}_{\Ta\times 1}\label{condition00}
\end{eqnarray}

Since, by assumption,  we  have that 
$\rank\left\{\miQ \miU_{\TT}([\ff_{1:\K},\ll_{r_\e+1:r'}]) \right\} =\K +\rank\left\{\miQ \miU_{\TT}(\ll_{r_\e+1:r'}) \right\}\\= \rank\left\{\miQ \miU_{\TT}(\ff_{1:\K})\right\}+\rank\left\{\miQ \miU_{\TT}(\ll_{r_\e+1:r'}) \right\}$, which using the result in \cite[eq. (2.19)]{Matsaglia74} implies that $\mathscr{C}(\miQ \miU_{\TT}(\ff_{1:\K}))\cap\mathscr{C}(\miQ \miU_{\TT}(\ll_{r_\e+1:r'}))=\emptyset$, then we have that the last line of \eqref{condition00} only could hold if $\begin{bmatrix}\boldsymbol{(\gamma}_\e-\boldsymbol{\beta}'_\e)^\top&\boldsymbol{\gamma}_{\n\e}^\top&\boldsymbol{\beta}'^\top_{\n\e}\end{bmatrix}^\top=\mathbf{0}_{\K+r'-r_\e\times 1}$. This implies that $\boldsymbol{\gamma}_\e= \boldsymbol{\beta}'_\e$, $\boldsymbol{\gamma}_{\n\e}=\mathbf{0}_{\K-r_\e\times 1}$ and  $\boldsymbol{\beta}'_{\n\e}=\mathbf{0}_{r'-r_\e\times 1}$ which requires  $\K=r_\e=r'$ from which it follows that  $\mihu(\ff_{1:\K})=\mil_u(\ll_{1:r})$. The proof of the converse is trivial.

\section{Proof of Thm. \ref{Thm3}} 
 \label{proofThm3}

We need to proof that if either condition \emph{i)}-\emph{ii)} or \emph{iii)}-\emph{iv)} of Thm. \ref{Thm3} hold, then $\rank\left\{\miQ \miU_{\TT}(\ff_{1:\K} ) \right\}= \K $ and also $\rank\left\{\miQ \miU_{\TT}([\ff_{1:\K}\ll_{1:r}] ) \right\}= \K+\rank\left\{\miQ \miU_{\TT}(\ll_{1:r}) \right\}$ assuming that none of the frequencies among the sets $\ff_{1:\K}$ and $\ll_{1:r}$ coincide. We start with condition \emph{i)}-\emph{ii)}. 

Given that from \emph{ii)}, $\miQ$  admits a left pseudo inverse, using \cite[Cor. 6.1 (eq (3.10))]{Matsaglia74} we have that, for any set of frequencies $\ff_{1:\K}$ satisfying  A.\ref{A1} and $\ll_{1:r}$ being any arbitrary set in $\T^{\ddl\times r}$, then $\rank\left\{\miQ \miU_{\TT}(\ff_{1:\K} ) \right\}=\rank\left\{\miU_{\TT}(\ff_{1:\K}) \right\}$, $\rank\left\{\miQ \miU_{\TT}([\ff_{1:\K}\ll_{1:r}] ) \right\}=\rank\left\{\miU_{\TT}([\ff_{1:\K}\ll_{1:r}] ) \right\}$ and $\rank\left\{\miQ \miU_{\TT}(\ll_{1:r}) \right\}=\rank\left\{\miU_{\TT}(\ll_{1:r})\right\}$.

Next using \emph{i)} for any value of $r$, from Cor. \ref{Prop3} then $\rank\left\{\miU_{\TT}(\ff_{1:\K} )\right\}=\K$, and we have:
 \begin{equation*}
 \rank\left\{\miQ \miU_{\TT}(\ff_{1:\K})\right\}=\rank\left\{\miU_{\TT}(\ff_{1:\K} )\right\}=\K
 \end{equation*}
 Furthermore, if none of the frequencies among the sets $\ff_{1:\K}$ and $\ll_{1:r}$ coincide (also element-wise due to $\ff_{1:\K}$ being drawn following A.\ref{A1}), using the condition  $\sum_{i=1}^{\ddl} \Tsf_i\geq\K+r+(\ddl-1)$ to call Cor. \ref{Prop3}:
 \begin{equation}
 \begin{split}
 &\rank\left\{\miQ \miU_{\TT}([\ff_{1:\K}\ll_{1:r}] ) \right\}=  \rank\left\{\miU_{\TT}([\ff_{1:\K}\ll_{1:r}] ) \right\}\\
 & = \K+r = \K+\rank\left\{\miQ\miU_{\TT}(\ll_{1:r})\right\}
 \end{split}
 \nonumber
 \label{eq_composition}
 \end{equation}

Finally we focus on condition \emph{iii)}-\emph{iv)}. To prove that $\rank\left\{\miQ \miU_{\TT}(\ff_{1:\K} ) \right\}= \K$, we first identify the SVD of the atom matrix $\miU_{\TT}(\ff_{1:\K})=\miV_{\text{lft}}\boldsymbol{\Sigma}\miV^\H_{\text{rgt}}$ which has $\rank$ $\K$, where $\miV_{\text{lft}}\in\C^{\Tu\times\Tu}$ and $\miV_{\text{rgt}}\in\C^{\K\times\K}$ are unitary matrices and $\boldsymbol{\Sigma}\in\R^{\Tu\times\K}$ is a rectangular matrix whose main diagonal contains the singular values of $\miU_{\TT}(\ff_{1:\K})$ and the rest of the elements are zero. Using \emph{iv)} it is easy to show that:
\begin{eqnarray}\rank\left\{\miQ \miU_{\TT}(\ff_{1:\K} ) \right\}&=& \rank\left\{\miQ_1 \miQ_2 \miV_{\text{lft}}\boldsymbol{\Sigma}\miV^\H_{\text{rgt}} \right\}\nonumber\\&=&\rank\left\{\miQ_2 \miV_{\text{lft}}\boldsymbol{\Sigma}\right\}=\rank\left\{\boldsymbol{\Sigma}\right\}=\K
\nonumber
\end{eqnarray}

Similarly we define the SVD of extended atom matrix $\miU_{\TT}([\ff_{1:\K}\ll_{1:r}]$ that by using \emph{iii)} and Cor. \ref{Prop3} has $\rank$ $\K+r$ as
$\miU_{\TT}([\ff_{1:\K}\ll_{1:r}]=\miZ_{\text{lft}}\boldsymbol{\Lambda}\miZ^\H_{\text{rgt}}$, where $\miZ_{\text{lft}}\in\C^{\Tu\times\Tu}$ and $\miZ_{\text{rgt}}\in\C^{\K+r\times\K+r}$ are unitary matrices and $\boldsymbol{\Lambda}\in\R^{\Tu\times\K+r}$ is a rectangular matrix whose main diagonal contains the singular values and the rest of the elements are zero. Finally, using \emph{iv)} is easy to show that:
\begin{eqnarray}\rank\left\{\miQ \miU_{\TT}([\ff_{1:\K}\ll_{1:r}]\right\}&=& \rank\left\{\miQ_1 \miQ_2 \miZ_{\text{lft}}\boldsymbol{\Lambda}\miZ^\H_{\text{rgt}} \right\}\nonumber
\\&=&\rank\left\{\miQ_2 \miZ_{\text{lft}}\boldsymbol{\Lambda}\right\}=\rank\left\{\boldsymbol{\Lambda}\right\}\nonumber\\&=&\K+r=\K+\rank\left\{\miQ \miU_{\TT}(\ll_{1:r}) \right\}
\nonumber
\end{eqnarray}

\section{Proof of Thm. \ref{Thm4}} 
 \label{proofThm4}

To proof Thm. \ref{Thm4} we follow a similar approach as in 
the proof Thm. \ref{Thm3}, taking into account now that $\miQ$ follows a structure as in \eqref{Q_def} with $\miQ=\left(\miP^\top\otimes\miI_\Na\right)\AL$. Given that $\AL\in\mathcal{A}_\TT(\Na\Ma,\Tu)$, using \eqref{struct2}, we can write $\AL \miU_{\TT}(\ff_{1:\K} )$ and $\AL\miU_{\TT}([\ff_{1:\K}\ll_{1:r}] )$ as follows: 
\begin{eqnarray}
\AL \miU_{\TT}(\ff_{1:\K} )
&=&
\Pint^{\mathsf{L}\H} \begin{bmatrix}
\miU_\SS(\mii_{1:\K}) \miD  \\
\miB
\end{bmatrix}
\\
\AL\miU_{\TT}([\ff_{1:\K}\ll_{1:r}] )
&=&
\Pint^{\mathsf{L}\H} \begin{bmatrix}
\miU_\SS([\mii_{1:\K}\ii_{1:r}] ) \miC \\  \miG
\end{bmatrix}\nonumber
\label{MAIPIU2}
\end{eqnarray}

Furthermore, using \emph{i)} or \emph{iii)} of Thm. \ref{Thm4}, it is easy to see then that $\rank\{\AL\miU_{\TT}([\ff_{1:\K}\ll_{1:r}] )\}=\K$ and that $\rank\{\AL\miU_{\TT}([\ff_{1:\K}\ll_{1:r}] )\}=\K+r$.
Finally using similar derivations and decompositions as for the proof of Thm. \ref{Thm3}
substituting now the $\miQ$ in proof of Thm. \ref{Thm3} by $\left(\miP^\top\otimes\miI_\Na\right)$, it is straightforward to show that $\rank\{\left(\miP^\top\otimes\miI_\Na\right)\AL\miU_{\TT}([\ff_{1:\K}\ll_{1:r}] )\}=\K$ and that $\rank\{\left(\miP^\top\otimes\miI_\Na\right)\AL\miU_{\TT}([\ff_{1:\K}\ll_{1:r}] )\}=\K+\rank\left\{\miQ \miU_{\TT}(\ll_{1:r}) \right\}$.

\section{Proof of Thm. \ref{Thm1}} 
 \label{proofThm1}

In the following, we will prove that given the assumptions of Theorem \ref{Thm1}, i.e. $ \K< \Tsf_{\ddl}$ and that the linear mapping that models the measurement process $\miQ$ satisfies C.1 of Thm. \ref{uniqueness}, denoting by $(r^\opt,\mil_u^\opt,\miT_\TT^\opt)$ the optimal solution  to \eqref{ec_Thm1}, is  unique in terms of $\mil_u^\opt=\mihu=\sum_{k=1}^K\gamma_k\miu_\TT(\ff_k)$ and in terms of the frequencies $\ff_{1:\K}$  identified using \refAlg1.
 
If is easy to show that a feasible solution  for \eqref{ec_Thm1} is
$\bigg(K,\sum_{k=1}^K\gamma_k\miu_\TT(\ff_k),\sum_{k=1}^K|\gamma_k|\miu_\TT(\ff_k)\miu_\TT(\ff_k)^\H\bigg)$, therefore $r^\circ\triangleq\rank \left \{ \miT_\TT^\opt\right\} \leq K< \Tsf_{\ddl}$.
Given also that  $\miT_{\TT}^\opt\in\mathcal{T}^{\ddl}_\TT$, we can apply Lemma \ref{Span_condition} to the optimal model $\mil_u^\opt$  and therefore we have that  $\mil_u^\opt\left(\ll_{1:r^\opt}\right)=\sum_{k=1}^{r^\opt}\beta_k\miu_\TT\left(\ll_{k}\right)$. 

The optimal model $\mil_u^\opt\left(\ll_{1:r^\opt}\right)$ is unique, i.e. $\mil_u^\opt\left(\ll_{1:r^\opt}\right)=\mihu(\ff_{1:\K})$, since $\miQ\mihu(\ff_{1:\K})=\miQ\mil_u^\opt\left(\ll_{1:r^\opt}\right)=\miy$ and the conditions of Thm. \ref{uniqueness} hold. Recall that from Thm. \ref{uniqueness} we have that $\K=r^\opt$, and  $\ff_{1:\K}=\ll_{1:r^\opt}$.

We finally show that  $\miT_{\TT}^\opt$ admits a unique Vandermonde decomposition in order to uniquely identify the frequencies $\ff_{1:\K}$. With this aim, let us consider the first $\Tsf_{\ddl}$ components of $\mil_u^\opt=\mihu$. They identify a $\Tsf_{\ddl}$-D vector that we  denote by $\mil_u^{\opt(1:\Tsf_{\ddl})}$ and $\mihu^{(1:\Tsf_{\ddl})}$, that necessarily needs to be equal, i.e. $\mil_u^{\opt(1:\Tsf_{\ddl})}=\mihu^{(1:\Tsf_{\ddl})}$. From \eqref{decomp}, we have that:
\begin{eqnarray}
\mil_u^{\opt(1:\Tsf_{\ddl})}&=&\sum_{i_{\ddl}=1}^{r^\opt_{\ddl}} \miu_{\Tsf_{\ddl}}(\nu^{\ddl}_{i_{\ddl}}) \sum_{i_1=1}^{r^\opt}\dots\sum_{i_{\ddl-1}=1}^{r^\opt_{\ddl-1}}\alpha_{i_1}p_{i_{\ddl}}\prod_{s=1}^{\ddl}o^{s}_{i_{s}i_{s-1}}\nonumber\\
&=&\sum_{i_{\ddl}=1}^{r^\opt_{\ddl}}\eta_{i_{\ddl}} \miu_{\Tsf_{\ddl}}(\nu^{\ddl}_{i_{\ddl}})=\mihu^{(1:\Tsf_{\ddl})}=\sum_{k=1}^\K\gamma_k \miu_{\Tsf_{\ddl}}(\ell^{\ddl}_k)
\nonumber\end{eqnarray}
which cannot be true unless $\K=r^\opt_{\ddl}$. Finally, since we also have that $\K=r^\circ\geq r^\opt_2\geq\dots\geq r^\opt_{\ddl}$, it is true then that $\K=r^\circ= r^\opt_2=\dots= r^\opt_{\ddl}$. 
From this it follows that $\rank\left\{\miT_{\TT}^\opt\right\}=\rank\left\{\miT_{\mathbf{0}\Tsf_{\ddl}}\right\}=\K$. This combined with the fact that $\K < \Tsf_{\ddl}$ and $\miT_{\TT}^\opt$ belongs to  $\mathcal{T}^{\ddl}_\TT$, by Lemma \ref{PrettyLemma}, $\miT_{\TT}^\opt$ admits a unique Vandermonde decomposition of order $\K$ from which the set of frequencies $\ff_{1:\K}$ can be uniquely determined.

\section{Proof of Thm. \ref{Thm2}} 
 \label{proofThm2}

We proof that given the assumptions of Thm. \ref{Thm2}, i.e. the measurement matrix $\miQ$, and the \atom set $\U^\F_\TT$, satisfies  Condition C.1 in Thm. \ref{uniqueness}  for all $r \leq \rank \left \{ \miT_{\TT}^\opt\right\}$ and if we have $\rank \left \{ \miT_{\TT}^\opt\right\}<  \Tsf_{\ddl}$ and $ \K<  \Tsf_{\ddl}$, then the optimal solution $(t^\opt,\mil_u^\opt,\miT_{\TT}^\opt)$ to \eqref{ec_Thm2} uniquely identifies the sparse channel model, and the frequencies.

Given that  $\miT_{\TT}^\opt\in\mathcal{T}^{\ddl}_\TT$ and that  by assumption $r^\opt\triangleq\rank \left \{ \miT_{\TT}^\opt\right\}<  \Tsf_{\ddl}$, we can apply Lemma \ref{Span_condition} to the optimal model $\mil_u^\opt$  and therefore we have that  $\mil_u^\opt\left(\ll_{1:r^\opt}\right)=\sum_{k=1}^{r^\opt}\beta_k\miu_\TT\left(\ll_{k}\right)$. To prove the uniqueness of this optimal model, i.e. that $\mil_u^\opt\left(\ll_{1:r^\opt}\right)=\mihu(\ff_{1:\K})$ and that $\K=r^\opt$, and  $\ff_{1:\K}=\ll_{1:r^\opt}$, we use C.1 of Thm.\ref{uniqueness}. 

To finally prove that  $\miT_{\TT}^\opt$ admits a unique Vandermonde decomposition in order to uniquely identify the frequencies $\ff_{1:\K}$ we follow a similar reasoning as in the Proof of Thm. \ref{Thm1} using that in this case, we also have that $\rank \left \{ \miT_{\TT}^\opt\right\}<  \Tsf_{\ddl}$.


\bibliographystyle{ieeetr}
\bibliography{Bibliography/Bell-Labs_bib,Bibliography/mimo,Bibliography/cell-free,Bibliography/AoA,Bibliography/compress-sensing,Bibliography/mmWave,Bibliography/hybridPrecod,Bibliography/multiuser_IT,Bibliography/mati,Bibliography/J-STSP-RAARSP-00205-2020_bib,Bibliography/channel_estimation,Bibliography/alvaro,Bibliography/Matrices,Bibliography/Spectral_analysis}

\begin{thebibliography}{10}

\bibitem{Akyildiz20}
I.~F. Akyildiz, A.~Kak, and S.~Nie, ``{6G} and beyond: The future of wireless
  communications systems,'' {\em IEEE Access}, vol.~8, pp.~133995--134030,
  2020.

\bibitem{Lee16}
J.~Lee, G.-T. Gil, and Y.~H. Lee, ``Channel estimation via {O}rthogonal
  {M}atching {P}ursuit for hybrid {MIMO} systems in millimeter wave
  communications,'' {\em IEEE Transactions on Communications}, vol.~64, no.~6,
  pp.~2370--2386, 2016.

\bibitem{Bajwa10}
W.~U. Bajwa, J.~Haupt, A.~M. Sayeed, and R.~Nowak, ``{Compressed Channel
  Sensing: A New Approach to Estimating Sparse Multipath Channels},'' {\em
  Proceedings of the IEEE}, vol.~98, no.~6, pp.~1058--1076, 2010.

\bibitem{Zhang24}
X.~Zhang, H.~Zhang, and Y.~C. Eldar, ``Near-field sparse channel representation
  and estimation in 6g wireless communications,'' {\em IEEE Transactions on
  Communications}, vol.~72, no.~1, pp.~450--464, 2024.

\bibitem{Gao19}
X.~Gao, L.~Dai, S.~Zhou, A.~M. Sayeed, and L.~Hanzo, ``Wideband beamspace
  channel estimation for millimeter-wave mimo systems relying on lens antenna
  arrays,'' {\em IEEE Transactions on Signal Processing}, vol.~67, no.~18,
  pp.~4809--4824, 2019.

\bibitem{Schniter14}
P.~Schniter and A.~Sayeed, ``Channel estimation and precoder design for
  millimeter-wave communications: The sparse way,'' in {\em 2014 48th Asilomar
  Conference on Signals, Systems and Computers}, pp.~273--277, 2014.

\bibitem{Assalini09}
A.~Assalini, E.~Dall'Anese, and S.~Pupolin, ``Linear mmse mimo channel
  estimation with imperfect channel covariance information,'' in {\em 2009 IEEE
  International Conference on Communications}, pp.~1--5, 2009.

\bibitem{Liao15}
W.~Liao, ``{MUSIC} for multidimensional spectral estimation: Stability and
  super-resolution,'' {\em IEEE Transactions on Signal Processing}, vol.~63,
  no.~23, pp.~6395--6406, 2015.

\bibitem{Swapna22}
S.~Swapna and M.~Ramarakula, ``Modified simultaneous weighted – {OMP} based
  channel estimation and hybrid precoding for massive {MIMO} systems,'' in {\em
  2022 International Conference on Wireless Communications Signal Processing
  and Networking (WiSPNET)}, pp.~221--225, 2022.

\bibitem{Chandrasekaran2012}
V.~Chandrasekaran, B.~Recht, P.~A. Parrilo, and A.~S. Willsky, ``The convex
  geometry of linear inverse problems,'' {\em Foundations of Computational
  Mathematics}, vol.~12, no.~6, pp.~805--849, 2012.

\bibitem{Chi-2020}
Y.~{Chi} and M.~{Ferreira Da Costa}, ``Harnessing sparsity over the continuum:
  Atomic norm minimization for superresolution,'' {\em IEEE Signal Processing
  Magazine}, vol.~37, pp.~39--57, Mar. 2020.

\bibitem{Heckel18}
R.~Heckel and M.~Soltanolkotabi, ``Generalized line spectral estimation via
  convex optimization,'' {\em IEEE Transactions on Information Theory},
  vol.~64, pp.~4001--4023, June 2018.

\bibitem{Yang16}
Z.~Yang, L.~Xie, and P.~Stoica, ``Vandermonde decomposition of multilevel
  {Toeplitz} matrices with application to multidimensional super-resolution,''
  {\em IEEE Transactions on Information Theory}, vol.~62, pp.~3685--3701, June
  2016.

\bibitem{Sanchez-Fernandez21}
M.~Sánchez-Fernández, V.~Jamali, J.~Llorca, and A.~M. Tulino, ``Gridless
  multidimensional angle-of-arrival estimation for arbitrary {3D} antenna
  arrays,'' {\em IEEE Transactions on Wireless Communications}, vol.~20, no.~7,
  pp.~4748--4764, 2021.

\bibitem{Vega21}
A.~Vega~Delgado, M.~Sánchez-Fernández, L.~Venturino, and A.~Tulino,
  ``Super-resolution in automotive pulse radars,'' {\em IEEE Journal of
  Selected Topics in Signal Processing}, vol.~15, no.~4, pp.~913--926, 2021.

\bibitem{Wu17b}
N.~Wu, F.~Zhu, and Q.~Liang, ``Evaluating spatial resolution and channel
  capacity of sparse cylindrical arrays for massive {MIMO},'' {\em IEEE
  Access}, vol.~5, pp.~23994--24003, 2017.

\bibitem{Hu18}
S.~Hu, F.~Rusek, and O.~Edfors, ``Beyond massive {MIMO}: The potential of
  positioning with large intelligent surfaces,'' {\em IEEE Transactions on
  Signal Processing}, vol.~66, pp.~1761--1774, April 2018.

\bibitem{CSbook}
S.~Foucart and H.~Rauhut, {\em A mathematical introduction to compressive
  sensing}.
\newblock Springer, 2013.

\bibitem{Tibshirani96}
R.~Tibshirani, ``Regression shrinkage and selection via the {Lasso},'' {\em
  Journal of the Royal Statistical Society. Series B (Methodological)},
  vol.~58, no.~1, pp.~267--288, 1996.

\bibitem{Chi19}
Y.~Chi and M.~F.~D. Costa, ``{Harnessing Sparsity Over the Continuum},'' {\em
  IEEE Signal Processing Magazine}, vol.~37, no.~2, pp.~39--57, 2020.

\bibitem{Caratheodory11}
C.~Carath\'{e}odory and L.~Fej\'{e}r, ``\"uber den {Zusammenhang} der
  {Extremen} von harmonischen {Funktionen} mit ihren {Koeffizienten} und \"uber
  den {Picard-Landauschen Satz},'' {\em Rendiconti del Circolo Matematico di
  Palermo (1884-1940)}, vol.~32, no.~1, p.~218–239, 1911.

\bibitem{Sidiropoulos01}
N.~D. Sidiropoulos, ``{Generalizing Carathéodory's Uniqueness of Harmonic
  Parameterization to $N$ Dimensions},'' {\em IEEE Transactions on Information
  Theory}, vol.~47, no.~4, pp.~1687--1690, 2001.

\bibitem{Tulino04b}
A.~M. Tulino and S.~Verdú, ``Random matrix theory and wireless
  communications,'' {\em Foundations and Trends® in Communications and
  Information Theory}, vol.~1, no.~1, pp.~1--182, 2004.

\bibitem{Tulino13}
A.~M. Tulino, G.~Caire, S.~Verdú, and S.~Shamai, ``Support recovery with
  sparsely sampled free random matrices,'' {\em IEEE Transactions on
  Information Theory}, vol.~59, no.~7, pp.~4243--4271, 2013.

\bibitem{Anderson14}
G.~W. Anderson and B.~Farrell, ``Asymptotically liberating sequences of random
  unitary matrices,'' {\em Advances in Mathematics}, vol.~255, pp.~381--413,
  2014.

\bibitem{Grossi20}
E.~{Grossi}, M.~{Lops}, and L.~{Venturino}, ``Adaptive detection and
  localization exploiting the {IEEE} 802.11ad standard,'' {\em IEEE
  Transactions on Wireless Communications}, vol.~19, no.~7, pp.~4394--4407,
  2020.

\bibitem{Meyer73}
C.~D. Meyer, Jr., ``{Generalized Inverses and Ranks of Block Matrices},'' {\em
  SIAM Journal on Applied Mathematics}, vol.~25, no.~4, pp.~597--602, 1973.

\bibitem{Gurvits_prueba}
L.~Gurvits and H.~Barnum, ``{Largest separable balls around the maximally mixed
  bipartite quantum state},'' {\em Physical Review A}, vol.~66, no.~6,
  p.~062311, 2002.

\bibitem{Matsaglia74}
G.~Matsaglia and G.~P.~H. Styan, ``{Equalities and Inequalities for Ranks of
  Matrices},'' vol.~2 of {\em Linear and Multilinear Algebra}, pp.~269--292,
  1974.

\end{thebibliography}

\end{document}